\documentclass[prd,aps
,tightenlines,
superscriptaddress,showpacs
,nofootinbib
,eqsecnum,amsfonts,amsmath,floatfix
]{revtex4}
\usepackage{latexsym}
\usepackage{amssymb}
\usepackage{amsfonts}
\usepackage{amsmath}
\usepackage{bm}
\usepackage[dvips]{graphicx}
\usepackage[usenames]{color}
\usepackage{ulem}
\usepackage{multirow}
\allowdisplaybreaks

\renewcommand{\a}{\alpha}
\renewcommand{\b}{\beta}
\renewcommand{\c}{\gamma}

\newcommand{\e}{\epsilon}

\newcommand{\m}{\mu}
\newcommand{\n}{\nu}
\renewcommand{\t}{\tau}
\newcommand{\z}{\omega}
\newcommand{\G}{\Gamma}
\newcommand{\simgt}{\lower.5ex\hbox{$\; \buildrel > \over \sim \;$}}
\newcommand{\simlt}{\lower.5ex\hbox{$\; \buildrel < \over \sim \;$}}

\begin{document}

\title{Spherical harmonic modes  of 5.5 post-Newtonian gravitational wave polarizations and associated factorized resummed waveforms for a particle in circular orbit around a Schwarzschild black hole}

  \author{Ryuichi \surname{Fujita}}
  \author{Bala R.~\surname{Iyer}}
  \affiliation{Raman Research Institute, Bangalore 560 080, India}

\begin{abstract}
Recent breakthroughs in numerical relativity enable one to 
examine the validity of  the post-Newtonian expansion in the late stages 
of inspiral. 
For the comparison between post-Newtonian (PN) expansion and 
numerical simulations, the waveforms in terms of the 
spin-weighted spherical harmonics are more useful than 
the  plus and cross polarizations, which are 
used for data analysis of gravitational waves. 
Factorized resummed waveforms 
achieve better agreement with numerical 
results than the conventional Taylor expanded post-Newtonian waveforms. 
In this paper, 
we revisit the post-Newtonian expansion
of gravitational waves for a test-particle of mass $\m$ in circular orbit
of radius $r_0$ around a Schwarzschild black hole of mass $M$ 
and derive the spherical 
harmonic components associated with the gravitational wave polarizations
up to order $v^{11}$  beyond Newtonian. 
Using the more accurate $h_{\ell m}$'s computed in this work, we provide
the more complete set of associated $\rho_{\ell m}$'s and $\delta_{\ell m}$'s 
that form important bricks in the factorized resummation of waveforms 
with potential applications for the construction of further improved waveforms 
for prototypical compact binary sources in the future. 
We also provide ready-to-use expressions of the 5.5PN gravitational waves polarizations 
$h_+$ and $h_\times$   in the test-particle limit for 
gravitational wave data analysis applications.
Additionally, we  provide closed analytical expressions for 
2.5PN $h_{\ell m}$, 2PN $\rho_{\ell m}$ and  3PN $\delta_{\ell m}$, 
for general multipolar orders $\ell$ and $m$ in the test-particle limit. 
Finally, we also examine the implications of the present analysis 
for compact binary sources in Laser Interferometer Space Antenna.
\end{abstract}
\date{\today}
  \pacs{
    04.25.Nx,   
    04.30.-w,   
    04.30.Db    
  }
\maketitle
\section{Introduction}
One of the most important sources of gravitational waves (GW) for 
the laser interferometer detectors is the inspiral and merger of 
a compact binary systems. To extract physical information of the 
source, accurate and efficient theoretical templates are needed 
to be matched with observed data. The early inspiral phase 
is accurately described by the analytic post-Newtonian (PN) approximation~\cite{dis,biops}, 
while the late inspiral and the subsequent merger phases are described by 
a full numerical solution of the Einstein equations. 

Since the recent breakthroughs in numerical relativity (NR)
~\cite{NR05p,NR05c,NR05b,boyleetal07}, 
a number of the simulations have computed gravitational waves through 
inspiral, merger and ringdown phases. 
Among them, comparisons between the PN and NR waveforms have been done 
very accurately (See e.g. recent reviews Ref.~\cite{Hannam09,Hinder10}).
One can use the comparison 
to investigate the region of validity of the  post-Newtonian approximation 
in the inspiral phase. 
Additionally, it is also important to investigate 
whether higher post-Newtonian terms broaden the region of validity 
because computational cost of NR simulations is very high. 
The comparisons show that we need to include high PN corrections~\cite{dn,betal09}. 

The PN approximation is not expected to model merger and ringdown due to
the  break-down of the adiabatic approximation or  also in some cases
due to the breakdown of the  monotonicity of frequency evolution.
 {\it Resummation methods} like  Pad\'e approximants~\cite{Damour:1997ub}
can be used to extend the
 numerical  validity of PN expansions (at least) up to the last stable orbit
(LSO).
The  effective-one-body (EOB) approach~\cite{bd} is a new resummation
to extend  validity of suitably resummed PN results
beyond the LSO, and up to the merger.
The EOB analytically provides the complete GW signal emitted by inspiralling,
 plunging, merging and ringing binary black holes.
By {\it flexing} it in the  parameters carefully chosen to 
characterize physical effects beyond what is currently analytically computed,
the EOB can be further improved and calibrated to Numerical Relativity 
simulations.
Improved EOB models~\cite{Damour:2007xr,DIN,dn,betal09,PBBCKPS09} are based on
 a {{\it multiplicative}} decomposition of the multipolar
waveform $h_{lm}$ into a {{\it product}} of
the Newtonian waveform $h_{\ell m}^{\rm (N)}$
and a PN-correction factor ${\hat{h}^{(\epsilon)}_{\ell m}}$
which is a{ {\it product of four factors}}
${\hat{h}_{\ell m}^{(\epsilon)}} ={ \hat{S}_{\rm  eff}^{(\epsilon)}}
{ T_{\ell m}}
 e^{i\delta_{\ell m}} {\rho_{\ell m}^\ell},$
with structure  $1+{\cal O}(x)$.
The choice of factors,  based on a  physical understanding of
the main  effects influencing the final waveform, facilitates
a graded improvement of the analytical waveform and possibility
of refinement by match to improved numerical relativity results.
The comparison of NR waveforms with the analytical EOB 
waveforms is currently a very active  area of research. In 
particular, the comparison via the factorized resummation of $h_{\ell m}$ has
been very successful and benefits from inclusion of higher order
multipoles (higher $\ell$) and  {\it hybridisation} using test-particle
results for higher PN orders and this provides the dominant motivation 
for the present investigation.

In this paper, we derive the post-Newtonian expansion
of gravitational waves for a test-particle of mass $\m$
in circular orbit around a Schwarzschild black hole of mass $M$. 
Gravitational waves can be computed using the black hole perturbation 
formalism.  The perturbation of the Schwarzschild black hole can be treated
by two different methods.   The first deals with  metric perturbations 
of the Schwarzschild black hole and the second with the perturbation of curvature tensors. 
For the Schwarzschild case, the master equation for the metric perturbations was derived by 
Regge and Wheeler for the odd parity mode, and Zerilli for the even parity 
mode~\cite{RW,Zerilli}. 
For the Kerr case however, we do not have such a master equation for the metric perturbations.
For the curvature perturbations the master equation is known for
both the cases and  was derived by  Bardeen and Press for a Schwarzschild black hole, 
and Teukolsky  for a Kerr black hole~\cite{BP,Teukolsky:1973ha}. 
Since in the future we would like to extend the present results to the case of a  Kerr black hole, 
in the current work we employ the Teukolsky equation to compute the 
gravitational waves for the Schwarzschild case also. 

The Teukolsky equation is the fundamental equation in 
black hole perturbation formalism. 
Although it is limited to the  test-particle limit, 
black hole perturbation formalism has the  big advantage 
that one can go to higher post-Newtonian orders systematically. 
For a particle in circular orbit around a Schwarzschild 
black hole~\cite{ref:TS,ref:TTS},  
the gravitational waveforms and energy flux to infinity are known up 
to $v^{8}$ and $v^{11}$ respectively. 
For a particle in circular and equatorial orbit around a Kerr 
black hole~\cite{Poisson_prd48,ref:TSTS}, however,
 (see for e.g. review Ref.~\cite{ST}) 
gravitational waveforms (energy flux to infinity) are known up 
to $v^{3}$ ($v^{8}$).
For the general mass ratio  nonspinning compact binaries
in quasicircular orbits the amplitude (orbital phase) of gravitational waves 
are known up to $v^{6}$ ($v^{7}$)
~\cite{djs,bde,bfij,bdei,K07,BFIS08,Favata09} 
(see e.g. review Ref.~\cite{post_Newton}). 
In this case for spinning precessing compact binaries
in quasicircular orbits, 
the amplitude (orbital phase) of gravitational waves 
are known up to $v^{3}$ ($v^{5}$)~\cite{K95,WW96,BBF06,FBB06,ABFO09}.
Lastly, for the nonprecessing case however, the
gravitational waveforms in amplitude are known through $v^3$ and $v^4$
for spin-orbit and spin(1)-spin(2) effects respectively.
In the case of the test-particle limit, 
there is a rather large gap of post-Newtonian order between waveforms 
and energy flux mainly because it has not been needed until recently
in connection with  the comparison and matching
of PN  analytical waveforms with waveforms from high accuracy
numerical simulations. 

In this work, taking account of the necessity for the comparison of 
waveforms between post-Newtonian approximation and numerical relativity, 
we improve on the accuracy  for gravitational waveforms and  consider 
gravitational waveforms also up to order $v^{11}$  beyond Newtonian,
i.e. 5.5PN. 
We derive 5.5PN waveforms projected 
onto spin-weighted spherical harmonics since they form the basis
for  the comparison of analytical computations
with the results of numerical simulations. 
The central object in this treatment is the function $Z_{\ell m\omega}$ and 
we provide its 2.5PN accurate analytical expression  for 
{\it arbitrary multipolar orders}  $\ell$ and $m$. 
To facilitate and improve existing works on the factorized resummation of
the gravitational waveform,
we also provide the $\rho_{\ell m}$ and $\delta_{\ell m}$
(see Sec.~\ref{sec:rholm}) to orders
consistent with our new improved 5.5PN GW polarizations.
En route to the above results, our present work extends the 1PN results
for even $h_{\ell m}$'s~\cite{K07} and odd $h_{\ell m}$'s~\cite{DIN}
by   providing  {\it closed analytical expressions} for 
$h_{\ell m}$, $\rho_{\ell m}$ and  $\delta_{\ell m}$ 
up to $O(v^{5})$, $O(v^{4})$ and $O(v^{6})$ respectively 
for {\it general multipolar orders} $\ell$ and $m$. 
Finally, we also examine the implications of the present analysis 
for compact binary sources in Laser Interferometer Space Antenna (LISA).

This paper is organized as follows. 
In Sec.~\ref{sec:formula}, we describe  the general formalism 
and relevant formulas that underlie the present work.
In Sec.~\ref{sec:MST}, we describe the Mano, Suzuki and Takasugi 
method~\cite{ref:MST,ref:MSTR} which is used in this paper 
to solve the Teukolsky equation.  
In particular, employing this formalism, in Sec.~\ref{sec:zlmw2.5PN}, 
we derive the 2.5PN accurate solution of the Teukolsky solution 
$Z_{\ell m \omega}$ for arbitrary multipolar orders $\ell$ and $m$. 
In Sec.~\ref{sec:hlm}, we  compute the gravitational waveforms expressed 
as spherical harmonic modes at 5.5PN  order.
In Sec.~\ref{sec:rholm}, the $\rho_{\ell m}$ and $\delta_{\ell m}$ 
needed for the construction of factorized resummed waveforms at 5PN are 
computed.
For general multipolar orders $\ell$ and $m$, in Sec.~\ref{sec:hlm}, 
we  exhibit  closed analytical forms  for $h_{\ell m}$ at 2.5PN, 
while in Sec.~\ref{sec:rho2PN} we provide ready-to-use expressions for
$\rho_{\ell m}$ at 2PN and $\delta_{\ell m}$ at 3PN. 
In Sec.~\ref{sec:zetalm}, we provide general formulas to compute  
5.5PN polarization modes  starting from  the explicit expressions 
of spherical harmonic  modes at 5.5PN in Sec.~\ref{sec:hlm} 
and Sec.~\ref{sec:rholm}.
In Sec.~\ref{sec:numerical}, we compare the results from our 5.5PN
approximation with the results from a numerical calculation, 
obtained by solving the Teukolsky equation~\cite{FT1,FT2}. 
Section~\ref{sec:summary} is devoted to a summary of the paper. 
The paper ends with three Appendices. 
In Appendices~\ref{sec:55PN} and \ref{sec:rho5PN}, 
we list $\hat{H}_{\ell m}$ and $\rho_{\ell m}$ 
for higher values of $\ell$ (consistent with 5.5PN GW polarizations)
than listed in the main text. 
And finally, in Appendix~\ref{sec:55PNpol} we list the complete
5.5PN GW polarizations $H_{+,\,\times}$ in the test-particle limit.
Throughout this paper, we use the units of $c=G=1$.
\section{General formulation}
\label{sec:formula}
In the Teukolsky formalism, 
the gravitational perturbation of a Kerr black hole is described in terms of 
the Newman-Penrose variables, $\Psi_0$ and $\Psi_4$, which satisfy  a master equation. In this section we recall the relevant equations needed in this
work following the notation in~\cite{ST}.
The Weyl scalar $\Psi_4$
is related to the amplitude of the gravitational wave at infinity as
\begin{eqnarray}
\Psi_4\rightarrow\frac{1}{2}(\ddot{h}_{+}-i\,\ddot{h}_{\times }),\,\,\,{\rm for}\,\,\,r\rightarrow\infty, 
\end{eqnarray}
where dot, $\dot{\,}$, denotes time derivative, $d/dt$.

The master equation for $\Psi_4$
can be separated into radial and angular parts if we
expand $\Psi_4$ in Fourier harmonic modes as
\begin{eqnarray}
\rho^{-4} \Psi_4=\displaystyle \sum_{\ell,m}\int_{-\infty}^{\infty} d\omega 
e^{-i\omega t} R_{\ell m\omega}(r)\ _{-2}S_{\ell m}^{a\omega}(\theta,\varphi),
\end{eqnarray}
where $\rho=(r-i a \cos\theta)^{-1}$, 
$M$ and $aM$ are the mass and angular momentum of the black hole, 
respectively 
and the angular function $_{-2}S_{\ell m}^{a\omega}(\theta,\varphi)$ 
is the spin-weighted spheroidal harmonic with spin $s=-2$, 
\begin{eqnarray}
\ _{-2}S_{\ell m}^{a\omega}(\theta,\varphi) = \frac{1}{\sqrt{2\pi}}\ _{-2}\tilde{S}_{\ell m}^{a\omega}(\theta) e^{i m \varphi}, 
\end{eqnarray}
where $\ _{-2}\tilde{S}_{\ell m}^{a\omega}(\theta)$ satisfies 
the angular Teukolsky equation and 
which is normalized as 
\begin{eqnarray}
\int_0^{2\pi} d\varphi\int_0^{\pi} d\theta\sin\theta | \ _{-2}S_{\ell m}^{a\omega}(\theta,\varphi)|^2 =1. 
\end{eqnarray}

In the following, we focus on the Schwarzschild black hole case. 
The spin-weighted spheroidal harmonic 
$_{-2}S_{\ell m}^{a\omega}(\theta,\varphi)$ 
is then  reduced to the spin-weighted spherical harmonic 
$_{-2}Y_{\ell m}(\theta,\varphi)$, 
whose definition is given by~\cite{K07} 
\begin{widetext}
\begin{eqnarray}
_{s}Y_{\ell m}(\theta,\varphi)&=&(-1)^m e^{i m \varphi}\sqrt{\frac{(2\ell+1)(\ell+m)!(\ell-m)!}{4\pi(\ell+s)!(\ell-s)!}}\sin^{2\ell}\left(\frac{\theta}{2}\right)\,\cr
&&
\times\sum_{k=k_1}^{k_2}{{\ell-s}\choose{k}}{{\ell+s}\choose{k+s-m}}(-1)^{\ell-k-s}\cot^{2k+s-m}\left(\frac{\theta}{2}\right)\,,
\end{eqnarray}
\end{widetext}
where $k_1={\rm max}(0,m-s)$ and $k_2={\rm min}(\ell+m,\ell-s)$. 

It is straightforward to compute the spin-weighted spherical harmonic, 
so we  need to focus only on how to solve the radial Teukolsky equation, 
which in Schwarzschild coordinates with $s=-2$ reads~\cite{ref:TTS}, 
\begin{equation}
 \left[\Delta^2{d\over dr}\left({1\over \Delta}{d\over dr}\right)+U(r)\right]
  R_{\ell m\omega}(r) = T_{\ell m\omega}(r),
\label{eq:Teu}
\end{equation}
with $\Delta=r(r-2M)$ and 
\begin{equation}
 U(r)={r^2\over\Delta}\left[\omega^2 r^2
-4i\omega(r-3M)\right]-(\ell-1)(\ell +2), 
\end{equation}
where $T_{\ell m\omega}$ is the source term which is a contraction of 
the energy momentum tensor of the small particle and the  null tetrad chosen. 

We solve Eq.~(\ref{eq:Teu}) 
using the Green function method. For this purpose, we 
need a homogeneous solution $R_{\ell m\omega}^{in}$ of Eq.~(\ref{eq:Teu}) 
which 
satisfies the boundary conditions 
\begin{equation}
R_{\ell m\omega}^{\rm in}=\left\{
  \begin{array}{lcc}
    B_{\ell m\omega}^{\rm trans}\Delta^2 e^{-i\omega r^{*}} & \hbox{for} 
    & r^{*}\rightarrow -\infty, \\
    r^3 B_{\ell m\omega}^{\rm ref} e^{i\omega r^{*}} +
    r^{-1} B_{\ell m\omega}^{\rm inc}  e^{-i\omega r^{*}} & \hbox{for} 
    & r^{*}\rightarrow +\infty,
   \end{array}
\right.
\label{eq:Rin_asymp}
\end{equation}
where $r^{*}=r+2M\ln(r/2M -1)$. 
Then the outgoing-wave solution of Eq.~(\ref{eq:Teu})
at infinity with appropriate boundary conditions at horizon is given by 
\begin{subequations}
\begin{align}
 R_{\ell m\omega}(r\rightarrow\infty)  &= {r^3 e^{i\omega r^{*}}\over 
    2i\omega B_{\ell m\omega}^{\rm inc}}\int_{2M}^{\infty} dr {R_{\ell m\omega}^{\rm in} T_{\ell m\omega}(r)\over \Delta^{2}},\\
   &\equiv r^3 e^{i\omega r^{*}}\tilde Z_{\ell m\omega}.
\label{eq:Zinf_b}
\end{align}
\label{eq:Zinf}
\end{subequations}
In the case of a circular orbit, the frequency spectrum of $T_{\ell m\omega}$ 
becomes discrete. Then $\tilde Z_{\ell m\omega}$ 
in Eq.~(\ref{eq:Zinf_b}) takes the form 
\begin{equation}
 \tilde Z_{\ell m\omega}=Z_{\ell m\omega}\delta(\omega-m\Omega),
\label{eq:tildeZ}
\end{equation}
where
\begin{widetext}
\begin{eqnarray}
Z_{\ell m\omega}& = &{\m\pi\over i\omega (r_0/M)^2 B_{\ell m\omega}^{{\rm inc}}}
  \Biggl\{\Biggl[
      -_0 b_{\ell m} -2i _{-1}b_{\ell m}
             \left(1+{i\over 2}{\omega r_0^2\over (r_0-2M)}\right)
    + i _{-2} b_{\ell m}{\omega r_0\over (1-2M/r_0)^{2}}
     \left(1-\frac{M}{r_0}+{1\over 2}i\omega r_0\right)\Biggr] R_{\ell m\omega}^{{\rm in}}
\cr &&
    +\left[i_{-1}b_{\ell m}-_{-2}b_{\ell m}
         \left(1+i{\omega r_0^2\over r_0-2M}\right)\right] 
           r_0 {R^{{\rm in}}_{\ell m\omega}}'(r_0)
      +{1\over 2} {}_{-2} b_{\ell m} r_0^2 {R_{\ell m\omega}^{{\rm in}}}''(r_0)
      \Biggr\}\quad,
\label{eq:Z8q0e0}
\end{eqnarray}
\end{widetext}
and prime, $'$, denotes $d/dr$ with $_s b_{\ell m}$ are defined by 
\begin{subequations}
\begin{align}
_0 b_{\ell m}  = & {1\over2}\left[(\ell-1)\ell(\ell+1)(\ell+2)\right]^{1/2}\cr
&\times {}_{0} Y_{\ell m}\left({\pi\over 2},\,0\right){\tilde E r_0\over r_0-2M},
\\
_{-1} b_{\ell m}  = & \left[(\ell-1)(\ell+2)\right]^{1/2}
    {}_{-1} Y_{\ell m}\left({\pi\over 2},\,0\right)\frac{\tilde L}{r_0},
\\
_{-2} b_{\ell m}  = & _{-2} Y_{\ell m}\left({\pi\over 2},\,0\right)
            \tilde L\Omega. 
\end{align}
\label{eq:blm}
\end{subequations}
Here, $\Omega$, 
$\tilde E$ and $\tilde L$ are the angular frequency, the specific energy and 
the angular momentum of the particle respectively, which are given by 
\begin{equation}
\Omega=\sqrt{\frac{M}{r_0^3}},\,\,\,
\tilde E=\frac{r_{0}-2M}{\sqrt{r_0(r_0-3M)}},\,\,\,
\tilde L=\frac{\sqrt{M r_{0}}}{\sqrt{1-3M/r_0}},
\end{equation}
where $r_0$ is the orbital radius. 

In terms of the amplitudes $Z_{\ell m\omega}$, the gravitational wave 
luminosity and the gravitational waveforms are respectively given by
\begin{equation}
{dE\over dt}=\sum_{\ell=2}^{\infty} \sum_{m=-\ell}^{\ell}
   \frac{\vert Z_{\ell m\omega}\vert^2}{4\pi \omega^2},
\label{eq:flux}
\end{equation}
and 
\begin{subequations}
\begin{align}
h_{+}-i\,h_{\times }&=-\frac{2}{r}\,\sum _{\ell,m} \frac{Z_{\ell m\omega}}{\omega^2}\,_{-2}Y_{\ell m}(\theta,\varphi )\,e^{i \omega (r^{*}-t)},\\
&\equiv
\sum_{\ell,m}(h_{lm}^+-i h_{\ell m}^\times),
\end{align}
\label{eq:wave}
\end{subequations}
where $\omega =m\Omega$ is the frequency of gravitational waves\footnote{Beware that $\omega$ is not the orbital frequency used in 
standard post-Newtonian approximation (e.g. Ref.~\cite{BFIS08}).
Here, following test-particle literature, we have $\dot{\phi}=\Omega$.}, 
$(\theta,\varphi)$ are the angles defining the location of the observer 
relative to the source 
and $h_{lm}^+$ and $ h_{\ell m}^\times$ are real. 
We calculate the gravitational waveforms in the post-Newtonian expansion, 
that is, in the expansion with respect to $v=(M/r_0)^{1/2}$. 
In order to compute $Z_{\ell m\omega}$, we need 
the series expansion of the ingoing-wave Teukolsky function 
$R_{\ell m\omega}^{{\rm in}}$ in terms of 
$\epsilon=2M\omega=2 M m \Omega=O(v^3)$ 
and $z=\omega r=O(v)$ and 
the asymptotic amplitudes $B^{{\rm inc}}_{\ell m\omega}$ 
in terms of $\epsilon$. We use the formalism developed by Mano, Suzuki and 
Takasugi~\cite{ref:MST} to compute them, and give a brief review of
the same for
the convenience of the reader in the following Sec.~\ref{sec:MST}. 
\section{The Mano, Suzuki and Takasugi method for 
analytic solutions of the homogeneous Teukolsky equation}
\label{sec:MST}
In the formalism developed by Mano, Suzuki and Takasugi, 
the homogeneous solutions of the Teukolsky equation 
are expressed in terms of two kinds of series of special 
functions: hypergeometric functions and Coulomb wave functions
~\cite{ref:MST,ref:MSTR}.
The series of hypergeometric functions is convergent at the horizon and 
the series of Coulomb wave functions at infinity. 
The matching of the two kinds of solutions is done analytically 
in the overlapping region of convergence. 
One can thus obtain analytic expressions of the asymptotic amplitudes 
of the homogeneous solutions without numerical integration. 
This enables one to compute the gravitational wave flux outward at
 infinity and 
into the horizon very accurately~\cite{FT1,FT2,FHT}. 
Furthermore, the formalism is very powerful 
for the calculation of the post-Newtonian expansion of the Teukolsky 
equation since the series expansion is closely related to 
the low frequency expansion. 
Using the formalism, the energy flux absorbed into the horizon was calculated 
up to relative 4PN (i.e. absolute  6.5PN) order 
for a particle in circular and equatorial orbit around a
 Kerr black 
hole in Ref.~\cite{TMT}. Gravitational wave flux to infinity was also 
computed up to 2.5PN order for slightly eccentric and inclined 
orbit around a  Kerr black hole in Ref.~\cite{STHGN,Ganz}.
Although the Mano, Suzuki, Takasugi formalism can be applied to the case of 
a Kerr black hole, 
we assume that $q=0$ since we consider the case of 
the Schwarzschild black hole in the present paper. 
For more details of the formalism, 
we refer the reader to the  recent review Ref.~\cite{ST}. 
\subsection{Ingoing-wave solution of the radial Teukolsky equation}
A homogeneous solution of the Teukolsky equation 
which is expressed as a series of Coulomb wave functions $R_{{\rm C}}^{\n}$ 
is given by 
\begin{equation}
R_{{\rm C}}^{\n}=
z\left(1-{\e  \over{{z}}}\right)^{2-i\e}f_{\n}(z), 
\label{eq:Rc}
\end{equation}
where the function $f_{\nu}(z)$ is expressed in a series of 
Coulomb wave functions as 
\begin{equation}
f_{\n}(z)= \displaystyle\sum_{n=-\infty}^{\infty}
(-i)^n\frac{(\n-1-i\e)_n}{(\n+3+i\e)_n} a_n^{\n} F_{n+\n}(2\,i-\e,z),
\label{eq:series of Rc}
\end{equation}
with $z=\z r$, our notation $(a)_{n}=\Gamma(a+n)/\Gamma(a)$,
and where $F_{N}(\eta,z)$ is a Coulomb wave function defined by 
\begin{eqnarray}
F_{N}(\eta,z)&=&e^{-iz}2^{N}z^{N+1}\frac{\G(N+1-i\eta)}{\G(2N+2)}\cr
&&\times\Phi(N+1-i\eta,2N+2;2iz).
\label{eq:defcoulomb}
\end{eqnarray}
Here $\Phi(\a,\b;z)$ is the confluent hypergeometric function, 
which is regular at $z=0$ (see $\S$ 13 of Ref.~\cite{handbook}). 

The expansion coefficients $a_{n}^{\n}$ satisfy 
the three-term recurrence relation
\begin{eqnarray}
\alpha_n^\nu a_{n+1}^{\n}+\beta_n^{\nu} a_{n}^{\n}+\gamma_n^\nu a_{n-1}^{\n}=0,
\label{eq:3term}
\end{eqnarray}
where
\begin{widetext}
\begin{subequations}
\begin{eqnarray}
\a_n^\n&=&{i\e (n+\n-1+i\e)(n+\n-1-i\e)(n+\n+1+i\e)
\over{(n+\n+1)(2n+2\n+3)}},
\\
\b_n^\n&=&-\ell(\ell+1)+(n+\n)(n+\n+1)+2\e^2
+{\e^2 (4+\e^2) \over{(n+\n)(n+\n+1)}},\\
\c_n^\n&=&-{i\e  (n+\n+2+i\e)(n+\n+2-i\e)(n+\n-i\e)
\over{(n+\n)(2n+2\n-1)}}.
\end{eqnarray}
\label{eq:3term_abc}
\end{subequations}
\end{widetext}
We note that the parameter $\nu$, called the renormalized angular momentum, 
introduced in the above formulas does not exist in the Teukolsky equation. 
This parameter is determined so that the series 
converges and actually represents a solution of the Teukolsky equation. 

The series converges if $\nu$ satisfies the equation, 
\begin{eqnarray}
R_{n}L_{n-1}=1,
\label{eq:consistency}
\end{eqnarray}
where $R_{n}$ and $L_{n}$ are defined in terms of continued fractions 
(in the second lines of equations below) as 
\begin{eqnarray}
R_n&\equiv& {a_{n}^{\n}\over a_{n-1}^{\n}}
=-{\gamma_n^\nu\over {\beta_n^\nu+\alpha_n^\nu R_{n+1}}},\cr
&=&-{\gamma_{n}^\nu\over \beta_{n}^\nu-}\,
{\alpha_{n}^\n\gamma_{n+1}^\nu\over \beta_{n+1}^\nu-}\,
{\alpha_{n+1}^\n\gamma_{n+2}^\nu\over \beta_{n+2}^\nu-}\cdots,
\label{eq:Rncont}\\
L_n&\equiv& {a_{n}^{\n}\over a_{n+1}^{\n}}
=-{\alpha_n^\nu\over {\beta_n^\nu+\gamma_n^\nu L_{n-1}}},\cr
&=&-{\alpha_{n}^\nu\over \beta_{n}^\nu-}\,
{\alpha_{n-1}^\n\gamma_{n}^\nu\over \beta_{n-1}^\nu-}\,
{\alpha_{n-2}^\n\gamma_{n-1}^\nu\over \beta_{n-2}^\nu-}\cdots.
\label{eq:Lncont}
\end{eqnarray}
We can obtain two kinds of the expansion coefficients, $a_{n}$, 
by the continued fractions, $R_{n}$ and $L_{n}$. 
If we choose $\nu$ such that it satisfies 
Eq.~(\ref{eq:consistency}), for a certain $n$, 
the two types of the expansion coefficients coincide and 
the series of Coulomb wave function Eq.~(\ref{eq:series of Rc}) 
converges for $r>r_{+}$. 
From Eq.~(\ref{eq:3term_abc}), we can show $\a_{-n}^{-\n-1}=\c_{n}^{\n}$ 
and $\b_{-n}^{-\n-1}=\b_{n}^{\n}$. Accordingly, we find that 
$a_{-n}^{-\n-1}$ satisfies the same recurrence relation Eq.~(\ref{eq:3term}) 
and $R_{{\rm C}}^{-\n-1}$ is also a homogeneous solution of the Teukolsky 
equation that converges for $r>r_{+}$.

Matching the solution in series of Coulomb wave functions, 
which converges at infinity, with the one in series of hypergeometric 
functions, which converges at the horizon, 
we can obtain the ingoing-wave solution $R_{lm\omega}^{{\rm in}}$, 
which converges in the entire region as 
\begin{eqnarray}
R_{lm\omega}^{{\rm in}}=K_{\n}R_{{\rm C}}^{\n}+K_{-\n-1}R_{{\rm C}}^{-\n-1}, 
\label{eq:secondRin}
\end{eqnarray}
where 
\begin{widetext}
\begin{eqnarray}
K_{\n}
&=&\frac{e^{i\e}(2\e)^{-2-\n-N}2^{2}i^{N}\Gamma(3-2i\e)\Gamma(N+2\n+2)}
{\Gamma(N+\n+3+i\e)\Gamma(N+\n+1+i\e)\Gamma(N+\n-1+i\e)}\nonumber \\
&&\times\left(\sum_{n=N}^{\infty}(-1)^{n}\frac{\Gamma(n+N+2\n+1)}{(n-N)!}
\frac{\Gamma(n+\n-1+i\e)\Gamma(n+\n+1+i\e)}
{\Gamma(n+\n+3-i\e)\Gamma(n+\n+1-i\e)}a_{n}^{\n}\right)\nonumber \\
&&\times\left(\sum_{n=-\infty}^{N}\frac{(-1)^{n}}{(N-n)!(N+2\n+2)_{n}}
\frac{(\n-1-i\e)_{n}}{(\n+3+i\e)_{n}}a_{n}^{\n}\right)^{-1},
\end{eqnarray}
\end{widetext}
and $N$ can be any integer.  
The factor $K_\n$ is a constant which is introduced to match the solutions 
in the overlap region of convergence.   It should be independent of the 
choice of $N$.

Comparing $R_{lm\omega}^{{\rm in}}$ in Eq.~(\ref{eq:Rin_asymp}) 
with Eq.~(\ref{eq:secondRin}) 
in the limit of $r^{*}\rightarrow \pm\infty$, 
we can obtain analytic expressions for the asymptotic amplitudes 
$B^{{\rm trans}}_{lm\omega}$, 
$B^{{\rm inc}}_{lm\omega}$ and $B^{{\rm ref}}_{lm\omega}$ 
defined in Eq.~(\ref{eq:Rin_asymp}) as
\begin{subequations}
\begin{eqnarray}
B^{{\rm trans}}_{lm\omega}
&=&\left(\frac{\e}{\omega}\right)^{-4}
\sum_{n=-\infty}^{\infty}a_{n}^{\n},\\
B^{{\rm inc}}_{lm\omega}
&=&\omega^{-1}\left[K_{\n}-ie^{-i\pi\n}
\frac{\sin\pi(\n+i\e)}{\sin\pi(\n-i\e)}K_{-\n-1}\right]\cr
&&\times A_{+}^{\n}e^{-i\e\ln\e},\\
B^{{\rm ref}}_{lm\omega}
&=&\omega^{3}[K_{\n}+ie^{i\pi\n}K_{-\n-1}]A_{-}^{\n}e^{i\e\ln\e},
\end{eqnarray}
\label{eq:asymp_amp}
\end{subequations}
where
\begin{subequations}
\begin{eqnarray}
A_{+}^{\n}&=&2^{-3-i\e}e^{-\frac{\pi\e}{2}}e^{\frac{\pi}{2}i(\n+3)}
\frac{\G(\n+3+i\e)}{\G(\n-1-i\e)}\cr
&&\times\sum_{n=-\infty}^{+\infty}a_{n}^{\n},
\\
A_{-}^{\n}&=&2^{1+i\e}e^{-\frac{\pi\e}{2}}e^{-\frac{\pi}{2}i(\n-1)}\cr
&&\times\sum_{n=-\infty}^{+\infty}(-1)^{n}\frac{(\n-1-i\e)_{n}}{(\n+3+i\e)_{n}}a_{n}^{\n}.
\end{eqnarray}
\end{subequations}
\subsection{Low frequency expansions of solutions}
\label{sec:MST_PN}
In this section, we show the relation of 
Mano, Suzuki and Takasugi formalism with the post-Newtonian expansion. 
In their formalism, 
we first solve Eq.~(\ref{eq:consistency}) to determine $\n$. 
Next, we derive the expansion coefficients $a_n^{\n}$ using 
the continued fractions Eq.~(\ref{eq:Rncont}) for $n>0$ 
and Eq.~(\ref{eq:Lncont}) for $n<0$ 
with the condition $a_0^{\n}=a_0^{-\n-1}=1$.
Then we can derive the ingoing-wave solution of the radial Teukolsky 
equation Eq~(\ref{eq:secondRin}) 
and the asymptotic amplitudes Eq~(\ref{eq:asymp_amp}). 

When we determine $\n$ in the practical calculation, 
we solve the alternative equation which is equivalent 
to Eq.~(\ref{eq:consistency}) for $n=1$ 
\begin{eqnarray}
\beta_0^{\nu}+\alpha_0^\nu R_{1}+\gamma_0^\nu L_{-1}=0,
\label{eq:determine_nu}
\end{eqnarray}
where $R_{1}$ and $L_{-1}$ are given by the continued fractions 
Eq.~(\ref{eq:Rncont}) and Eq.~(\ref{eq:Lncont}) respectively. 

In the limit of low frequency, 
the orders of $\a_n^\n$, $\c_n^\n$ and $\b_n^\n$ 
in $\e$ are $O(\e)$, $O(\e)$ and $O(1)$ respectively 
except for certain values of $n<0$ (see Ref.~\cite{ref:MST,ST}). 
However, it is straightforward to derive the low frequency expansion of 
$\n$ by solving Eq.~(\ref{eq:determine_nu}) order by order in $\e$. 
The low frequency expansion of $\n$ up to $O(\e^2)$ is given by
~\cite{ref:MST}. 
\begin{subequations}
\label{eq:nusol}
\begin{eqnarray}
\nu&=& \ell+\nu^{(2)}(\ell)\e^2 +O(\epsilon^3)
\label{eq:nusola},\\
\nu^{(2)}(\ell) &\equiv& {1\over{2\ell+1}}\left[-2-{4\over{\ell(\ell+1)}}\right.\cr
&&
+{[(\ell+1)^2-4]^2\over{(2\ell+1)(2\ell+2)(2\ell+3)}}\cr
&&
\left.-{(\ell^2-4)^2\over{(2\ell-1)2\ell(2\ell+1)}}\right].
\label{eq:nusolb}
\end{eqnarray}
\end{subequations}
We note that the above expression of $\n$ up to $O(\e^2)$ 
is independent of $m$. 

Combining Eq.~(\ref{eq:nusol}) with Eq.~(\ref{eq:Rncont}) for $n>0$ 
and Eq.~(\ref{eq:Lncont}) for $n<0$, 
we can derive the expansion coefficients 
$a_n^{\n}$ up to $O(\e^2)$ (which is valid for $l \ge \frac 32$) as
~\cite{ref:MST}, 
\begin{subequations}
\begin{align}
a_1^{\n}=&
\,i\,{\frac {\left(\ell+3\right)^{2}}{ 2\left(\ell+1\right) \left( 2\,\ell+1\right) }}\e+{\frac{\left(\ell+3\right)^{2}}{2\left(\ell+1\right)^{2} \left( 2\,\ell+1\right)}}{\e}^{2}+O(\e^3),\\
a_2^{\n}=&-{\frac{\left(\ell+3\right)^{2} \left(\ell+4\right) ^{2}}{ 4\left(\ell+1\right) \left(2\,\ell+1\right) \left(2\,\ell+3\right)^{2}}}{\e}^{2}+O(\e^3),\\
a_{-1}^{\n}=&\,i\,{\frac{\left(\ell-2\right)^{2}}{2\,\ell\,\left(2\,\ell+1\right) }}\e-{\frac{\left(\ell-2\right)^{2}}{2\,{\ell}^{2} \left(2\,\ell+1\right)}}{\e}^{2}+O(\e^3),\\
a_{-2}^{\n}=&-{\frac{\left(\ell-3\right)^{2}\left(\ell-2\right)^{2}}{4\,\ell\, \left(2\,\ell+1\right) \left(2\,\ell-1\right)^{2}}}{\e}^{2}+O(\e^3).
\end{align}
\label{eq:an_nu}
\end{subequations}
As one can find from Eq.~(\ref{eq:an_nu}), 
the leading order of $a_n^\n$ in $\e$ increases with $\mid n\mid$ 
for $\mid n\mid\le 2$ because $R_{\mid n\mid}\sim O(\e)$ 
and $L_{-\mid n\mid}\sim O(\e)$. 
Basically, this property of $a_n^\n$ holds for $\mid n\mid\ge 2$ 
although we have to be careful for $n<0$~\cite{ref:MST,ST}. 
Thus, we can derive the low frequency expansion of the asymptotic amplitudes 
Eq~(\ref{eq:asymp_amp}) using that of $a_n^\n$. 
Moreover, the Coulomb wave function, defined in Eq.~(\ref{eq:defcoulomb}), 
is a function of $z$. 
Since each term of the expansion depends on $\e\sim O(v^3)$ 
and $z\sim O(v)$, the ingoing-wave solution $R_{lm\omega}^{{\rm in}}$ 
in Eq.~(\ref{eq:secondRin}) is very useful for the post-Newtonian expansion. 
\subsection{2.5PN formulas for $Z_{\ell m\omega}$}
\label{sec:zlmw2.5PN}
In this section, we derive 2.5PN formulas for $Z_{\ell m\omega}$ 
Eq.~(\ref{eq:Z8q0e0}) using $O(\e^2)$ results in Sec.~\ref{sec:MST_PN}. 
In the computation of $Z_{\ell m\omega}$, Eq.~(\ref{eq:Z8q0e0}), 
we need to estimate the homogeneous Teukolsky solution 
$R_{lm\omega}^{{\rm in}}$ in Eq.~(\ref{eq:secondRin}) 
and asymptotic amplitude $B^{{\rm inc}}_{lm\omega}$ in 
Eq.~(\ref{eq:asymp_amp}). In $O(\e^2)$ calculation, we can neglect 
$K_{-\n-1}$ terms in Eqs.~(\ref{eq:secondRin}) and (\ref{eq:asymp_amp}) 
since $K_{-\n-1}/K_{\n}=O(\e^{2\ell})$ when we restrict $\ell\ge 3/2$ for 
the Schwarzschild case~\cite{ref:MST}. 
Then we can approximate them as 
$R_{lm\omega}^{{\rm in}}=K_{\n}R_{{\rm C}}^{\n}$ 
and $B^{{\rm inc}}_{lm\omega}=K_{\n}A_{+}^{\n}e^{-i\e\ln\e}/\omega$ 
in $O(\e^2)$ calculation. 
However, we note that one cannot derive 3PN formulas but 2.5PN formulas 
for $Z_{\ell m\omega}$ using $O(\e^2)$ results, 
Eqs.~(\ref{eq:nusol}) and (\ref{eq:an_nu}), in Sec.~\ref{sec:MST_PN}. 
This is because the Coulomb wave function Eq.~(\ref{eq:defcoulomb}), 
which is needed to compute $R_{{\rm C}}^{\n}$ in Eq.~(\ref{eq:Rc}), 
is proportional to $z^{N+1}=O(v^{N+1})$. 
Thus, the post-Newtonian order of a series of 
Coulomb wave functions for $n<0$ grows slower than that for $n>0$. 
One can estimate the post-Newtonian order of $a_n^{\n} F_{n+\n}(2\,i-\e,z)$, 
which appears in the homogeneous Teukolsky solution in terms of a series of 
Coulomb wave functions $R_{{\rm C}}^{\n}$ Eq.~(\ref{eq:Rc}), as 
\begin{subequations}
\begin{eqnarray}
{a_1^{\n} F_{1+\n}(2\,i-\e,z) \over a_0^{\n} F_{\n}(2\,i-\e,z)} &=& O(v^4),\\
{a_2^{\n} F_{2+\n}(2\,i-\e,z) \over a_0^{\n} F_{\n}(2\,i-\e,z)} &=& O(v^8), \\
{a_{-1}^{\n} F_{-1+\n}(2\,i-\e,z) \over a_0^{\n} F_{\n}(2\,i-\e,z)} &=& O(v^2),\\
{a_{-2}^{\n} F_{-2+\n}(2\,i-\e,z) \over a_0^{\n} F_{\n}(2\,i-\e,z)} &=& O(v^4). 
\end{eqnarray}
\end{subequations}
Then one finds that summation over $n$ from $-2$ to $2$ 
in Eq.~(\ref{eq:series of Rc}) produces 2.5PN results. 

From the expression of asymptotic amplitude 
$B^{{\rm inc}}_{lm\omega}$ in Eq.~(\ref{eq:asymp_amp}), one finds 
$Z_{\ell m\omega}\propto e^{i\e\ln 2\e}e^{-i\pi\n^{(2)}(\ell)\e^2/2}e^{\pi\e/2}$, 
where $\n^{(2)}(\ell)$ is defined in Eq.~(\ref{eq:nusol}). 
If we factor out the phase that arises from the integration 
in Eq.~(\ref{eq:Zinf}) 
in addition to phase from the asymptotic amplitude $B^{{\rm inc}}_{lm\omega}$, 
we can derive $Z_{\ell m\omega}$ up to 2.5PN as
\begin{widetext}
\begin{subequations}
\begin{eqnarray}
\label{eq:zlmw2.5PNa}
Z_{\ell m\omega} &=& 
e^{i\e\ln 2\e}e^{-i\pi\n^{(2)}(\ell)\e^2/2}e^{\frac{\pi\e}{2}}Z_{\ell m \omega}^{(0)} \left[1+Z_{\ell m}^{(2)}v^2+i\,m\,Z_{\ell m}^{(3)}v^3+Z_{\ell m}^{(4)}v^4+i\,m\,Z_{\ell m}^{(2)}Z_{\ell m}^{(3)}v^5+O(v^6)\right],\\
\label{eq:zlmw2.5PNb}
&=& e^{-im\psi_{\ell m}^{(\rm 3PN)}}e^{\frac{\pi\e}{2}}Z_{\ell m \omega}^{(0)} \left[1+Z_{\ell m}^{(2)}v^2+Z_{\ell m}^{(4)}v^4+O(v^6)\right],\\
\label{eq:zlmw2.5PNc}
&=& e^{-im\psi_{\ell m}^{(\rm 3PN)}}Z_{\ell m\omega}^{(0)}\left[1+Z_{\ell m}^{(2)}v^2+m\pi v^3+Z_{\ell m}^{(4)}v^4+m\pi Z_{\ell m}^{(2)}v^5+O(v^6)\right],
\end{eqnarray}
\label{eq:zlmw2.5PN}
\end{subequations}
\end{widetext}
where $Z_{\ell m \omega}^{(0)}$, $Z_{\ell m}^{(2)}$, $Z_{\ell m}^{(3)}$ and 
$Z_{\ell m}^{(4)}$ are real (see  Sec.~\ref{sec:zlmw2.5PN_even} and 
Sec.~\ref{sec:zlmw2.5PN_odd}) and, 
\begin{subequations}
\begin{align}
\label{eq:psi3pna}
\psi_{\ell m}^{(\rm 3PN)}=& -2\ln(4m\,v^3)v^3 - Z_{\ell m}^{(3)}v^3 + 2\,m\pi\nu^{(2)}(\ell)v^6,\\
\label{eq:psi3pnb}
=&2v^3\left(\Psi^{(0)}(\ell)-\frac{1}{\ell}+\frac{1}{2}+\frac{2}{\ell+1}-\ln(4m\,v^3)\right.\cr 
&\left.-3\frac{1+(-1)^{\ell+m}}{(\ell-1)\ell (\ell+1)(\ell+2)}\right)+2\,m\pi\nu^{(2)}(\ell)v^6, 
\end{align}
\label{eq:psi3pn}
\end{subequations}
where $\Psi^{(n)}(z)$ is the polygamma function and $\nu^{(2)}(\ell)$ 
calculated from Eq.~(\ref{eq:nusolb}). 
$\Psi^{(0)}(\ell)$ 
is related to the digamma function whose explicit value
can be calculated using
\begin{equation}
\Psi^{(0)}(\ell)=\sum_{k=1}^{\ell-1} \frac{1}{k} -\gamma,
\end{equation}
where $\gamma$ is the Euler constant. 
To go from Eq.~(\ref{eq:zlmw2.5PNa}) to Eq.~(\ref{eq:zlmw2.5PNb}), 
we move  the imaginary terms $i\,m\,Z_{\ell m}^{(3)}v^3$ 
from the amplitude of the post-Newtonian expansion to the phase. 
The $e^{\pi\e/2}$ in Eq.~(\ref{eq:zlmw2.5PNb})  is the motivation
for the chosen dependence in tail terms in the
factorized resummed waveforms in Ref~\cite{DIN} 
(See Eq.~(\ref{eq:Tlm}) in Sec.~\ref{sec:rholm}).
 With such overall factorization the remaining
series in  Eq.~(\ref{eq:zlmw2.5PNb}) is expected 
to achieve improved  convergence to numerical results 
since coefficients in this  series have smaller
post-Newtonian coefficients. 
To compare to  spherical harmonic modes in literature, we expand $e^{\pi\e/2}$ 
in Eq.~(\ref{eq:zlmw2.5PNb}) and obtain the alternative
form Eq.~(\ref{eq:zlmw2.5PNc}). 
Finally, to go from Eq.~(\ref{eq:psi3pna}) to 
Eq.~(\ref{eq:psi3pnb}), we have used the general formula of $Z_{\ell m}^{(3)}$ 
given in Sec.~\ref{sec:zlmw2.5PN_even} and Sec.~\ref{sec:zlmw2.5PN_odd}. 
Though this is a 2.5PN calculation, one may notice that we have 
included the 3PN phase term in Eq.~(\ref{eq:psi3pn}) derived from 
the asymptotic amplitude $B^{{\rm inc}}_{lm\omega}$ for completeness.
This is sufficient because no other 3PN phase terms were generated from 
the integration in Eq.~(\ref{eq:Zinf}) in our 5.5PN calculation 
in Sec.~\ref{sec:hlm}, Sec.~\ref{sec:rholm} and Sec.~\ref{sec:zetalm}. 
Thus, in the case of Schwarzschild black hole, there may not exist 
any further 3PN phase terms arising from 
the integration in Eq.~(\ref{eq:Zinf}).

Since the leading order of $Z_{\ell m\omega}$ depends on 
whether $_{0}b_{\ell m}\sim O(1)$, i.e. ${}_0Y_{\ell m}(\frac{\pi}{2},\,0)$, 
vanishes or not~\cite{ref:poisson}, 
we treat the case for $\ell+m={\rm even}$ in Sec.~\ref{sec:zlmw2.5PN_even}
and $\ell+m={\rm odd}$ in Sec.~\ref{sec:zlmw2.5PN_odd}. 
Also observe that the 1PN term of $Z_{\ell m\omega}$, 
i.e. $Z_{\ell m\omega}^{(2)}$, contains a linear term of $\ell$, 
which reinforces the suggestion in Ref.~\cite{DIN} 
to introduce the  $\ell$-th root of the amplitude 
for factorized resummation 
(See Sec.~\ref{sec:rholm} and Sec.~\ref{sec:rho2PN}). 
\subsubsection{$\ell+m={\rm even}$ case}
\label{sec:zlmw2.5PN_even}
\begin{subequations}
\begin{align}
Z_{\ell m \omega}^{(0)}=& -\frac{\m m^{\ell+2}\pi^{3/2}}{i^\ell 2^{\ell-1} \Gamma(\ell+3/2)} \sqrt{\frac{(\ell+2)(\ell+1)}{\ell(\ell-1)}}\cr
&\times {}_{0}Y_{\ell m}\left(\frac{\pi}{2},\,0\right)\frac{v^{\ell+2}}{(r_0/M)^2} ,\\
Z_{\ell m}^{(2)}=&
- \ell + {\displaystyle \frac {1}{2}} -{\frac {m^2(\ell+9)}{ 2\left( 2\,\ell+3 \right)  \left( \ell+1 \right) }},\\
Z_{\ell m}^{(3)}=& -2\left(\Psi^{(0)}(\ell)-\frac{1}{\ell}+\frac{1}{2}+\frac{2}{\ell+1}\right.\cr
&\left.-\frac{6}{(\ell-1)\ell (\ell+1)(\ell+2)}\right),\\
Z_{\ell m}^{(4)}=&
{\displaystyle \frac {\ell^{2}}{2}} - {\displaystyle \frac {5\,\ell}{4}} + 2 
+{\frac {17\ell-1}{8\,\left(2\ell-1\right)\left(\ell-1 \right)}}\cr
&+ m^{2}\left({\displaystyle \frac {1}{4}} -{\frac {2\,(3\,{\ell}^{3}+6\,{\ell}^{2}-\ell+4)}{ \left( \ell+1 \right) ^{2} \left( \ell-1 \right)  \left( \ell+2 \right) \ell}}\right)
\cr&
+{\frac {m^{4}({\ell}^{2}+19\ell+50)}{8\,\left(2\ell+5\right)\left( 2\ell+3\right)\left(\ell+2\right)\left(\ell+1\right)}}.
\end{align}
\label{eq:zlmw2.5PN_even}
\end{subequations}
\subsubsection{$\ell+m={\rm odd}$ case}
\label{sec:zlmw2.5PN_odd}
\begin{subequations}
\begin{align}
Z_{\ell m \omega}^{(0)}=& -\frac{\m m^{\ell+2} \pi^{3/2}}{i^{\ell+1} 2^{\ell-2}\Gamma(\ell+3/2)}\sqrt{\frac{\ell+2}{\ell-1}}\cr
&\times{}_{-1}Y_{\ell m}\left(\frac{\pi}{2},\,0\right)\frac{v^{\ell+3}}{(r_0/M)^2}, \\
Z_{\ell m}^{(2)}=& - \ell + 
{\displaystyle \frac {1}{2}} + {\displaystyle \frac {2}{\ell}} - {\frac {m^{2}(\ell+4)}{2 \left( \ell+2 \right)  \left( 2\,\ell+3 \right) }},\\
Z_{\ell m}^{(3)}=& -2\left(\Psi^{(0)}(\ell)-\frac{1}{\ell}+\frac{1}{2}+\frac{2}{\ell+1}\right),\\
Z_{\ell m}^{(4)}=&
{\displaystyle \frac {\ell^{2}}{2}} - {\displaystyle \frac {5\,\ell}{4}} - {\displaystyle \frac {1}{2}} + {\frac {29\ell+8}{8\ell\left(2\ell-1\right)}}\cr 
&+ m^{2}\left({\displaystyle \frac {1}{4}} - {\frac {6\,{\ell}^{3}+33\,{\ell}^{2}+55\ell+36}{2\,\left(\ell+1 \right)\ell\left(\ell+2 \right)\left( 2\ell+3\right)}}\right)
\cr
&
+{\frac{m^{4}(\ell+6)}{8\,\left(2\ell+5\right)\left(2\ell+3\right)\left(\ell+2\right)}}.
\end{align}
\label{eq:zlmw2.5PN_odd}
\end{subequations}

Using both the 2.5PN formulas for $Z_{\ell m\omega}$ 
in this section and Eq.~(\ref{eq:wave}), one has a general formula for 
computation of 5.5PN waveforms for $\ell\ge 8$. However, for completeness, 
we list those modes in Appendix~\ref{sec:55PN}. 

We conclude this rather technical section by recapitulating the
main steps in 
the formalism  developed by Mano, Suzuki and Takasugi, 
to  analytically  compute  homogeneous solutions of the Teukolsky equation. 
The most important task in the formalism is to determine 
the renormalized angular momentum $\n$, which is introduced 
so that the series of two types of special function, 
hypergeometric functions and Coulomb wave functions converge. 
$\n$ is determined by solving the continued fraction 
equation Eq.~(\ref{eq:determine_nu}). We then compute the expansion 
coefficients $a_{n}^{\n}$ using Eq.~(\ref{eq:Rncont}) for $n>0$ 
and Eq.~(\ref{eq:Lncont}) $n<0$ with the condition 
$a_{0}^{\n}=a_{0}^{-\n-1}=1$. The asymptotic amplitude 
of the homogeneous solution $B^{{\rm inc}}_{lm\omega}$
is subsequently calculated and 
the homogeneous solution $R_{lm\omega}^{{\rm in}}$ constructed using 
Eq.~(\ref{eq:asymp_amp}) and Eq.~(\ref{eq:secondRin}) respectively. 
Finally, we compute $Z_{\ell m\omega}$ using Eq.~(\ref{eq:Z8q0e0}), 
which enables one to compute gravitational wave flux to infinity and 
gravitational waveforms by Eq.~(\ref{eq:flux}) and 
Eq.~(\ref{eq:wave}) respectively. 

In the coming sections, Sec.~\ref{sec:hlm}, Sec.~\ref{sec:rholm} and 
Sec.~\ref{sec:zetalm}, we derive the 5.5PN waveforms computing 
$Z_{\ell m\omega}$ following the above steps. These $Z_{\ell m\omega}$ not only lead to the  5.5PN energy flux obtained in Ref.~\cite{ref:TTS} as required 
but also contain new terms which are needed for the calculation of 
5.5PN waveforms. 
\section{Spherical harmonic modes}
\label{sec:hlm}
In this section, we project the waveforms onto spin-weighted 
spherical harmonics, and compute $h_{\ell m}$ up to $O(v^{11})$ 
which are useful 
for the comparison between the post-Newtonian and numerical results. 
For the comparisons between the post-Newtonian expansion and 
numerical simulations, 
we decompose 
$h_{+}$ and $h_{\times}$ 
into the modes of spin-weighted spherical harmonics as
\begin{eqnarray}
h_{+} - i\,h_{\times}
=\sum_{\ell,m} h_{\ell m} \;_{-2}Y_{\ell m}(\Theta, \Phi),
\label{eq:hlmdef}
\end{eqnarray}
where $(\Theta, \Phi)$ are the angles defining the direction of propagation 
of gravitational waves.
Using the orthonormality condition of spin-weighted spherical harmonics, 
$h_{\ell m}$ can be derived as 
\begin{eqnarray}
h_{\ell m}=\int \sin\Theta d\Theta d\Phi\,(h_{+} - i\,h_{\times})\;_{-2}\bar{Y}_{\ell m}(\Theta, \Phi)\label{eq:hlm_int_a}.
\label{eq:hlm_int}
\end{eqnarray}
Recall that the polarizations in Eq.~(\ref{eq:wave}) are the functions of 
both the orbital phase $\Omega t$ and 
the angles $(\theta,\varphi)$ defining the observer relative to the source. 
Therefore to obtain the polarizations 
corresponding to the direction of propagation of 
gravitational waves, $(\Theta,\Phi)$,  in Eq.~(\ref{eq:hlm_int}) we have to replace 
$(\theta,\Omega t+\varphi)$ in $h_{+}$ and $h_{\times}$ by 
$(\Theta,\Omega t+\varphi-\Phi)$~\cite{BFIS08}. 
(In Ref.~\cite{BFIS08}, $(\theta,\varphi)$ is defined as $(i,\pi/2)$). 
Then we obtain $h_{\ell m}$ as 
\begin{subequations}
\begin{align}
h_{\ell m}
=&-\frac{2}{r}\,\sum _{\ell',m'}\frac{Z_{\ell' m'\omega'}\,e^{i m'\Omega r^{*}}}{(m'\Omega)^2}\int \sin\Theta d\Theta d\Phi\,\cr
&\times e^{-i m'(\Omega\,t-\Phi)}\,_{-2}Y_{\ell' m'}(\Theta,\varphi)\;_{-2}\bar{Y}_{\ell m}(\Theta,\Phi),\label{eq:hlm2zlmw_a}\\
=&-\frac{2}{r}\,\frac{Z_{\ell m\omega}\,e^{i\,m\Omega (r^{*}-t)}\,e^{i\,m\,\varphi}}{(m\Omega)^2},
\label{eq:hlm2zlmw_b}
\end{align}
\label{eq:hlm2zlmw}
\end{subequations}
where 
$Z_{\ell m\omega}$ is given in Sec.~\ref{sec:formula}.
To go from  Eq.~(\ref{eq:hlm2zlmw_a}) to Eq.~(\ref{eq:hlm2zlmw_b}), 
we have used 
the orthonormality condition of spin-weighted spherical harmonics. 

Following in spirit but generalizing suitably the notation 
defined in Ref.~\cite{BFIS08} we write\footnote
{Recall that there is an overall  difference of sign 
in $h_{\ell m}$ 
between Ref.~\cite{BFIS08} and Ref.~\cite{K07} due to a different choice 
of the polarization triad~\cite{K07}. The sign of $h_{\ell m}$ 
in Eq.~(\ref{eq:hlm2zlmw_b}) matches with that in  Ref.~\cite{K07} 
and is consistent with Eq.~(\ref{eq:hlm_fact_a}). 
In the test-particle limit, 
$m_1=\m$, $m_2=M$,
 $\Delta=(m_1-m_2)/(m_1+m_2)$ in Ref.~\cite{BFIS08} reduces to $-1$. 
For comparison between black hole perturbation theory and 
standard post-Newtonian theory, we have to substitute 
$\ln x_0 = 17/18-2\ln 2-2\gamma/3$ into the phase in Ref.~\cite{BFIS08} 
in order to match the phase up to 1.5PN 
since Schwarzschild coordinates are used in the test-particle limit and 
harmonic coordinates in the generic mass ratio case as pointed out in 
Ref.~\cite{K07}. 
Lastly unlike in PN works beyond 2.5PN where $\dot{\Omega}$ due to the 
radiation reaction must be included~\cite{KBI07,BFIS08}, 
here in the test-particle case such terms 
are absent since they are higher order in mass ratio, $\m/M$. 
}

\begin{subequations}
\begin{eqnarray}
\label{eq:hlm_fact_a}
h_{\ell m}&=&-\frac{2\,\m\,v^2}{r}H_{\ell m},\\
\label{eq:hlm_fact_b}
H_{\ell m}&=&\sqrt{\frac{16\pi}{5}}\hat{H}_{\ell m}e^{-im\psi_{\ell m}}. 
\end{eqnarray}
\label{eq:hlm_fact}
\end{subequations}
Note that the phase in Eq.~(\ref{eq:hlm_fact_b}) is more general 
in that it is multipole -- i.e. $(\ell,m)$ -- dependent, 
while the phase  in Ref.~\cite{BFIS08} is independent of $(\ell,m)$  
equivalent to the 1.5PN-accurate $\psi_{2,\,2}^{\rm 1.5PN}$ of this section. 

Using the 3PN phase of $Z_{\ell m\omega}$ given in Eq.~(\ref{eq:psi3pn}) 
for any multipolar order  $\ell$ and $m$ as 
\begin{align}
\psi_{\ell m}^{(\rm 3PN)}=&2v^3\left(\Psi^{(0)}(\ell)-\frac{1}{\ell}+\frac{1}{2}+\frac{2}{\ell+1}-\ln(4m\,v^3)\right.\cr
&\left. -3\frac{1+(-1)^{\ell+m}}{(\ell-1)\ell (\ell+1)(\ell+2)}\right)+2\,m\pi\nu^{(2)}(\ell)v^6, 
\end{align}
where 
$\n^{(2)}(\ell)$ is given in Eq.~(\ref{eq:nusol}), 
the phase of the waveforms up to 3PN is given by 
\begin{eqnarray}
\psi_{\ell m} = \Omega(t-r^*)-\varphi+\psi_{\ell m}^{(\rm 3PN)},
\label{eq:psi3pn_fact}
\end{eqnarray}

Using Eq.~(\ref{eq:psi3pn_fact}), we derive $H_{2,\,2}$   to be
\begin{widetext}
\begin{subequations}
\begin{eqnarray}
{H_{2, \,2}}&=&\sqrt{\frac{16\pi}{5}}
{\rm exp}\left[-2\,i\,\psi_{2,\, 2}^{(\rm 3PN)}\right]
\left[1 - {\displaystyle \frac {107}{42}} \,v^{2} + 2\,\pi
 \,v^{3} - {\displaystyle \frac {2173}{1512}} \,v^{4} - 
{\displaystyle \frac {107}{21}} \,\pi \,v^{5} \right.\cr
&&
 + v^{6}\,\left\{ - {\displaystyle \frac {856}{105}} \,{\rm 
eulerlog}(2, \,v) + {\displaystyle \frac {27027409}{646800}}  + 
{\displaystyle \frac {2}{3}} \,\pi ^{2}\right\} - {\displaystyle \frac {
2173}{756}} \,\pi \,v^{7} \cr
&&
 + v^{8}\,\left\{ - {\displaystyle \frac {846557506853}{
12713500800}}  - {\displaystyle \frac {107}{63}} \,\pi ^{2} + 
{\displaystyle \frac {45796}{2205}} \,{\rm eulerlog}(2, \,v)\right\}
 \cr
&&
+ v^{9}\left\{
{\displaystyle \frac {27027409}{323400}} \,\pi  
- {\displaystyle \frac {4}{3}} \,\pi ^{3} 
- {\displaystyle \frac {1712}{105}} \,\pi \,{\rm eulerlog}(2, \,v)
+ i\,\left({\displaystyle \frac {1712}{315}} \,\pi ^{2} 
- {\displaystyle \frac {64}{3}} \,\zeta (3) 
- {\displaystyle \frac {259}{81}}\right) 
\right\} \cr
&&
 + v^{10}\,\left\{ - {\displaystyle \frac {866305477369}{
9153720576}}  - {\displaystyle \frac {2173}{2268}} \,\pi ^{2} + 
{\displaystyle \frac {232511}{19845}} \,{\rm eulerlog}(2, \,v)
\right\} \cr
&&
+ v^{11}\left\{
- {\displaystyle \frac {846557506853}{6356750400}} \,\pi  
+ {\displaystyle \frac {214}{63}} \,\pi ^{3} 
+ {\displaystyle \frac {91592}{2205}} \,\pi \,{\rm eulerlog}(2, \,v) 
\right.\cr
&&
\left.\left. 
+ i\,\left({\displaystyle \frac {3424}{63}} \,\zeta (3) 
+ {\displaystyle \frac {3959}{486}} 
- {\displaystyle \frac {91592}{6615}} \,\pi ^{2} \right)
\right\}\right] 
\label{eq:H22a}\\
&=&
\sqrt{\frac{16\pi}{5}}
{\rm exp}\left[-2\,i\,\left\{\psi_{2,\, 2}^{(\rm 3PN)}
+{v}^{9}\left( {\frac {32}{3}}\,\zeta  \left( 3 \right) -{\frac {856}{315}}\,{\pi }^{2}+{\frac {259}{162}} \right)\right\}\right]\cr
&&
\times\left[
1 - {\displaystyle \frac {107}{42}} \,v^{2} + 2\,\pi \,v^{3} - {\displaystyle \frac {2173}{1512}} \,v^{4} - {\displaystyle \frac {107}{21}} \,\pi \,v^{5} \right.\cr
&&
 + v^{6}\,\left( - {\displaystyle \frac {856}{105}} \,{\rm eulerlog}\left(2, \,v\right) + {\displaystyle \frac {27027409}{646800}}  + {\displaystyle \frac {2}{3}} \,\pi ^{2}\right) - {\displaystyle \frac {2173}{756}} \,\pi \,v^{7} \cr
&&
 + v^{8}\,\left( - {\displaystyle \frac {846557506853}{12713500800}}  - {\displaystyle \frac {107}{63}} \,\pi ^{2} + {\displaystyle \frac {45796}{2205}} \,{\rm eulerlog}\left(2, \,v\right)\right) \cr
&&
 + v^{9}\,\left({\displaystyle \frac {27027409}{323400}} \,\pi  - {\displaystyle \frac {4}{3}} \,\pi ^{3} - {\displaystyle \frac {1712}{105}} \,\pi \,{\rm eulerlog}\left(2, \,v\right)\right) \cr
&&
 + v^{10}\,\left( - {\displaystyle \frac {866305477369}{9153720576}}  - {\displaystyle \frac {2173}{2268}} \,\pi ^{2} + {\displaystyle \frac {232511}{19845}} \,{\rm eulerlog}\left(2, \,v\right)\right) \cr
&&
\left. + v^{11}\,\left( - {\displaystyle \frac {846557506853}{6356750400}} \,\pi  + {\displaystyle \frac {214}{63}} \,\pi ^{3} + {\displaystyle \frac {91592}{2205}} \,\pi \,{\rm eulerlog}\left(2, \,v\right)\right) \right],
\label{eq:H22b}
\end{eqnarray}
\end{subequations}
\end{widetext}
where ${\rm eulerlog}(2, \,v) \equiv \gamma + \ln(4v)$.

Note from Eq.~(\ref{eq:H22a}) that even after factoring out  the 3PN-accurate 
phase $\psi_{2,\,2}^{\rm(3PN)}$, the remaining part of
 $H_{2,\,2}$ is still complex and not real. 
However, with a more accurate choice of $\psi_{2,2}$ involving a $O(v^9)$
term all the imaginary terms (including that at $O(v^{11})$) in the
rest of $H_{2,\,2}$ 
can be absorbed into this phase as can be seen 
in Eq.~(\ref{eq:H22b}).  
Thus, with this improved phase,   the remaining Taylor expansion
of $H_{2,\,2}$ becomes real. 
Thus it is useful to introduce $O(v^9)$ correction to the phase for 
$(\ell,m)=(2,2)$ mode. 
From the 2.5PN formulas for $Z_{\ell m\omega}$ 
in Sec.~\ref{sec:zlmw2.5PN}, 
we find that the leading order of $\hat{H}_{\ell m}$ is 
$O(v^{\ell-2})$ for $\ell+m$ is even and $O(v^{\ell-1})$ for $\ell+m$ is odd. 
Thus, we find that it is also useful to introduce $O(v^9)$ correction 
to the phase for $\ell=2, 3$, $(\ell,m)=(4,4)$ and $(\ell,m)=(4,2)$ modes 
in the computation of 5.5PN waveforms. 
The treatment to go from Eq.~(\ref{eq:H22a}) to Eq.~(\ref{eq:H22b})
 is quite general and the phase $\psi_{\ell m}$ for $2\le\ell\le 4$
 up to $O(v^{9})$ can be similarly derived.  We thus have, 
\begin{widetext}
\begin{subequations}
\begin{eqnarray}
\psi_{2, \,2} &=& \Omega(t-r^*)-\varphi+\left( {\frac {17}{6}}-2\,\gamma-2\,\ln \left( 8\,{v}^{3} \right) \right) {v}^{3}-{\frac {214}{105}}\,\pi \,{v}^{6}+ \left( {\frac {32}{3}}\,\zeta  \left( 3 \right) -{\frac {856}{315}}\,{\pi }^{2}+{\frac {259}{162}} \right) {v}^{9},\\
\psi_{2, \,1} &=& \Omega(t-r^*)-\varphi+\left( \frac{10}{3}-2\,\gamma-2\,\ln  \left( 4\,{v}^{3} \right)  \right) {v}^{3}-{\frac {107}{105}}\,\pi \,{v}^{6}+ \left( -{\frac {214}{315}}\,{\pi }^{2}+\frac{8}{3} \,\zeta \left( 3 \right) +{\frac {29}{81}} \right) {v}^{9},\\
\psi_{3, \,3} &=& \Omega(t-r^*)-\varphi+\left({\frac {127}{30}}-2\,\gamma-2\,\ln  \left( 12\,{v}^{3} \right)  \right) {v}^{3}-{\frac {13}{7}}\,\pi \,{v}^{6}+ \left( -{\frac {26}{7}}\,{\pi }^{2}+24\,\zeta  \left( 3 \right) -{\frac {23173}{9000}} \right) {v}^{9},\\
\psi_{3, \,2} &=& \Omega(t-r^*)-\varphi+\left(\frac{13}{3}-2\,\gamma-2\,\ln\left( 8\,{v}^{3} \right)  \right) {v}^{3}-{\frac {26}{21}}\,\pi \,{v}^{6}+ \left( -{\frac {464}{405}}-{\frac {104}{63}}\,{\pi }^{2}+{\frac {32}{3}}\,\zeta  \left( 3 \right) \right) {v}^{9},\\
\psi_{3, \,1} &=& \Omega(t-r^*)-\varphi+\left( {\frac{127}{30}}-2\,\gamma-2\,\ln  \left( 4\,{v}^{3} \right) \right) {v}^{3}-{\frac {13}{21}}\,\pi \,{v}^{6}+ \left( \frac{8}{3}\,\zeta \left( 3 \right) -{\frac {26}{63}}\,{\pi }^{2}-{\frac {23173}{81000}}\right) {v}^{9},\\
\psi_{4, \,4} &=& \Omega(t-r^*)-\varphi+\left( {\frac{74}{15}}-2\,\gamma-2\,\ln\left( 16\,{v}^{3} \right) \right) {v}^{3}-{\frac {6284}{3465}}\,\pi \,{v}^{6}+ \left( {\frac {128}{3}}\,\zeta  \left( 3 \right) -{\frac {50272}{10395}}\,{\pi }^{2}-{\frac {136528}{10125}} \right) {v}^{9},\\
\psi_{4,\,3} &=& \Omega(t-r^*)-\varphi+\left( {\frac {149}{30}}-2\,\gamma-2\,\ln  \left( 12\,{v}^{3}\right)  \right) {v}^{3}-{\frac {1571}{1155}}\,\pi \,{v}^{6},\\
\psi_{4,\,2} &=& \Omega(t-r^*)-\varphi+\left( {\frac {74}{15}}-2\,\gamma-2\,\ln  \left( 8\,{v}^{3} \right) \right) {v}^{3}-{\frac {3142}{3465}}\,\pi \,{v}^{6}+ \left( -{\frac {12568}{10395}}\,{\pi }^{2}+{\frac {32}{3}}\,\zeta  \left( 3 \right) -{\frac {34132}{10125}} \right) {v}^{9},\\
\psi_{4,\,1} &=& \Omega(t-r^*)-\varphi+\left( {\frac {149}{30}}-2\,\gamma-2\,\ln  \left( 4\,{v}^{3} \right) \right) {v}^{3}-{\frac {1571}{3465}}\,\pi \,{v}^{6},
\end{eqnarray}
\label{eq:psi_lm}
\end{subequations}
\end{widetext}
where $\zeta(n)$ is the zeta function. 
The $O(v^{9})$ terms in the above equations are one of the new results
derived in this paper while the $O(v^{6})$ terms are consistent 
with Ref.~\cite{ref:TS} as required. 
Note that we show the phase $\psi_{\ell m}$ up to $O(v^{6})$ 
for $(\ell,m)=(4,3)$ and $(4,1)$ modes. These corrections to the phase 
$\psi_{\ell m}$ up to $O(v^{3})$, $O(v^{6})$ and $O(v^{9})$ represents 
phase shift in the waveforms due to the tail effects. 

Using $\psi_{\ell m}$ in the above Eq.~(\ref{eq:psi_lm}) and 
$Z_{\ell m\omega}$ which can be derived by the method 
in Sec.~\ref{sec:MST}, 
the amplitudes $\hat{H}_{\ell m}$ up to $O(v^{11})$ for $2\le\ell\le 4$
are derived as 
\begin{widetext}
\begin{subequations}
\begin{eqnarray}
{\hat{H}_{2, \,2}}&=&
1 - {\displaystyle \frac {107}{42}} \,v^{2} + 2\,\pi \,v^{3} - {\displaystyle \frac {2173}{1512}} \,v^{4} - {\displaystyle \frac {107}{21}} \,\pi \,v^{5} \cr
&&
 + v^{6}\,\left( - {\displaystyle \frac {856}{105}} \,{\rm eulerlog}\left(2, \,v\right) + {\displaystyle \frac {27027409}{646800}}  + {\displaystyle \frac {2}{3}} \,\pi ^{2}\right) - {\displaystyle \frac {2173}{756}} \,\pi \,v^{7} \cr
&&
 + v^{8}\,\left( - {\displaystyle \frac {846557506853}{12713500800}}  - {\displaystyle \frac {107}{63}} \,\pi ^{2} + {\displaystyle \frac {45796}{2205}} \,{\rm eulerlog}\left(2, \,v\right)\right) \cr
&&
 + v^{9}\,\left({\displaystyle \frac {27027409}{323400}} \,\pi  - {\displaystyle \frac {4}{3}} \,\pi ^{3} - {\displaystyle \frac {1712}{105}} \,\pi \,{\rm eulerlog}\left(2, \,v\right)\right) \cr
&&
 + v^{10}\,\left( - {\displaystyle \frac {866305477369}{9153720576}}  - {\displaystyle \frac {2173}{2268}} \,\pi ^{2} + {\displaystyle \frac {232511}{19845}} \,{\rm eulerlog}\left(2, \,v\right)\right) \cr
&&
 + v^{11}\,\left( - {\displaystyle \frac {846557506853}{6356750400}} \,\pi  + {\displaystyle \frac {214}{63}} \,\pi ^{3} + {\displaystyle \frac {91592}{2205}} \,\pi \,{\rm eulerlog}\left(2, \,v\right)\right) ,\\
{\hat{H}_{2, \,1}}&=&
-{\displaystyle \frac {1}{3}} \,i\left(v - {\displaystyle \frac {17}{28}} \,v^{3} + \pi \,v^{4} - {\displaystyle \frac {43}{126}} \,v^{5} - {\displaystyle \frac {17}{28}} \,\pi \,v^{6} \right.\cr
&&
 + v^{7}\,\left({\displaystyle \frac {30811367}{2910600}}  - {\displaystyle \frac {214}{105}} \,{\rm eulerlog}\left(1, \,v\right) + {\displaystyle \frac {1}{6}} \,\pi ^{2}\right) - {\displaystyle \frac {43}{126}}\,\pi \,v^{8}  \cr
&&
 + v^{9}\,\left({\displaystyle \frac {46023611}{10594584}}  - {\displaystyle \frac {17}{168}} \,\pi ^{2} + {\displaystyle \frac {1819}{1470}} \,{\rm eulerlog}\left(1, \,v\right)\right) \cr
&&
 + v^{10}\,\left({\displaystyle \frac {30811367}{2910600}} \,\pi  - {\displaystyle \frac {214}{105}} \,\pi \,{\rm eulerlog}\left(1, \,v\right) - {\displaystyle \frac {1}{6}} \,\pi ^{3}\right) \cr
&&
\left. + v^{11}\,\left({\displaystyle \frac {4601}{6615}} \,{\rm eulerlog}\left(1, \,v\right) + {\displaystyle \frac {13504725881}{1191890700}}  - {\displaystyle \frac {43}{756}} \,\pi ^{2}\right)\right) ,\\
{\hat{H}_{3, \,3}}&=&
{\displaystyle \frac {3}{56}} \,\sqrt{210}\,i\left(v - 4\,v^{3} + 3\,\pi \,v^{4} + {\displaystyle \frac {123}{110}} \,v^{5} - 12\,\pi \,v^{6} \right.\cr
&&
 + v^{7}\,\left({\displaystyle \frac {3}{2}} \,\pi ^{2} + {\displaystyle \frac {109301083}{1401400}}  - {\displaystyle \frac {78}{7}} \,{\rm eulerlog}\left(3, \,v\right)\right) + {\displaystyle \frac {369}{110}} \,\pi\,v^{8}  \cr
&&
 + v^{9}\,\left({\displaystyle \frac {312}{7}} \,{\rm eulerlog}\left(3, \,v\right) - {\displaystyle \frac {84974767}{350350}}  - 6\,\pi ^{2}\right) \cr
&&
 + v^{10}\,\left({\displaystyle \frac {327903249}{1401400}} \,\pi  - {\displaystyle \frac {234}{7}} \,\pi \,{\rm eulerlog}\left(3, \,v\right) - {\displaystyle \frac {9}{2}} \,\pi ^{3}\right) \cr
&&
\left. + v^{11}\,\left( - {\displaystyle \frac {96055340127}{2620618000}}  - {\displaystyle \frac {4797}{385}} \,{\rm eulerlog}\left(3, \,v\right) + {\displaystyle \frac {369}{220}} \,\pi ^{2}\right)\right) ,\\
{\hat{H}_{3, \,2}}&=&
{\displaystyle \frac {\sqrt{35}}{21}} \left(v^{2} - {\displaystyle \frac {193}{90}} \,v^{4} + 2\,\pi \,v^{5} - {\displaystyle \frac {1451}{3960}} \,v^{6} - {\displaystyle \frac {193}{45}} \,\pi \,v^{7} \right.\cr
&&
 + v^{8}\,\left({\displaystyle \frac {2501041027}{75675600}}  + {\displaystyle \frac {2}{3}} \,\pi ^{2} - {\displaystyle \frac {104}{21}} \,{\rm eulerlog}\left(2, \,v\right)\right) - {\displaystyle \frac {1451}{1980}} \,\pi \,v^{9} \cr
&&
 + v^{10}\,\left( - {\displaystyle \frac {23891243939}{495331200}}  - {\displaystyle \frac {193}{135}} \,\pi ^{2} + {\displaystyle \frac {10036}{945}} \,{\rm eulerlog}\left(2, \,v\right)\right) \cr
&&
\left. + v^{11}\,\left({\displaystyle \frac {2501041027}{37837800}} \,\pi  - {\displaystyle \frac {4}{3}} \,\pi ^{3} - {\displaystyle \frac {208}{21}} \,\pi \,{\rm eulerlog}\left(2, \,v\right)\right)\right) ,\\
{\hat{H}_{3, \,1}}&=&
-{\displaystyle \frac {\sqrt{14}}{168}} \,i\left(v - {\displaystyle \frac {8}{3}} \,v^{3} + \pi \,v^{4} + {\displaystyle \frac {607}{198}} \,v^{5} - {\displaystyle \frac {8}{3}} \,\pi \,v^{6} \right.\cr
&&
 + v^{7}\,\left({\displaystyle \frac {305915969}{37837800}}  + {\displaystyle \frac {1}{6}} \,\pi ^{2} - {\displaystyle \frac {26}{21}} \,{\rm eulerlog}\left(1, \,v\right)\right) + {\displaystyle \frac {607}{198}} \,\pi\,v^{8}  \cr
&&
 + v^{9}\,\left( - {\displaystyle \frac {4}{9}} \,\pi ^{2} - {\displaystyle \frac {15638341}{1091475}}  + {\displaystyle \frac {208}{63}} \,{\rm eulerlog}\left(1, \,v\right)\right) \cr
&&
 + v^{10}\,\left({\displaystyle \frac {305915969}{37837800}} \,\pi  - {\displaystyle \frac {26}{21}} \,\pi \,{\rm eulerlog}\left(1, \,v\right) - {\displaystyle \frac {1}{6}} \,\pi ^{3}\right) \cr
&&
\left. + v^{11}\,\left({\displaystyle \frac {3262398379463}{127362034800}}  - {\displaystyle \frac {7891}{2079}} \,{\rm eulerlog}\left(1, \,v\right) + {\displaystyle \frac {607}{1188}} \,\pi ^{2}\right)\right) ,\\
{\hat{H}_{4, \,4}}&=&
 - {\displaystyle \frac {8}{63}} \sqrt{35}\left(v^{2} - {\displaystyle \frac {593}{110}} \,v^{4} + 4\,\pi \,v^{5} + {\displaystyle \frac {1068671}{200200}} \,v^{6} - {\displaystyle \frac {1186}{55}} \,\pi \,v^{7} \right.\cr
&&
 + v^{8}\,\left({\displaystyle \frac {302024067749}{2497294800}}  + {\displaystyle \frac {8}{3}} \,\pi ^{2} - {\displaystyle \frac {50272}{3465}} \,{\rm eulerlog}\left(4, \,v\right)\right) + {\displaystyle \frac {1068671}{50050}} \,\pi \,v^{9} \cr
&&
 + v^{10}\,\left( - {\displaystyle \frac {796354151819507}{1436905008000}}  - {\displaystyle \frac {2372}{165}} \,\pi ^{2} + {\displaystyle \frac {14905648}{190575}} \,{\rm eulerlog}\left(4, \,v\right)\right) \cr
&&
\left. + v^{11}\,\left({\displaystyle \frac {302024067749}{624323700}} \,\pi  - {\displaystyle \frac {201088}{3465}} \,\pi \,{\rm eulerlog}\left(4, \,v\right) - {\displaystyle \frac {32}{3}} \,\pi ^{3}\right)\right) ,\\
{\hat{H}_{4, \,3}}&=&
{\displaystyle \frac {9}{280}} \,\sqrt{70}\,i\left(v^{3} - {\displaystyle \frac {39}{11}} \,v^{5} + 3\,\pi \,v^{6} + {\displaystyle \frac {7206}{5005}} \,v^{7} - {\displaystyle \frac {117}{11}} \,\pi\,v^{8}  \right.\cr
&&
 + v^{9}\,\left({\displaystyle \frac {9204203473}{138738600}}  + {\displaystyle \frac {3}{2}} \,\pi ^{2} - {\displaystyle \frac {3142}{385}} \,{\rm eulerlog}\left(3, \,v\right)\right) + {\displaystyle \frac {21618}{5005}} \,\pi\,v^{10}  \cr
&&
\left. + v^{11}\,\left( - {\displaystyle \frac {1658233837937}{8648039400}}  - {\displaystyle \frac {117}{22}} \,\pi ^{2} + {\displaystyle \frac {122538}{4235}} \,{\rm eulerlog}\left(3, \,v\right)\right)\right) ,\\
{\hat{H}_{4, \,2}}&=&
{\displaystyle \frac {\sqrt{5}}{63}} \left(v^{2} - {\displaystyle \frac {437}{110}} \,v^{4} + 2\,\pi \,v^{5} + {\displaystyle \frac {1038039}{200200}} \,v^{6} - {\displaystyle \frac {437}{55}} \,\pi \,v^{7} \right.\cr
&&
 + v^{8}\,\left({\displaystyle \frac {67008495809}{2497294800}}  + {\displaystyle \frac {2}{3}} \,\pi ^{2} - {\displaystyle \frac {12568}{3465}} \,{\rm eulerlog}\left(2, \,v\right)\right) + {\displaystyle \frac {1038039}{100100}} \,\pi \,v^{9} \cr
&&
 + v^{10}\,\left( - {\displaystyle \frac {1826104347159431}{18679765104000}}  + {\displaystyle \frac {2746108}{190575}} \,{\rm eulerlog}\left(2, \,v\right) - {\displaystyle \frac {437}{165}} \,\pi ^{2}\right) \cr
&&
\left. + v^{11}\,\left({\displaystyle \frac {67008495809}{1248647400}} \,\pi  - {\displaystyle \frac {25136}{3465}} \,\pi \,{\rm eulerlog}\left(2, \,v\right) - {\displaystyle \frac {4}{3}} \,\pi ^{3}\right)\right) ,\\
{\hat{H}_{4, \,1}}&=&
-{\displaystyle \frac {\sqrt{10}}{840}} \,i\left(v^{3} - {\displaystyle \frac {101}{33}} \,v^{5} + \pi \,v^{6} + {\displaystyle \frac {42982}{15015}} \,v^{7} - {\displaystyle \frac {101}{33}} \,\pi\,v^{8}  \right.\cr
&&
 + v^{9}\,\left({\displaystyle \frac {9092103793}{1248647400}}  - {\displaystyle \frac {3142}{3465}} \,{\rm eulerlog}\left(1, \,v\right) + {\displaystyle \frac {1}{6}} \,\pi ^{2}\right) + {\displaystyle \frac {42982}{15015}} \,\pi\,v^{10}  \cr
&&
\left. + v^{11}\,\left({\displaystyle \frac {317342}{114345}} \,{\rm eulerlog}\left(1, \,v\right) - {\displaystyle \frac {11770664091577}{700491191400}}  - {\displaystyle \frac {101}{198}} \,\pi ^{2}\right)\right) ,
\end{eqnarray}
\end{subequations}
\end{widetext}
\begin{eqnarray}
{\rm where}\;
{\rm eulerlog}(m, \,v) = \gamma + \ln(2mv).
\end{eqnarray}
The results in this paper 
are consistent with 3PN results in the test-particle limit 
in Ref.~\cite{BFIS08}. 
The $O(v^{7})$ to $O(v^{11})$ terms in the above equations are 
one of the new results derived in this paper. 
We also note that the spherical harmonic modes in this paper are simpler 
than Ref.~\cite{BFIS08} since we completely factored out the phase. 
We show 5.5PN expressions of $\hat{H}_{\ell m}$ for $5\le\ell\le 13$ 
in Appendix~\ref{sec:55PN}. 
\subsection*{2.5PN formulas for $h_{\ell m}$}
\label{sec:hlm2.5PN}
In this section, we derive 2.5PN formulas for spherical harmonic modes 
$h_{\ell m}$ in the test-particle limit using 2.5PN formula of 
$Z_{\ell m\omega}$ in Eq.~(\ref{eq:zlmw2.5PN}) in Sec.~\ref{sec:zlmw2.5PN}. 
Once we have the 2.5PN formula of $Z_{\ell m\omega}$ in 
Eq.~(\ref{eq:zlmw2.5PN}), we can derive 2.5PN $h_{\ell m}$ 
from Eq.~(\ref{eq:hlm2zlmw_b}) as 
\begin{widetext}
\begin{subequations}
\begin{eqnarray}
\label{eq:hlm2.5PNa}
h_{\ell m} &=& -\frac{2}{r}\,\frac{Z_{\ell m\omega}\,e^{i\,m\Omega (r^{*}-t)}\,e^{i\,m\,\varphi}}{\omega^2},\\
\label{eq:hlm2.5PNb}
&=& -{2\,Z_{\ell m\omega}^{(0)}\over r\,\omega^2}
e^{-i\,m\,[\Omega(t-r^*)-\varphi+\psi_{\ell m}^{\rm (3PN)}]}
\left[1+Z_{\ell m}^{(2)}v^2+m\pi v^3+Z_{\ell m}^{(4)}v^4+m\pi Z_{\ell m}^{(2)}v^5+O(v^6)\right],
\end{eqnarray}
\label{eq:hlm2.5PN}
\end{subequations}
\end{widetext}
where $Z_{\ell m \omega}^{(0)}$, $Z_{\ell m}^{(2)}$ and $Z_{\ell m}^{(4)}$ 
are given in Eq.~(\ref{eq:zlmw2.5PN_even}) for $\ell+m={\rm even}$ case 
and Eq.~(\ref{eq:zlmw2.5PN_odd}) for $\ell+m={\rm odd}$ case. If we define 
$h_{\ell m}^{(0)}\equiv -2\,Z_{\ell m\omega}^{(0)}/(r\,\omega^2)$, 
$h_{\ell m}^{(0)}$ is given as 
\begin{subequations}
\begin{eqnarray}
h_{\ell m}^{(0)}&=& \frac{2}{r}\frac{\m m^{\ell}\pi^{3/2}}{i^\ell 2^{\ell-1} \Gamma(\ell+3/2)} \sqrt{\frac{(\ell+2)(\ell+1)}{\ell(\ell-1)}}\cr 
&&\times {}_{0}Y_{\ell m}\left(\frac{\pi}{2},\,0\right)v^\ell ,
\;(\ell+m={\rm even}),\\
h_{\ell m}^{(0)}&=& \frac{2}{r}\frac{\m m^{\ell} \pi^{3/2}}{i^{\ell+1} 2^{\ell-2}\Gamma(\ell+3/2)}\sqrt{\frac{\ell+2}{\ell-1}}\cr
&&\times {}_{-1}Y_{\ell m}\left(\frac{\pi}{2},\,0\right)v^{\ell+1},
\; (\ell+m={\rm odd}).
\end{eqnarray}
\end{subequations}
\section{Resummed waveforms}
\label{sec:rholm}
Recently, Damour, Iyer and Nagar~\cite{DIN} suggested 
a factorized resummed waveform 
which improves agreement with the results of numerical simulations. 
They decomposed the waveforms into five factors as 
\begin{eqnarray}
h_{\ell m}=h_{\ell m}^{({\rm N},\e_p)}\,\hat{S}_{\rm eff}^{(\e_p)}\,T_{\ell m}\,
e^{i\delta_{\ell m}}(\rho_{\ell m})^\ell\,.
\label{eq:rholm}
\end{eqnarray}
Here, $\e_p$ denotes the parity of the multipolar waveforms. In the case of 
circular orbits, $\e_p=0$ when $\ell+m$ is even, and $\e_p=1$ when $\ell+m$ is 
odd. 

The first factor $h_{\ell m}^{({\rm N},\e_p)}$ represents 
the Newtonian contribution to waveforms. 
\begin{equation}
h_{\ell m}^{({\rm N},\e_p)}=\frac{GM\nu}{c^2\,r}n_{\ell m}^{(\e_p)}\,c_{\ell+\e_p}(\nu)\,v^{(\ell+\e_p)}\,Y^{\ell-\e_p,-m}\,\left(\frac{\pi}{2},\phi\right)\,,
\label{eq:rholmN}
\end{equation}
where $\phi$ is the orbital phase and $n_{\ell m}^{(\epsilon_p)}$ are
\begin{subequations}
\begin{align}
n^{(0)}_{\ell m}=&(im)^\ell\frac{8\pi}{(2\ell+1)!!}\sqrt{\frac{(\ell+1)(\ell+2)}{\ell(\ell-1)}}\,, \\
n^{(1)}_{\ell m}=&-(im)^\ell\frac{16\pi i}{(2\ell+1)!!}\sqrt{\frac{(2\ell+1)(\ell+2)(\ell^2-m^2)}{(2\ell-1)(\ell+1)\ell(\ell-1)}}\,,
\end{align}
\end{subequations}
and $c_{\ell+\e_p}(\nu)$ are functions 
of the symmetric mass ratio $\nu\equiv \mu\,M/(M+\mu)^2$, defined by 
\begin{eqnarray}
c_{\ell+\e_p}(\nu)&=&\left(\frac{1}{2}-\frac{1}{2}\sqrt{1-4\nu}\right)^{\ell+\e_p-1} \cr
&&+(-)^{\ell+\e_p}\left(\frac{1}{2}+\frac{1}{2}\sqrt{1-4\nu}\right)^{\ell+\e_p-1}.
\end{eqnarray}

The second factor in Eq.~(\ref{eq:rholm}), $\hat{S}_{\rm eff}^{(\e_p)}$, 
is motivated by the effective source term for partial waves in the perturbation
formalism and  its replacement  by  analogous quantities that characterize 
the effective-one-body (EOB) dynamics~\cite{DIN}, 
\begin{equation}
\hat{S}_{\rm eff}^{(\e_p)}=\left\{
  \begin{array}{lc}
    \tilde E\,, & \hbox{for}\,\,\,\e_p=0\,, \\
    v\tilde L/M\,, & \hbox{for}\,\,\,\e_p=1\,. 
  \end{array}
\right.
\end{equation}

The third factor in Eq.~(\ref{eq:rholm}), $T_{\ell m}$, is 
the resummed tail factor which resums the leading logarithms of the tail 
effects~\cite{Damour:2007xr,Damour:2007yf,DIN}
\begin{equation}
T_{\ell m}=\frac{\Gamma(\ell+1-2i\hat{\hat{k}})}{\Gamma(\ell+1)}\,e^{\pi\hat{\hat{k}}}\,e^{2i\hat{\hat{k}}\,\ln(2kr_{0s})}\,,
\label{eq:Tlm}
\end{equation}
where $k=m\,\Omega$ and $\hat{\hat{k}}=M\,k$. 
Here we denote by $r_{0s}$ what was denoted by $r_0$ in~\cite{DIN}. 
As pointed out in Sec.~\ref{sec:zlmw2.5PN}, 
$Z_{\ell m\omega}\propto e^{\pi\e/2}e^{i\e\ln 2\e}$ and this 
motivates the idea to introduce $e^{\pi\hat{\hat{k}}}$ 
in addition to $e^{2i\hat{\hat{k}}\,\ln(2kr_{0s})}$ in the resummed tail factor
to improve the convergence of the residual PN series. 

The fourth factor in Eq.~(\ref{eq:rholm}), $\delta_{\ell m}$, 
is a supplementary phase of the resummed tail factor, 
$T_{\ell m}$. 
If we decompose $T_{\ell m}$ as 
$T_{\ell m}=\mid T_{\ell m}\mid e^{i\t_{\ell m}}$, 
the phase of $T_{\ell m}$, i.e. $\t_{\ell m}$, 
can be derived by the post-Newtonian expansion of $T_{\ell m}$ up to 
the required order. 
Using Eqs.~(\ref{eq:hlm_fact}) and (\ref{eq:rholm}), 
the difference between the phase $\psi_{\ell m}$ 
and the phase of the resummed tail factor $T_{\ell m}$ is included into 
$\delta_{\ell m}$ up to 4.5PN as 
\begin{subequations}
\begin{eqnarray}
\delta_{\ell m}(r_{0s}) &=& -m\psi_{\ell m}+m\phi-\t_{\ell m}(r_{0s})\\
&=&-m\psi_{\ell m}+m\Omega(t-r^{*})\cr
&&-m\,v^3\left(2\ln\left({2m\,r_{0s}\,v^3\over M}\right)-2\Psi^{(0)}(\ell)-\frac{2}{\ell}\right)\cr
&&-(mv^3)^3\left(\frac{4}{3}\Psi^{(2)}(\ell)+\frac{8}{3\ell^3}\right)\\
&=&2mv^3\left(\frac{2}{\ell}-\frac{2}{\ell+1}+3\frac{1+(-1)^{\ell+m}}{(\ell-1)\ell (\ell+1)(\ell+2)}\right.\cr
&&\left.-2\ln\left({r_{0s}\over 2\,M\,e^{-1/2}}\right)\right)\cr
&&-2\pi\nu^{(2)}(\ell)(mv^3)^2-m\psi_{\ell m}^{(9)}v^9\cr
&&-(mv^3)^3\left(\frac{4}{3}\Psi^{(2)}(\ell)+\frac{8}{3\ell^3}\right),
\end{eqnarray}
\label{eq:psi2delta_45pn}
\end{subequations}
where $\psi_{\ell m}^{(9)}$ is $O(v^9)$ term of $\psi_{\ell m}$ in 
Eq.~(\ref{eq:hlm_fact}) and $\phi=\Omega(t-r^{*})$ 
comes from the spherical harmonics,  
$Y^{\ell-\e_p,-m}\,\left(\frac{\pi}{2},\phi\right)\propto e^{-im\phi}$, 
in the Newtonian contribution to $h_{\ell m}$ in Eq.~(\ref{eq:rholmN}). 
Note that $-\Omega r^{*}$ in $\phi$ is introduced 
to cancel out $-\Omega r^{*}$ in $\psi_{\ell m}$ and may be interpreted 
as the initial value of $\phi$.

The choice\footnote{Note that in the generic mass case, where one works in harmonic 
coordinates~\cite{K07,BFIS08,DIN} $\ln x_0$ is chosen to be 
$\ln x_0 = 11/18-2/3\gamma-4/3\ln 2+2/3\ln \left[G(m_1+m_2)/c^2/r_0^{\rm h}\right]$, where $r_0^{\rm h}$ is a freely-specifiable constant  
related to the choice of the origin of the retarded time in radiative coordinates relative  to harmonic coordinates. 
We denote by $r_0^{\rm h}$ the $r_0$ in Ref.~\cite{K07,BFIS08} to avoid conflict with the notation used in perturbation works 
where $r_0$ as in this paper denotes the orbital radius. 
With  $r_0^{\rm h}=2\,M/\sqrt{e}$ the above expression reduces to   $\ln x_0 = 17/18-2\ln 2-2\gamma/3$, 
the relation used to match PN results in terms of $\ln x_0$ to  black hole perturbation results in Schwarzschild coordinates. 
}  of $r_{0s}=2\,M/\sqrt{e}$ in Schwarzschild coordinate will 
reproduce the phase $\delta_{\ell m}$ in Ref.~\cite{DIN}.
In this paper, we choose $r_{0s}=2\,M/\sqrt{e}$ to 
reproduce $\delta_{\ell m}$ in Ref.~\cite{DIN} and define as 
$\delta_{\ell m}(r_{0s})\equiv\delta_{\ell m}$. 
Using Eqs.~(\ref{eq:psi_lm}) and (\ref{eq:psi2delta_45pn}), 
we can derive 4.5PN expression of $\delta_{\ell m}$ for $2\le\ell\le 4$ as 
\begin{subequations}
\begin{align}
\delta_{2,\,2} =& \frac{7}{3}\,{v}^{3}+{\frac {428}{105}}\,\pi \,{v}^{6}+ \left( -{\frac {2203}{81}}+{\frac {1712}{315}}\,{\pi }^{2} \right) {v}^{9},\\
\delta_{2,\,1} =& \frac{2}{3}\,{v}^{3}+{\frac {107}{105}}\,\pi \,{v}^{6}+ \left( -{\frac {272}{81}}+{\frac {214}{315}}\,{\pi }^{2} \right) {v}^{9},\\
\delta_{3,\,3} =& {\frac {13}{10}}\,{v}^{3}+{\frac {39}{7}}\,\pi \,{v}^{6}+ \left( -{\frac {227827}{3000}}+{\frac {78}{7}}\,{\pi }^{2} \right) {v}^{9},\\
\delta_{3,\,2} =& \frac{2}{3}\,{v}^{3}+{\frac {52}{21}}\,\pi \,{v}^{6}+ \left( -{\frac {9112}{405}}+{\frac {208}{63}}\,{\pi }^{2} \right) {v}^{9}, \\
\delta_{3,\,1} =& {\frac {13}{30}}\,{v}^{3}+{\frac {13}{21}}\,\pi \,{v}^{6}+ \left( -{\frac {227827}{81000}}+{\frac {26}{63}}\,{\pi }^{2} \right) {v}^{9},\\
\delta_{4,\,4} =& {\frac {14}{15}}\,{v}^{3}+{\frac {25136}{3465}}\,\pi \,{v}^{6}+ \left( -{\frac {55144}{375}}+{\frac {201088}{10395}}\,{\pi }^{2} \right) {v}^{9},\\
\delta_{4,\,3} =& \frac{3}{5}\,{v}^{3}+{\frac {1571}{385}}\,\pi \,{v}^{6},\\
\delta_{4,\,2} =& {\frac {7}{15}}\,{v}^{3}+{\frac {6284}{3465}}\,\pi \,{v}^{6}+ \left( -{\frac {6893}{375}}+{\frac {25136}{10395}}\,{\pi }^{2} \right) {v}^{9},\\
\delta_{4,\,1} =& \frac{1}{5}\,{v}^{3}+{\frac {1571}{3465}}\,\pi \,{v}^{6}.
\end{align}
\label{eq:delta_lm}
\end{subequations}
Note the new $O(v^6)$ corrections for $(\ell,m)=(2,1)$, $\ell=3$ and $\ell=4$ 
modes and the new $O(v^9)$ corrections in Eq.~(\ref{eq:delta_lm}) 
beyond those available from Ref.~\cite{DIN}. 
This is because, in the case of the test-particle limit, 
Ref.~\cite{DIN} used the results in Ref.~\cite{ref:TTS}, 
that provided only the 5.5PN energy flux but not the 5.5PN GW polarizations.
Having computed the 5.5PN waveforms in this work we are able to improve on
this accuracy. 

According to Ref.~\cite{DIN}, the decomposition of the post-Newtonian 
waveforms into five factors improves the convergence of the waveforms 
since the coefficients of the post-Newtonian expansion become smaller. 
However, the convergence of the amplitude, 
$\mid h_{\ell m}/(h_{\ell m}^{({\rm N},\e_p)}\,\hat{S}_{\rm eff}^{(\e_p)}\,T_{\ell m}\,e^{i\delta_{\ell m}})\mid$, was not good enough around 
the innermost stable circular orbit (ISCO). 
To alleviate this problem, 
Ref.~\cite{DIN} introduced the $\ell$-th root of the amplitude,
 $\rho_{\ell m}$, to deal with the  linear dependence on $\ell$ in the
 1PN terms of the amplitude (Notice the related  linear dependence on
 $\ell$ in the 1PN terms of 
$Z_{\ell m\omega}$ derived in Sec.~\ref{sec:zlmw2.5PN}). 
Then, the coefficients of the post-Newtonian expansion become smaller and 
give much better improvement even around ISCO. 
As shown in Sec.~\ref{sec:numerical}, 
factorized resummed waveforms achieve about 5 times better agreement 
with numerical calculation than Taylor expanded waveforms.

Using Eqs.~(\ref{eq:hlm2zlmw}) and (\ref{eq:rholm}), 
$\rho_{\ell m}$ can be derived as $\rho_{\ell m} =(h_{\ell m}/(h_{\ell m}^{({\rm N},\e_p)}\,\hat{S}_{\rm eff}^{(\e_p)}\,T_{\ell m}\,e^{i\delta_{\ell m}}))^{1/\ell}$. 
Then, we can derive 5PN expressions of $\rho_{\ell m}$ for $2\le\ell\le 4$ as 
\begin{widetext}
\begin{subequations}
\begin{eqnarray}
{\rho _{2, \,2}}&=&
1 - {\displaystyle \frac {43}{42}} \,v^{2} - {\displaystyle \frac {20555}{10584}} \,v^{4} + \left({\displaystyle \frac {1556919113}{122245200}}  - {\displaystyle \frac {428}{105}} \,{\rm eulerlog}\left(2, \,v\right)\right)\,v^{6} \cr
&&
 + \left({\displaystyle \frac {9202}{2205}} \,{\rm eulerlog}\left(2, \,v\right) - {\displaystyle \frac {387216563023}{160190110080}} \right)\,v^{8} \cr
&&
 + \left({\displaystyle \frac {439877}{55566}} \,{\rm eulerlog}\left(2, \,v\right) - {\displaystyle \frac {16094530514677}{533967033600}} \right)\,v^{10},\\
{\rho _{2, \,1}}&=&
1 - {\displaystyle \frac {59}{56}} \,v^{2} - {\displaystyle \frac {47009}{56448}} \,v^{4} + \left({\displaystyle \frac {7613184941}{2607897600}}  - {\displaystyle \frac {107}{105}} \,{\rm eulerlog}\left(1, \,v\right)\right)\,v^{6} \cr
&&
 + \left( - {\displaystyle \frac {1168617463883}{911303737344}}  + {\displaystyle \frac {6313}{5880}} \,{\rm eulerlog}\left(1, \,v\right)\right)\,v^{8} \cr
&&
 + \left( - {\displaystyle \frac {63735873771463}{16569158860800}}  + {\displaystyle \frac {5029963}{5927040}} \,{\rm eulerlog}\left(1, \,v\right)\right)\,v^{10},\\
{\rho _{3, \,3}}&=&
1 - {\displaystyle \frac {7}{6}} \,v^{2} - {\displaystyle \frac {6719}{3960}} \,v^{4} + \left({\displaystyle \frac {3203101567}{227026800}}  - {\displaystyle \frac {26}{7}} \,{\rm eulerlog}\left(3, \,v\right)\right)\,v^{6} \cr
&&
 + \left({\displaystyle \frac {13}{3}} \,{\rm eulerlog}\left(3, \,v\right) - {\displaystyle \frac {57566572157}{8562153600}} \right)\,v^{8} \cr
&&
 + \left( - {\displaystyle \frac {903823148417327}{30566888352000}}  + {\displaystyle \frac {87347}{13860}} \,{\rm eulerlog}\left(3, \,v\right)\right)\,v^{10},\\
{\rho _{3, \,2}}&=&
1 - {\displaystyle \frac {164}{135}} \,v^{2} - {\displaystyle \frac {180566}{200475}} \,v^{4} + \left({\displaystyle \frac {5849948554}{940355325}}  - {\displaystyle \frac {104}{63}} \,{\rm eulerlog}\left(2, \,v\right)\right)\,v^{6} \cr
&&
 + \left( - {\displaystyle \frac {10607269449358}{3072140846775}}  + {\displaystyle \frac {17056}{8505}} \,{\rm eulerlog}\left(2, \,v\right)\right)\,v^{8},\\
{\rho _{3, \,1}}&=&
1 - {\displaystyle \frac {13}{18}} \,v^{2} + {\displaystyle \frac {101}{7128}} \,v^{4} + \left({\displaystyle \frac {11706720301}{6129723600}}  - {\displaystyle \frac {26}{63}} \,{\rm eulerlog}\left(1, \,v\right)\right)\,v^{6} \cr
&&
 + \left({\displaystyle \frac {169}{567}} \,{\rm eulerlog}\left(1, \,v\right) + {\displaystyle \frac {2606097992581}{4854741091200}} \right)\,v^{8} \cr
&&
 + \left({\displaystyle \frac {430750057673539}{297110154781440}}  - {\displaystyle \frac {1313}{224532}} \,{\rm eulerlog}\left(1, \,v\right)\right)\,v^{10},\\
{\rho _{4, \,4}}&=&
1 - {\displaystyle \frac {269}{220}} \,v^{2} - {\displaystyle \frac {14210377}{8808800}} \,v^{4} 
 + \left( - {\displaystyle \frac {12568}{3465}} \,{\rm eulerlog}\left(4, \,v\right) + {\displaystyle \frac {16600939332793}{1098809712000}} \right)\,v^{6} \cr
&&
 + \left({\displaystyle \frac {845198}{190575}} \,{\rm eulerlog}\left(4, \,v\right) - {\displaystyle \frac {172066910136202271}{19426955708160000}} \right)\,v^{8},\\
{\rho _{4, \,3}}&=&
1 - {\displaystyle \frac {111}{88}} \,v^{2} - {\displaystyle \frac {6894273}{7047040}} \,v^{4} 
 + \left( - {\displaystyle \frac {1571}{770}} \,{\rm eulerlog}\left(3, \,v\right) + {\displaystyle \frac {1664224207351}{195343948800}} \right)\,v^{6} \cr
&&
 + \left( - {\displaystyle \frac {2465107182496333}{460490801971200}}  + {\displaystyle \frac {174381}{67760}} \,{\rm eulerlog}\left(3, \,v\right)\right)\,v^{8},\\
{\rho _{4, \,2}}&=&
1 - {\displaystyle \frac {191}{220}} \,v^{2} - {\displaystyle \frac {3190529}{8808800}} \,v^{4} 
 + \left( - {\displaystyle \frac {3142}{3465}} \,{\rm eulerlog}\left(2, \,v\right) + {\displaystyle \frac {848238724511}{219761942400}} \right)\,v^{6} \cr
&&
 + \left({\displaystyle \frac {300061}{381150}} \,{\rm eulerlog}\left(2, \,v\right) - {\displaystyle \frac {12864377174485679}{19426955708160000}} \right)\,v^{8},\\
{\rho _{4, \,1}}&=&
1 - {\displaystyle \frac {301}{264}} \,v^{2} - {\displaystyle \frac {7775491}{21141120}} \,v^{4} 
 + \left( - {\displaystyle \frac {1571}{6930}} \,{\rm eulerlog}\left(1, \,v\right) + {\displaystyle \frac {1227423222031}{1758095539200}} \right)\,v^{6} \cr
&&
 + \left( - {\displaystyle \frac {29584392078751453}{37299754959667200}}  + {\displaystyle \frac {67553}{261360}} \,{\rm eulerlog}\left(1, \,v\right)\right)\,v^{8}.
\end{eqnarray}
\end{subequations}
\end{widetext}
We note that in the factorized resummed waveforms, all the $\rho_{\ell m}$'s 
contain only even powers of $v$~\cite{DIN}. Thus, 5.5PN waveforms produce 
5PN expressions of $\rho_{\ell m}$. 
We show 5PN expressions of $\rho_{\ell m}$ for $5\le\ell\le 7$ 
in Appendix~\ref{sec:rho5PN}, but not for $8\le\ell\le 13$ 
since one can derive them using 2PN expression of $\rho_{\ell m}$ 
in Sec.~\ref{sec:rho2PN}. 5PN expressions of $\rho_{\ell m}$ in this paper 
are consistent  at lower PN orders to expressions derived in Ref.~\cite{DIN}. 
\subsection{2PN formulas for $\rho_{\ell m}$}
\label{sec:rho2PN}
In this section, we derive 2PN formulas for resummed waveforms 
$\rho_{\ell m}$ Eq.~(\ref{eq:rholm}) using 2.5PN formula of 
$Z_{\ell m\omega}$ Eq.~(\ref{eq:zlmw2.5PN}) in Sec.~\ref{sec:zlmw2.5PN}. 
As explained in Sec.~\ref{sec:zlmw2.5PN}, 
we do not have to derive 5PN $\rho_{\ell m}$ for $\ell\ge 8$ 
since one can derive them using the general formulas in this section. 

Once we have the 2.5PN formula of $Z_{\ell m\omega}$ Eq.~(\ref{eq:zlmw2.5PN}), 
we can derive 2PN $\rho_{\ell m}$ and 3PN $\delta_{\ell m}$ 
from Eqs.~(\ref{eq:hlm2zlmw}) and (\ref{eq:rholm}) as
\begin{subequations}
\begin{eqnarray}
\rho_{\ell m} &=&\left(\frac{h_{\ell m}}{h_{\ell m}^{({\rm N},\e_p)}\,\hat{S}_{\rm eff}^{(\e_p)}\,T_{\ell m}\,e^{i\delta_{\ell m}}}\right)^{1/\ell},\\ 
&=& 1 +\rho_{\ell m}^{(2)} v^2 +\rho_{\ell m}^{(4)} v^4 +O(v^6),
\end{eqnarray}
\label{eq:rho2PN}
\end{subequations}
and 
\begin{subequations}
\begin{eqnarray}
\delta_{\ell m} &=& -m\psi_{\ell m}+m\Omega(t-r^{*})-\t_{\ell m},\\
&=& -m\psi_{\ell m}+m\Omega(t-r^{*})\cr
&&-m\,v^3\left(2\ln(4m\,v^3)-1-2\Psi(\ell)-\frac{2}{\ell}\right),\\
&=&2mv^3\left(\frac{2}{\ell}-\frac{2}{\ell+1}+3\frac{1+(-1)^{\ell+m}}{(\ell-1)\ell (\ell+1)(\ell+2)}\right)\cr
&&-2\,(mv^3)^2\pi\nu^{(2)}(\ell), 
\end{eqnarray}
\label{eq:psi2delta}
\end{subequations}
where $\t_{\ell m}$ is the phase of the tail term $T_{\ell m}$, 
defined as $T_{\ell m}=\mid T_{\ell m}\mid e^{i\t_{\ell m}}$. 
As noted in Sec.~\ref{sec:rholm}, factorized resummed waveforms 
$\rho_{\ell m}$ have only even powers of $v$~\cite{DIN}. 
As mentioned earlier in Sec.~\ref{sec:zlmw2.5PN}, 
although this is a 2.5PN calculation, 
we give the 3PN formula for $\delta_{\ell m}$ since there may not exist 
any further 3PN phase contributions. 

Observe that, in the following subsections, 
1PN terms of $\rho_{\ell m}$, i.e. $\rho_{\ell m}^{(2)}$, 
are consistent with that of Ref.~\cite{DIN} and 
2PN terms $\rho_{\ell m}^{(4)}$ are our new results. 
\subsubsection{$\ell+m={\rm even}$ case}
\begin{subequations}
\begin{eqnarray}
\rho_{\ell m}^{(2)} &=& -1+\frac{1}{\ell}-{\frac {{m}^{2} \left( \ell+9 \right) }{2\ell \left( \ell+1 \right)  \left( 2\,\ell+3 \right) }},\\
\rho_{\ell m}^{(4)} &=& -\frac{1}{4} + {\frac {5\,{\ell}^{3}+5\,{\ell}^{2}-4\ell+2}{4\,\left(2\ell-1\right)\left(\ell-1\right){\ell}^{2}}}\cr
&& 
- {\frac {m^2(5\,{\ell}^{5}+20\,{\ell}^{4}+60\,{\ell}^{3}+86\,{\ell}^{2}+39\ell+30)}{2\,{\ell}^{2} \left(2\ell+3\right)\left(\ell+1\right)^{2}\left(\ell+2\right)\left(\ell-1\right)}}\cr
&&-{\frac{m^4(91\,{\ell}^{3}+268\,{\ell}^{2}-249\ell-810)}{8\,{\ell}^{2}\left(2\ell+3\right)^{2}\left(\ell+1\right)^{2}\left(\ell+2\right)\left(2\ell+5\right)}}.
\end{eqnarray}
\label{eq:rho2PNeven}
\end{subequations}
\subsubsection{$\ell+m={\rm odd}$ case}
\begin{subequations}
\begin{eqnarray}
\rho_{\ell m}^{(2)} &=& -1-\frac{1}{\ell}+\frac{2}{\ell^2}-{\frac {{m}^{2} \left( \ell+4 \right) }{2\ell\left( \ell+2 \right)  \left( 2\,\ell+3 \right) }},\\
\rho_{\ell m}^{(4)} &=& -\frac{1}{4} + \frac{{\ell}^{4}-6\,{\ell}^{3}-18\,{\ell}^{2}+32\ell-8}{4\,\left(2\ell-1\right){\ell}^{4}}\cr
&&
 - {\frac{m^2(5\,{\ell}^{4}+20\,{\ell}^{3}+33\,{\ell}^{2}+34\ell+8)}{2\,{\ell}^{3}\left(2\ell+3\right)\left(\ell+2\right)\left(\ell+1\right)}}\cr
&&
-{\frac{m^4(3\,{\ell}^{2}-28\ell-80)}{8\,{\ell}^{2}\left(2\ell+3\right)^{2}\left(\ell+2\right)^{2}\left(2\ell+5\right)}}.
\end{eqnarray}
\label{eq:rho2PNodd}
\end{subequations}
\section{$+$ and $\times$ polarizations}
\label{sec:zetalm}
In the previous sections we have explicitly listed the 5.5PN $h_{\ell m}$ 
and 5PN $\rho_{\ell m}$, that are directly useful
for a comparison of analytical PN results with NR simulations. In this section, we present the general formulas
using which the $+$ and $\times$ polarizations can be obtained from the formulas for $h_{\ell m}$ listed earlier. 
To compare the results in literature in the test-particle limit, 
we use the same notation as in Refs~\cite{ref:poisson,ref:TS} 
to derive plus and cross polarizations of gravitational waveforms.
With $h_{\ell m}^{+,\times}$  defined as in Eq.~(\ref{eq:wave}), we have
\begin{eqnarray}
h_{\ell m}^{+,\times} + h_{\ell,-m}^{+,\times}
=-\left({\mu\over r}\right)\left({M\over r_0}\right)
\zeta_{\ell m}^{+,\times}.
\label{eq:zetadef}
\end{eqnarray}
For gravitational waves propagating towards the observer located
at $(\theta,\varphi)$ relative to the source, we have $(\Theta,\Phi)=(\theta,\varphi)$.
In this case, from Eq.~(\ref{eq:wave}) and (\ref{eq:hlmdef}) 
it  follows that $h_{\ell m}^+-i h_{\ell m}^\times=h_{\ell m} \;_{-2}Y_{\ell m}(\theta,\varphi)$. 
Then we find 
\begin{subequations}
\begin{eqnarray}
-\left({\mu\over r}\right)\left({M\over r_0}\right)\zeta_{\ell m}^+ &=&
{\rm Re}\left[h_{\ell m} \;_{-2}Y_{\ell m}(\theta,\varphi)\right.\cr
&&\left.+h_{\ell,-m} \;_{-2}Y_{\ell,-m}(\theta,\varphi)\right],\\
-\left({\mu\over r}\right)\left({M\over r_0}\right)\zeta_{\ell m}^\times &=&
-{\rm Im}\left[h_{\ell m} \;_{-2}Y_{\ell m}(\theta,\varphi)\right.\cr
&&\left.+h_{\ell,-m} \;_{-2}Y_{\ell,-m}(\theta,\varphi)\right].
\end{eqnarray}
\end{subequations}

Using spherical harmonic modes of gravitational waveforms $\hat{H}_{\ell m}$ 
and phase factor $\psi_{\ell m}$ introduced in Sec.~\ref{sec:hlm}, 
we derive $\zeta_{\ell m}^{+,\times}$ for the even $m$ case as 
\begin{subequations}
\begin{eqnarray}
\zeta_{\ell m}^{+} &=& 8\sqrt{\frac{\pi}{5}}\left[ {}_{-2}Y_{\ell m}(\theta,\,0)+{}_{-2}Y_{\ell m}(\pi-\theta,\,0)\right]\cr
&&\times\cos(m\psi_{\ell m})\hat{H}_{\ell m},\\
\zeta_{\ell m}^{\times} &=& 8i\sqrt{\frac{\pi}{5}}\left[ {}_{-2}Y_{\ell m}(\theta,\,0)-{}_{-2}Y_{\ell m}(\pi-\theta,\,0)\right]\cr
&&\times\left[-i\sin(m\psi_{\ell m})\right]\hat{H}_{\ell m}, 
\end{eqnarray}
\label{eq:zeta1}
\end{subequations}
and for the odd $m$  case as 
\begin{subequations}
\begin{eqnarray}
\zeta_{\ell m}^{+} &=& 8\sqrt{\frac{\pi}{5}}\left[ {}_{-2}Y_{\ell m}(\theta,\,0)+{}_{-2}Y_{\ell m}(\pi-\theta,\,0)\right]\cr
&&\times\left[-i\sin(m\psi_{\ell m})\right]\hat{H}_{\ell m},\\
\zeta_{\ell m}^{\times} &=& 8i\sqrt{\frac{\pi}{5}}\left[ {}_{-2}Y_{\ell m}(\theta,\,0)-{}_{-2}Y_{\ell m}(\pi-\theta,\,0)\right]\cr
&&\times\cos(m\psi_{\ell m})\hat{H}_{\ell m}.
\end{eqnarray}
\label{eq:zeta2}
\end{subequations}
Here we used the known properties that 
$Z_{\ell,-m,-\omega}=(-1)^{\ell}\bar{Z}_{\ell m\omega}$ and 
${}_{s}Y_{\ell,-m}(\theta,\varphi)=(-1)^{\ell+s}{}_{s}\bar{Y}_{\ell m}(\pi-\theta,\varphi)$, where the bar denotes complex conjugate. 
We note that the signs of $\zeta_{\ell m}^{+}$ 
in Refs~\cite{ref:poisson,ref:TS} are opposite to each other, 
as pointed out in Ref.~\cite{K07}. 
The signs of resulting $\zeta_{\ell m}^{+}$ 
in this paper are same as the ones in Ref.~\cite{ref:poisson}.

Using the above relations, we compared $\zeta_{\ell m}^{+,\times}$ to 
the 4PN expressions in Ref.~\cite{ref:TS}. 
We find agreement in almost all the terms except four
corresponding to $\zeta^{+}_{8,7}$, $\zeta^{\times}_{8,7}$ and 
$\zeta^{+}_{10,6}$ in addition to a misprint of the sign 
of $\zeta_{7,3}^{\times}$ in Ref.~\cite{ref:TS}, 
pointed out in Refs.~\cite{ABIQ04,BFIS08}. 
Our resulting  expressions for 
$\zeta^{+}_{8,7}$, $\zeta^{\times}_{8,7}$ and $\zeta^{+}_{10,6}$ are given by,
\begin{widetext}
\begin{eqnarray}
{\zeta^{+}_{8, \,7}}&=&
-{\frac {823543}{829440}}\, \left( 5+7\,\cos \left( 2\,\theta\right)\right)\sin\left(\theta\right)^{5}\,\sin(7\,\psi_{8,7})
\left(v^{7} - {\displaystyle \frac {3343}{380}} \,v^{9} + 7\,\pi\,v^{10} + {\displaystyle \frac {42607}{1710}} \,v^{11}\right),\\
{\zeta^{\times}_{8, \,7}}&=&
{\frac {823543}{414720}}\, \left( \cos \left( 3\,\theta \right) +5\,
\cos \left( \theta \right)  \right)\sin\left(\theta\right)^{5}\,\cos(7\,\psi_{8,7})\left(v^{7} - {\displaystyle \frac {3343}{380}} \,v^{9} + 7\,\pi\,v^{10} + {\displaystyle \frac {42607}{1710}} \,v^{11}\right),\\
{\zeta^{+}_{10, \,6}}&=&
-{\frac {2187}{115763200}}\, ( 19318+31299\,\cos \left( 2\,\theta \right) +16218\,\cos \left( 4\,\theta \right)+4845\,\cos \left( 6\,\theta \right))\sin(\theta)^{4}\,\cos(6\,\psi_{10,6})
\cr&&
\times\left(v^{8} - {\displaystyle \frac {5491}{506}} \,v^{10} + 6\,\pi\,v^{11}\right).
\end{eqnarray}
\end{widetext}
The differences from Ref.~\cite{ref:TS} are as follows : 
the sign of $\zeta^{+}_{8,7}$ is changed, 
$\zeta^{\times}_{8,7}$ is multiplied by 
$-[207360(\cos(3\theta)+5\cos(\theta))]/[414720(2+\cos(2\theta))]$ and 
$\zeta^{+}_{10,6}$ is multiplied by $-4087/4096$. 

Once we obtain the 5.5PN $\zeta_{\ell m}^{+,\times}$, the 5.5PN plus and cross 
polarizations measured by  a general observer located at $(\theta, \varphi)$ 
are derived  using
\begin{subequations}
\begin{eqnarray}
h_{+} &=& -\left({\mu\over r}\right)\left({M\over r_0}\right)\sum_{\ell=2}^{13}\sum_{m=1}^{\ell} \zeta_{\ell m}^{+},\\
h_{\times} &=& -\left({\mu\over r}\right)\left({M\over r_0}\right)\sum_{\ell=2}^{13}\sum_{m=1}^{\ell} \zeta_{\ell m}^{\times}.
\end{eqnarray}
\label{eq:hp_hx}
\end{subequations}
In standard terminology used for GW polarizations~\cite{K07}, the above results for the polarization corresponds to
the choice $(\vec{N}=\vec{e}_R,\vec{P}=\vec{e}_\Theta,\vec{Q}=\vec{e}_\Phi)$.
The standard PN expressions~\cite{BFIS08} in the test-particle limit
corresponds to $(\vec{P}=-\vec{e}_\Phi,\vec{Q}=\vec{e}_\Theta)$
evaluated at $(\theta, \varphi)=(i,\pi/2)$ leading to
an overall sign difference. This also shows up in the overall sign difference 
in $h_{\ell m}$. 
 Ready-to-use 5.5PN expressions of GW polarization modes
$h_+$ and $h_\times$ in the test-particle limit are
listed in Appendix~\ref{sec:55PNpol} for possible use in GW data analysis
applications.

\section{Comparison with numerical results}
\label{sec:numerical}
To assess quantitative implications of our present work, 
in this section 
we perform two different types of comparisons.
First, we compare the formulas of our post-Newtonian 
expansion with results obtained by the numerical solution of the
Teukolsky equation. Secondly, for the case of LISA we investigate the
adequacy of the present ${\cal{O}}(v^{11})$ waveform for dephasing accuracy
of about a fraction of a cycle. 

The numerical calculation is based on 
the high precision code which deals with the gravitational waves 
from a particle in a circular orbit around a black hole~\cite{FT1,FT2}. 
The numerical method uses the formalism developed by Mano, Suzuki and 
Takasugi, which is the same formalism used in the post-Newtonian expansion 
in this paper, to compute the homogeneous solution of the Teukolsky equation. 
The precision of the energy flux of each mode $(\ell,m)$ can be 
achieved to about 14 significant figures in the double precision calculation. 
Thus the dominant errors for the energy flux in the code are due to 
truncation of summation over $(\ell,m)$-mode in Eq.~(\ref{eq:flux}). 
In the numerical calculation, we set maximum value of the summation 
over $\ell$ in Eq.~(\ref{eq:flux}) as $\ell=20$. Then we find the relative 
error of the energy flux at $r_0=6M$, ISCO of Schwarzschild black hole, 
is less than $10^{-10}$. Here we define the relative error as 
$F[\ell=20]/F[2\le\ell\le 20]$, where $F=dE/dt$.  References~\cite{FT1,FT2}, 
did not compute the energy absorption into the black hole horizon, but 
we include it in the computation of the energy flux, 
using in this paper the more general code for eccentric and inclined orbits 
developed in Ref.~\cite{FHT}.

\begin{figure*}[htb]
\begin{center}
\includegraphics[scale=1.0]{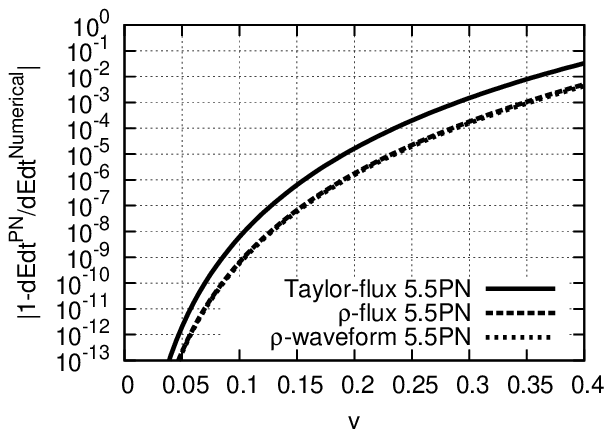}$\quad$
\includegraphics[scale=1.0]{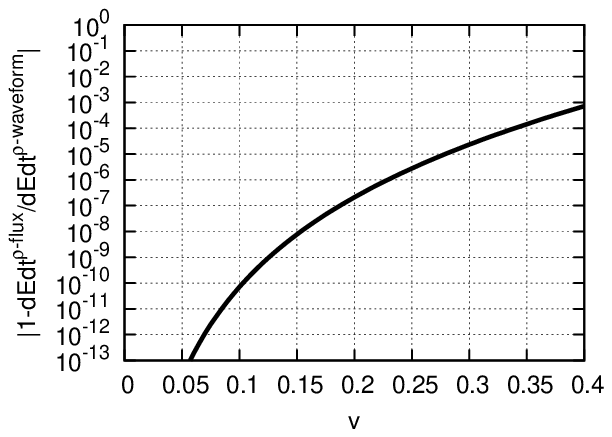}
\caption{(Left) Absolute values of difference of energy flux between numerical 
and post-Newtonian results as a function of the orbital velocity $v$. 
The numerical calculation is performed using the high precision code for 
energy flux in Ref.~\cite{FT1,FT2}. 
For the numerical calculation, we set the maximum value of $\ell$ to $20$ in 
Eq.~(\ref{eq:flux}) that leads to a relative accuracy  less than $10^{-10}$ 
in the numerical calculation. 
Taylor-flux 5.5PN computes energy flux using  Taylor expanded 
post-Newtonian approximation shown in Eq.~(3.1) in Ref.~\cite{ref:TTS}. 
$\rho$-flux 5.5PN uses resummed waveforms $\rho_{\ell m}$ in Ref.~\cite{DIN}, 
which derived $\rho_{\ell m}$ using the results of the 5.5PN energy flux, 
while $\rho$-waveform 5.5PN uses $\rho_{\ell m}$ obtained 
in Sec.~\ref{sec:rholm} using the new 5.5PN waveform.
The difference for the case of $\rho$-waveform 5.5PN and 
$\rho$-flux 5.5PN is not visible in this figure. 
(Right) Absolute values of difference of flux between $\rho$-waveform 5.5PN 
and $\rho$-flux 5.5PN results. 
}\label{fig:flux}
\end{center}
\end{figure*}

In the left panel of Fig.~\ref{fig:flux}, we show 
the absolute values of the relative error of energy flux 
between the 5.5PN approximation and numerical results 
as a function of the orbital velocity $v$. 
We compare three different post-Newtonian schemes, referred to as 
Taylor-flux 5.5PN, $\rho$-flux 5.5PN and $\rho$-waveform 5.5PN. 
Taylor-flux 5.5PN uses the result of Taylor expanded post-Newtonian 
energy flux, which is shown in Eq.~(3.1) in Ref.~\cite{ref:TTS}. 
$\rho$-flux 5.5PN uses resummed waveforms $\rho_{\ell m}$ in Ref.~\cite{DIN}, 
which computed $\rho_{\ell m}$ using the results of 5.5PN energy flux, 
while $\rho$-waveform 5.5PN uses resummed waveforms $\rho_{\ell m}$ obtained 
in Sec.~\ref{sec:rholm}. Thus, $\rho_{\ell m}$ in 
$\rho$-flux 5.5PN is computed up to $O[v^{10-2\,(\ell-2+\e_p)}]$ relative to 
the lowest order of itself, while $\rho_{\ell m}$ in $\rho$-waveform 5.5PN 
is computed up to $O[v^{10-(\ell-2)}]$ for $\ell$ is even and 
$O[v^{10-(\ell-3+2\e_p)}]$ for $\ell$ is odd. 
However, we do not find any visible difference between 
$\rho$-flux 5.5PN and $\rho$-waveform 5.5PN in the left panel of 
Fig.~\ref{fig:flux}. This may be because they do not have 
any difference in the dominant mode $(\ell,m)=(2,2)$, but 
a few correction terms in the nondominant mode $(\ell,m)=(2,1)$ 
and $\ell\ge 3$. 
In the right panel of Fig.~\ref{fig:flux}, we show the absolute values of 
the relative error of energy flux between $\rho$-flux 5.5PN and 
$\rho$-waveform 5.5PN. The relative error of energy flux between $\rho$-flux 5.5PN and 
$\rho$-waveform 5.5PN is less than $10^{-3}$ even in the region around ISCO. 
Thus we do not compare the results from $\rho$-flux 5.5PN to the ones from 
$\rho$-waveform 5.5PN in the following calculations. 

In order to estimate the validity of the 5.5PN expansion, 
we compare the phase after two years evolution with the numerical calculation. 
Here, we examine two systems, which were also studied in 
Ref.~\cite{EOB_EMRI} for the comparison of the phase between 
EOB waveforms and numerical ones. 
Both systems sweep  different frequency regions in the LISA band
during their two years quasicircular inspiral. 
One of the systems, named system-I, has masses $(M,\m)=(10^5,10)M_{\odot}$ 
and starts inspiral from $r_0\simeq 29.34M$ to $r_0\simeq 16.1M$, 
whose frequency corresponds to from $f_{\rm GW}\simeq 4\times 10^{-3}$Hz to 
$f_{\rm GW}\simeq 10^{-2}$Hz. 
The other system, named system-II, has masses $(M,\m)=(10^6,10)M_{\odot}$ 
and starts inspiral from $r_0\simeq 10.6M$ to $r_0\simeq 6.0M$, 
whose frequency corresponds to from $f_{\rm GW}\simeq 1.8\times 10^{-3}$Hz to 
$f_{\rm GW}\simeq 4.4\times 10^{-3}$Hz. 
System-I(II) represents the early (late) inspiral 
phase of  an extreme mass ratio inspiral 
in the frequency band of LISA.

For the numerical calculation of the phase, 
we adopt the one described in Ref.~\cite{Hughes2001}, 
which is also used in Ref.~\cite{EOB_EMRI}. 
The numerical calculation is implemented as follows. 
First, we prepare $10^3$ points data over the range from $v=0.01$ to 
$v=0.408$. In this work, the data contains $v$, $d r_0/dt$ and 
$\Psi_{\ell m}$, where $\Psi_{\ell m}$ is the phase of $Z_{\ell m\omega}$, 
$d r_0/dt = (\partial r_0/\partial\tilde{E})d\tilde{E}/dt = (r_0-6M)/(2\sqrt{r_0(r_0-3M)})d\tilde{E}/dt$ and 
$d\tilde{E}/dt$ is derived from the energy balance equation 
$d\tilde{E}/dt=-dE/dt$ and Eq.~(\ref{eq:flux}). 
Then, from the $10^3$ points data of ($v$, $d r_0/dt$, $\Psi_{\ell m}$), 
we compute ($r_0(t)$, $\Psi_{\ell m}(t)$) 
using cubic spline interpolation~\cite{Recipes}. 
Though one can use stepping algorithm such as Runge-Kutta method, 
we use the interpolation for the computation of ($r_0(t)$, $\Psi_{\ell m}(t)$) 
to save computational time. Then one can compute the phase of the waveforms 
by $m\int_0^{t}\Omega(t')dt'-\Psi_{\ell m}(t)$, where $\Omega(t)=\sqrt{M/r_0^3(t)}$.

\begin{figure*}[htb]
\begin{center}
\includegraphics[scale=1.0]{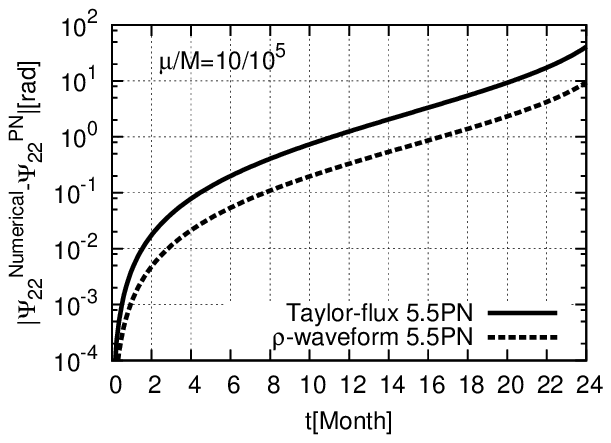}$\quad$
\includegraphics[scale=1.0]{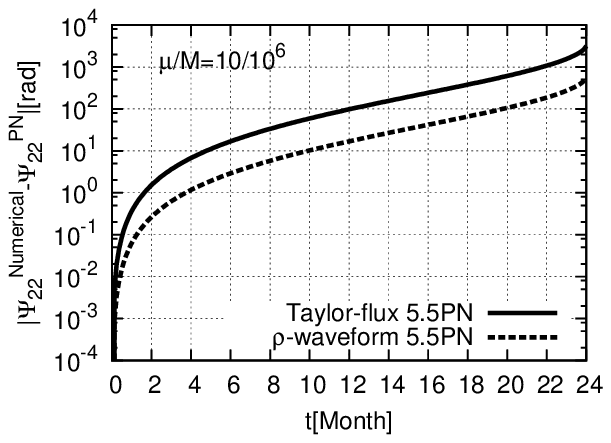}
\caption{Absolute values of the phase difference between the post-Newtonian expansion and the numerical results for $\ell=m=2$ mode as a function of the time 
in the  month. 
The left panel shows the dephase due to two years inspiral 
for $(M,\m)=(10^5,10)M_{\odot}$. 
The inspiral  is considered between $r_0\simeq 29.34M$ and $r_0\simeq 16.1M$, 
and corresponds to frequencies from $f_{\rm GW}\simeq 4\times 10^{-3}$Hz to 
$f_{\rm GW}\simeq 10^{-2}$Hz. 
The right panel shows the dephase due to 
two years inspiral for $(M,\m)=(10^6,10)M_{\odot}$
for inspiral from $r_0\simeq 10.6M$ to $r_0\simeq 6.0M$, 
with associated frequencies from   $f_{\rm GW}\simeq 1.8\times 10^{-3}$Hz to 
$f_{\rm GW}\simeq 4.4\times 10^{-3}$Hz. 
The left (right) panel represents the early (late) inspiral 
phase of an extreme mass ratio inspiral for LISA band.
The details of each post-Newtonian approximation is the same as in 
Fig.~\ref{fig:flux}. 
}\label{fig:phase22}
\end{center}
\end{figure*}

In Fig.~\ref{fig:phase22}, we show the absolute values of the phase 
difference of the dominant mode $(\ell,m)=(2,2)$ between 
5.5 post-Newtonian approximations and numerical results. 
We find that after two years evolution the dephasing is $\sim 40$ (3000) rads 
for system-I (system-II) when using the Taylor-flux 5.5PN~\cite{ref:TTS}, 
and $\sim 10$ ($530$) rads for system-I (system-II) when 
using $\rho$-waveform 5.5PN. These results are consistent with Ref.~\cite{EOB_EMRI}. 
Though $\rho$-waveform 5.5PN achieves about 5 times better phasing than 
Taylor-flux 5.5PN, the accuracy of the phasing is  not enough to 
extract physical parameters of the source of gravitational waves by 
the data analysis of LISA because we have to  reduce
the dephase to within 1 rad during the observation~\cite{ref:three}. 
In Ref.~\cite{EOB_EMRI}, the EOB model with 6PN factorized resummation,  
which is calibrated to numerical results, reduced the phase errors to
 $\sim$ rad. 
The EOB model with the calibrated 6.5PN Pad\'e approximation 
reduced the phase errors to less than 0.1 rad. 
Thus, for parameter estimation in LISA, we need  post-Newtonian terms higher
than 5.5PN and  other resummation techniques like the EOB. 
\section{Summary and conclusion}
\label{sec:summary}
In this work we have revisited   the post-Newtonian expansion
of gravitational waves for a test-particle of mass $\m$ in circular orbit
of radius $r_0$ around a Schwarzschild black hole of mass $M$
and provided   5.5PN GW polarizations ready for use in GW data analysis
applications. 
Taking account of the need to compare post-Newtonian waveforms with 
numerical relativity waveforms, we have derived 
the spherical harmonic components associated with the gravitational
 wave polarizations up to order $v^{11}$ beyond Newtonian.  
We have derived  more accurate factorized 
post-Newtonian waveforms at higher multipolar orders, 
extending work in Ref.~\cite{DIN} 
to obtain better agreement with 
numerical results than conventional Taylor expanded post-Newtonian waveforms. 
In addition to $h_{\ell m}$ modes corresponding to
5.5 post-Newtonian waveforms, 
we have derived a general expression for 2.5PN accurate $Z_{\ell m\omega}$, 
needed to obtain spherical modes and polarization modes, 
for general multipolar orders $\ell$ and $m$. 
We also provide general analytical results for
spherical harmonic modes at 2.5PN, 
2PN factorized waveforms $\rho_{\ell m}$ and 
3PN phase $\delta_{\ell m}$ for  arbitrary multipolar orders 
$\ell$ and $m$. 
Thanks to these 2.5PN or 2PN expressions of waveforms, 
we do not have to provide explicit expressions
of  waveforms for $\ell\ge 8$ modes in 
the 5.5PN calculation since we can compute them using their general 2.5PN or 
2PN formulas. 

To investigate the validity of post-Newtonian approximation up to $v^{11}$, 
we have compared the phase with numerical calculation 
of black hole perturbation for two years inspiral. 
We have found that the phase difference became larger than 10 rad, 
though the resummed waveforms achieve better agreement with numerical 
results than conventional Taylor expanded post-Newtonian waveforms. 
Thus we need higher post-Newtonian order corrections than 5.5PN 
in order to extract physical information from  LISA data analysis.
In Ref.~\cite{EOB_EMRI}, the EOB model with 
6PN and 6.5PN approximation, which are calibrated to numerical results, 
reduced the phase to $\simlt$ rad. These results motivate one to 
derive the factorized waveforms at post-Newtonian orders  higher than 
$v^{12}$. 

Another extension of the present investigation,
is the computation of gravitational waveforms for a particle in 
a circular orbit around a Kerr black hole. 
Unlike a Schwarzschild black hole, the waveforms in the Teukolsky formalism 
are expressed in terms of spin-weighted spheroidal harmonics in the case 
of a Kerr black hole. 
Thus, the calculation of the spin-weighted spherical harmonic components 
from the plus and cross polarizations for the Kerr case 
involves an additional transformation 
between the spin-weighted spheroidal harmonics and 
the spin-weighted spherical harmonics. 
It is also important to extend the factorized resummation in Ref.~\cite{DIN} 
to the case of a Kerr black hole~\cite{ref:spin_resum} 
and also noncircular orbits. 
These and similar related issues are left to  future investigations.
\begin{center}
\Large{\bf Acknowledgments}
\end{center}
RF would like to thank Nico Yunes, Alessandra Buonanno and Yi Pan for 
useful discussions. RF is also grateful to Theoretical Astrophysics Group at 
Osaka University for hospitality during the intermediate stages of editing 
this manuscript. 
A part of the results in this paper was obtained by mathematics software 
Maple and Mathematica.
\appendix
\section{5.5PN formulas for $\hat{H}_{\ell m}$ for $5\le\ell\le 13$}
\label{sec:55PN}
In this appendix, we give 5.5PN gravitational waveforms 
of spherical harmonic modes for $5\le\ell\le 13$, 
which are not shown completely in Sec.~\ref{sec:hlm}. 
We do not give the phase factor 
$\psi_{\ell m}$ Eq.~(\ref{eq:psi3pn_fact}) since one can derive it from 
the general 3PN formula. As we noted in Sec.~\ref{sec:zlmw2.5PN}, 
we can also derive ${\hat{H}_{\ell m}}$ for $\ell\ge 8$ 
using the general 2.5PN formula of $Z_{\ell m\omega}$ in Sec.~\ref{sec:zlmw2.5PN}, 
but we show them explicitly for ready reference. 
\begin{widetext}
\begin{subequations}
\begin{eqnarray}
{\hat{H}_{5, \,5}}&=&
-{\displaystyle \frac {625}{6336}}\,\sqrt{66} \,i\left(v^{3} - {\displaystyle \frac {263}{39}} \,v^{5} + 5\,\pi \,v^{6} + {\displaystyle \frac {9185}{819}} \,v^{7} - {\displaystyle \frac {1315}{39}} \,\pi\,v^{8}  \right.\cr
&&
 + v^{9}\,\left({\displaystyle \frac {25}{6}} \,\pi ^{2} + {\displaystyle \frac {154743717347}{919836918}}  - {\displaystyle \frac {7730}{429}} \,{\rm eulerlog}\left(5, \,v\right)\right) + {\displaystyle \frac {45925}{819}}\,\pi \,v^{10}  \cr
&&
\left. + v^{11}\,\left( - {\displaystyle \frac {20684766695623}{20047034007}}  + {\displaystyle \frac {2032990}{16731}} \,{\rm eulerlog}\left(5, \,v\right) - {\displaystyle \frac {6575}{234}} \,\pi ^{2}\right)\right) ,\\
{\hat{H}_{5, \,4}}&=&
 - {\displaystyle \frac {32}{1485}} \sqrt{165}\left(v^{4} - {\displaystyle \frac {4451}{910}} \,v^{6} + 4\,\pi \,v^{7} + {\displaystyle \frac {10715}{2184}} \,v^{8} - {\displaystyle \frac {8902}{455}} \,\pi \,v^{9} \right.\cr
&&
\left. + v^{10}\,\left({\displaystyle \frac {19767269662573}{183967383600}}  + {\displaystyle \frac {8}{3}} \,\pi ^{2} - {\displaystyle \frac {24736}{2145}} \,{\rm eulerlog}\left(4, \,v\right)\right) + {\displaystyle \frac {10715}{546}} \,\pi \,v^{11}\right) ,\\
{\hat{H}_{5, \,3}}&=&
{\displaystyle \frac {9}{3520}} \,\sqrt{330}\,i\left(v^{3} - {\displaystyle \frac {69}{13}} \,v^{5} + 3\,\pi \,v^{6} + {\displaystyle \frac {12463}{1365}} \,v^{7} - {\displaystyle \frac {207}{13}}\,\pi \,v^{8}  \right.\cr
&&
 + v^{9}\,\left({\displaystyle \frac {45961635793}{851700850}}  - {\displaystyle \frac {4638}{715}} \,{\rm eulerlog}\left(3, \,v\right) + {\displaystyle \frac {3}{2}} \,\pi ^{2}\right) + {\displaystyle \frac {12463}{455}}\,\pi \,v^{10}  \cr
&&
\left. + v^{11}\,\left( - {\displaystyle \frac {29896692069163}{105185054975}}  - {\displaystyle \frac {207}{26}} \,\pi ^{2} + {\displaystyle \frac {320022}{9295}} \,{\rm eulerlog}\left(3, \,v\right)\right)\right) ,\\
{\hat{H}_{5, \,2}}&=&
{\displaystyle \frac {2}{1485}} \sqrt{55}\left(v^{4} - {\displaystyle \frac {3911}{910}} \,v^{6} + 2\,\pi \,v^{7} + {\displaystyle \frac {63439}{10920}} \,v^{8} - {\displaystyle \frac {3911}{455}} \,\pi \,v^{9} \right.\cr
&&
\left. + v^{10}\,\left({\displaystyle \frac {4450644105337}{183967383600}}  + {\displaystyle \frac {2}{3}} \,\pi ^{2} - {\displaystyle \frac {6184}{2145}} \,{\rm eulerlog}\left(2, \,v\right)\right) + {\displaystyle \frac {63439}{5460}} \,\pi \,v^{11}\right) ,\\
{\hat{H}_{5, \,1}}&=&
-{\displaystyle \frac {\sqrt{385}}{110880}} \,i\left(v^{3} - {\displaystyle \frac {179}{39}} \,v^{5} + \pi \,v^{6} + {\displaystyle \frac {5023}{585}} \,v^{7} - {\displaystyle \frac {179}{39}} \,\pi\,v^{8}  \right.\cr
&&
 + v^{9}\,\left( - {\displaystyle \frac {12574241461}{22995922950}}  + {\displaystyle \frac {1}{6}} \,\pi ^{2} - {\displaystyle \frac {1546}{2145}} \,{\rm eulerlog}\left(1, \,v\right)\right) + {\displaystyle \frac {5023}{585}} \,\pi\,v^{10} \cr
&&
\left. + v^{11}\,\left( - {\displaystyle \frac {197070394320953}{8519989452975}}  + {\displaystyle \frac {276734}{83655}} \,{\rm eulerlog}\left(1, \,v\right) - {\displaystyle \frac {179}{234}} \,\pi ^{2}\right)\right) ,
\end{eqnarray}
\end{subequations}
\begin{subequations}
\begin{eqnarray}
{\hat{H}_{6, \,6}}&=&
{\displaystyle \frac {54}{715}} \sqrt{143}\left(v^{4} - {\displaystyle \frac {113}{14}} \,v^{6} + 6\,\pi \,v^{7} + {\displaystyle \frac {1372317}{73304}} \,v^{8} - {\displaystyle \frac {339}{7}} \,\pi \,v^{9} \right.\cr
&&
\left. + v^{10}\,\left({\displaystyle \frac {993676618159}{4569124560}}  - {\displaystyle \frac {21624}{1001}} \,{\rm eulerlog}\left(6, \,v\right) + 6\,\pi ^{2}\right) + {\displaystyle \frac {4116951}{36652}} \,\pi \,v^{11}\right) ,\\
{\hat{H}_{6, \,5}}&=&
-{\displaystyle \frac {3125}{216216}} \,\sqrt{429}\,i\left(v^{5} - {\displaystyle \frac {149}{24}} \,v^{7} + 5\,\pi\,v^{8}  + {\displaystyle \frac {156857}{15708}} \,v^{9} - {\displaystyle \frac {745}{24}} \,\pi\,v^{10}  \right.\cr
&&
\left. + v^{11}\,\left({\displaystyle \frac {43028942343011}{279630423072}}  - {\displaystyle \frac {45050}{3003}} \,{\rm eulerlog}\left(5, \,v\right) + {\displaystyle \frac {25}{6}} \,\pi ^{2}\right)\right) ,\\
{\hat{H}_{6, \,4}}&=&
 - {\displaystyle \frac {128}{19305}} \sqrt{78}\left(v^{4} - {\displaystyle \frac {93}{14}} \,v^{6} + 4\,\pi \,v^{7} + {\displaystyle \frac {3261767}{219912}} \,v^{8} - {\displaystyle \frac {186}{7}} \,\pi \,v^{9} \right.\cr
&&
\left. + v^{10}\,\left({\displaystyle \frac {60337719431417}{699076057680}}  - {\displaystyle \frac {28832}{3003}} \,{\rm eulerlog}\left(4, \,v\right) + {\displaystyle \frac {8}{3}} \,\pi ^{2}\right) + {\displaystyle \frac {3261767}{54978}} \,\pi \,v^{11}\right) ,\\
{\hat{H}_{6, \,3}}&=&
{\displaystyle \frac {81}{40040}} \,\sqrt{65}\,i\left(v^{5} - {\displaystyle \frac {133}{24}} \,v^{7} + 3\,\pi\,v^{8}  + {\displaystyle \frac {794009}{78540}} \,v^{9} - {\displaystyle \frac {133}{8}}\,\pi \,v^{10}  \right.\cr
&&
\left. + v^{11}\,\left({\displaystyle \frac {3}{2}} \,\pi ^{2} + {\displaystyle \frac {37970972674259}{776751175200}}  - {\displaystyle \frac {5406}{1001}} \,{\rm eulerlog}\left(3, \,v\right)\right)\right) ,\\
{\hat{H}_{6, \,2}}&=&
{\displaystyle \frac {2}{19305}} \sqrt{65}\left(v^{4} - {\displaystyle \frac {81}{14}} \,v^{6} + 2\,\pi \,v^{7} + {\displaystyle \frac {14482483}{1099560}} \,v^{8} - {\displaystyle \frac {81}{7}} \,\pi \,v^{9} \right.\cr
&&
\left. + v^{10}\,\left({\displaystyle \frac {1019185435937}{99868008240}}  - {\displaystyle \frac {7208}{3003}} \,{\rm eulerlog}\left(2, \,v\right) + {\displaystyle \frac {2}{3}} \,\pi ^{2}\right) + {\displaystyle \frac {14482483}{549780}} \,\pi \,v^{11}\right) ,\\
{\hat{H}_{6, \,1}}&=&
-{\displaystyle \frac {\sqrt{26}}{216216}} \,i\left(v^{5} - {\displaystyle \frac {125}{24}} \,v^{7} + \pi\,v^{8}  + {\displaystyle \frac {809959}{78540}} \,v^{9} - {\displaystyle \frac {125}{24}} \,\pi\,v^{10}  \right.\cr
&&
\left. + v^{11}\,\left( - {\displaystyle \frac {1063981210373}{411221210400}}  + {\displaystyle \frac {1}{6}} \,\pi ^{2} - {\displaystyle \frac {1802}{3003}} \,{\rm eulerlog}\left(1, \,v\right)\right)\right) ,
\end{eqnarray}
\end{subequations}
\begin{subequations}
\begin{eqnarray}
{\hat{H}_{7, \,7}}&=&
{\displaystyle \frac {16807}{1235520}} \,\sqrt{6006}\,i\left(v^{5} - {\displaystyle \frac {319}{34}} \,v^{7} + 7\,\pi\,v^{8}  + {\displaystyle \frac {2805191}{100776}} \,v^{9} \right.\cr
&&
\left. - {\displaystyle \frac {2233}{34}} \,\pi\,v^{10}  + v^{11}\,\left({\displaystyle \frac {49}{6}} \,\pi ^{2} - {\displaystyle \frac {83636}{3315}} \,{\rm eulerlog}\left(7, \,v\right) + {\displaystyle \frac {35250285887269}{132292686240}} \right)\right) ,\\
{\hat{H}_{7, \,6}}&=&
{\displaystyle \frac {81}{5005}} \,\sqrt{429}\,\left(v^{6} - {\displaystyle \frac {1787}{238}} \,v^{8} + 6\,\pi \,v^{9} + {\displaystyle \frac {3917323}{235144}} \,v^{10} - {\displaystyle \frac {5361}{119}} \,\pi \,v^{11}\right),\\
{\hat{H}_{7, \,5}}&=&
-{\displaystyle \frac {15625}{1729728}} \,\sqrt{66}\,i\left(v^{5} - {\displaystyle \frac {271}{34}} \,v^{7} + 5\,\pi\,v^{8}  + {\displaystyle \frac {15689017}{705432}} \,v^{9} - {\displaystyle \frac {1355}{34}} \,\pi\,v^{10}  \right.\cr
&&
\left. + v^{11}\,\left({\displaystyle \frac {157114390081769}{1296468325152}}  - {\displaystyle \frac {59740}{4641}} \,{\rm eulerlog}\left(5, \,v\right) + {\displaystyle \frac {25}{6}} \,\pi ^{2}\right)\right) ,\\
{\hat{H}_{7, \,4}}&=&
 - {\displaystyle \frac {128}{45045}} \,\sqrt{66}\left(v^{6} - {\displaystyle \frac {14543}{2142}} \,v^{8} + 4\,\pi \,v^{9} + {\displaystyle \frac {4794461}{302328}} \,v^{10} - {\displaystyle \frac {29086}{1071}} \,\pi \,v^{11}\right) ,\\
{\hat{H}_{7, \,3}}&=&
{\displaystyle \frac {243}{320320}} \,\sqrt{6}\,i\left(v^{5} - {\displaystyle \frac {239}{34}} \,v^{7} + 3\,\pi\,v^{8}  + {\displaystyle \frac {13620457}{705432}} \,v^{9} - {\displaystyle \frac {717}{34}} \,\pi\,v^{10}  \right.\cr
&&
\left. + v^{11}\,\left( - {\displaystyle \frac {35844}{7735}} \,{\rm eulerlog}\left(3, \,v\right) + {\displaystyle \frac {18990037711669}{720260180640}}  + {\displaystyle \frac {3}{2}} \,\pi ^{2}\right)\right) ,\\
{\hat{H}_{7, \,2}}&=&
{\displaystyle \frac {\sqrt{3}}{9009}}\left(v^{6} - {\displaystyle \frac {13619}{2142}} \,v^{8} + 2\,\pi \,v^{9} + {\displaystyle \frac {32998691}{2116296}} \,v^{10} - {\displaystyle \frac {13619}{1071}} \,\pi \,v^{11}\right),\\
{\hat{H}_{7, \,1}}&=&
-{\displaystyle \frac {\sqrt{2}}{1729728}} \,i\left(v^{5} - {\displaystyle \frac {223}{34}} \,v^{7} + \pi\,v^{8}  + {\displaystyle \frac {4251691}{235144}} \,v^{9} - {\displaystyle \frac {223}{34}} \,\pi\,v^{10}  \right.\cr
&&
\left. + v^{11}\,\left( - {\displaystyle \frac {131624253235451}{6482341625760}}  + {\displaystyle \frac {1}{6}} \,\pi ^{2} - {\displaystyle \frac {11948}{23205}} \,{\rm eulerlog}\left(1, \,v\right)\right)\right) ,
\end{eqnarray}
\end{subequations}
\begin{subequations}
\begin{eqnarray}
{\hat{H}_{8, \,8}}&=&
 - {\displaystyle \frac {16384}{5360355}}
\sqrt{170170}\left(v^{6} - {\displaystyle \frac {3653}{342}} \,v^{8} + 8\,\pi \,v^{9} + {\displaystyle \frac {124891}{3240}} \,v^{10} - {\displaystyle \frac {14612}{171}} \,\pi \,v^{11}\right),\\
{\hat{H}_{8, \,7}}&=&
{\displaystyle \frac {117649}{126023040}} \,\sqrt{170170}\,i\,\left(v^{7} - {\displaystyle \frac {3343}{380}} \,v^{9} + 7\,\pi\,v^{10}  + {\displaystyle \frac {42607}{1710}} \,v^{11}\right), \\
{\hat{H}_{8, \,6}}&=&
{\displaystyle \frac {243}{595595}} \,\sqrt{51051}\,\left(v^{6} - {\displaystyle \frac {353}{38}} \,v^{8} + 6\,\pi \,v^{9} + {\displaystyle \frac {2255}{72}} \,v^{10} - {\displaystyle \frac {1059}{19}} \,\pi \,v^{11}\right), \\
{\hat{H}_{8, \,5}}&=&
-{\displaystyle \frac {78125}{176432256}} \,\sqrt{4862}\,i\,\left(v^{7} - {\displaystyle \frac {611}{76}} \,v^{9} + 5\,\pi\,v^{10}  + {\displaystyle \frac {7901}{342}} \,v^{11}\right), \\
{\hat{H}_{8, \,4}}&=&
 - {\displaystyle \frac {128}{765765}} \,\sqrt{374}\,\left(v^{6} - {\displaystyle \frac {2837}{342}} \,v^{8} + 4\,\pi \,v^{9} + {\displaystyle \frac {122681}{4536}} \,v^{10} - {\displaystyle \frac {5674}{171}} \,\pi \,v^{11}\right), \\
{\hat{H}_{8, \,3}}&=&
{\displaystyle \frac {81}{10890880}} \,\sqrt{5610}\,i\,\left(v^{7} - {\displaystyle \frac {2863}{380}} \,v^{9} + 3\,\pi\,v^{10}  + {\displaystyle \frac {4213}{190}} \,v^{11}\right), \\
{\hat{H}_{8, \,2}}&=&
{\displaystyle \frac {\sqrt{85}}{765765}} \,\left(v^{6} - {\displaystyle \frac {2633}{342}} \,v^{8} + 2\,\pi \,v^{9} + {\displaystyle \frac {563317}{22680}} \,v^{10} - {\displaystyle \frac {2633}{171}} \,\pi \,v^{11}\right),\\
{\hat{H}_{8, \,1}}&=&
-{\displaystyle \frac {\sqrt{238}}{176432256}} \,i\,\left(v^{7} - {\displaystyle \frac {2767}{380}} \,v^{9} + \pi\,v^{10}  + {\displaystyle \frac {37267}{1710}} \,v^{11}\right), 
\end{eqnarray}
\end{subequations}
\begin{subequations}
\begin{eqnarray}
{\hat{H}_{9, \,9}}&=&
-{\displaystyle \frac {1594323}{150492160}} \,\sqrt{20995}\,i\,\left(v^{7} - {\displaystyle \frac {419}{35}} \,v^{9} + 9\,\pi\,v^{10}  + {\displaystyle \frac {38269131}{752675}} \,v^{11}\right), \\
{\hat{H}_{9, \,8}}&=&
 - {\displaystyle \frac {131072}{59520825}} \,\sqrt{41990}\,\left(v^{8} - {\displaystyle \frac {13969}{1386}} \,v^{10} + 8\,\pi \,v^{11}\right), \\
{\hat{H}_{9, \,7}}&=&
{\displaystyle \frac {5764801}{1741409280}} \,\sqrt{1235}\,i\,\left(v^{7} - {\displaystyle \frac {53}{5}} \,v^{9} + 7\,\pi\,v^{10}  + {\displaystyle \frac {40687373}{967725}} \,v^{11}\right), \\
{\hat{H}_{9, \,6}}&=&
{\displaystyle \frac {486}{734825}} \,\sqrt{3705}\,\left(v^{8} - {\displaystyle \frac {12877}{1386}} \,v^{10} + 6\,\pi \,v^{11}\right), \\
{\hat{H}_{9, \,5}}&=&
-{\displaystyle \frac {390625}{1218986496}} \,\sqrt{247}\,i\,\left(v^{7} - {\displaystyle \frac {67}{7}} \,v^{9} + 5\,\pi\,v^{10}  + {\displaystyle \frac {9862901}{270963}} \,v^{11}\right), \\
{\hat{H}_{9, \,4}}&=&
 - {\displaystyle \frac {512}{59520825}} \,\sqrt{17290}\,\left(v^{8} - {\displaystyle \frac {12097}{1386}} \,v^{10} + 4\,\pi \,v^{11}\right), \\
{\hat{H}_{9, \,3}}&=&
{\displaystyle \frac {81}{75246080}} \,\sqrt{1995}\,i\,\left(v^{7} - {\displaystyle \frac {311}{35}} \,v^{9} + 3\,\pi\,v^{10}  + {\displaystyle \frac {24906689}{752675}} \,v^{11}\right), \\
{\hat{H}_{9, \,2}}&=&
{\displaystyle \frac {2}{8502975}} \,\sqrt{95}\,\left(v^{8} - {\displaystyle \frac {11629}{1386}} \,v^{10} + 2\,\pi \,v^{11}\right), \\
{\hat{H}_{9, \,1}}&=&
-{\displaystyle \frac {\sqrt{2090}}{19155502080}} \,i\,\left(v^{7} - {\displaystyle \frac {299}{35}} \,v^{9} + \pi\,v^{10}  + {\displaystyle \frac {19443469}{615825}} \,v^{11}\right), 
\end{eqnarray}
\end{subequations}
\begin{subequations}
\begin{eqnarray}
{\hat{H}_{10, \,10}}&=&
{\displaystyle \frac {390625}{49997493}} \,\sqrt{58786}\,\left(v^{8} - {\displaystyle \frac {6707}{506}} \,v^{10} + 10\,\pi \,v^{11}\right), \\
{\hat{H}_{10, \,9}}&=&
-{\displaystyle \frac {14348907}{14484870400}} \,\sqrt{293930}\,i\,\left(v^{9} - {\displaystyle \frac {5223}{460}} \,v^{11}\right), \\
{\hat{H}_{10, \,8}}&=&
 - {\displaystyle \frac {2097152}{1249937325}} \,\sqrt{7735}\,\left(v^{8} - {\displaystyle \frac {6023}{506}} \,v^{10} + 8\,\pi \,v^{11}\right), \\
{\hat{H}_{10, \,7}}&=&
{\displaystyle \frac {5764801}{23944377600}} \,\sqrt{46410}\,i\,\left(v^{9} - {\displaystyle \frac {14549}{1380}} \,v^{11}\right), \\
{\hat{H}_{10, \,6}}&=&
{\displaystyle \frac {729}{5143775}} \,\sqrt{2730}\,\left(v^{8} - {\displaystyle \frac {5491}{506}} \,v^{10} + 6\,\pi \,v^{11}\right), \\
{\hat{H}_{10, \,5}}&=&
-{\displaystyle \frac {1953125}{23465490048}} \,\sqrt{546}\,i\,\left(v^{9} - {\displaystyle \frac {13709}{1380}} \,v^{11}\right), \\
{\hat{H}_{10, \,4}}&=&
 - {\displaystyle \frac {4096}{1249937325}} \,\sqrt{1365}\,\left(v^{8} - {\displaystyle \frac {5111}{506}} \,v^{10} + 4\,\pi \,v^{11}\right), \\
{\hat{H}_{10, \,3}}&=&
{\displaystyle \frac {243}{1034633600}} \,\sqrt{2730}\,i\,\left(v^{9} - {\displaystyle \frac {4383}{460}} \,v^{11}\right), \\
{\hat{H}_{10, \,2}}&=&
{\displaystyle \frac {2}{178562475}} \,\sqrt{105}\,\left(v^{8} - {\displaystyle \frac {4883}{506}} \,v^{10} + 2\,\pi \,v^{11}\right), \\
{\hat{H}_{10, \,1}}&=&
-{\displaystyle \frac {\sqrt{35}}{27935107200}} \,i\,\left(v^{9} - {\displaystyle \frac {12869}{1380}} \,v^{11}\right), 
\end{eqnarray}
\end{subequations}
\begin{subequations}
\begin{eqnarray}
{\hat{H}_{11, \,11}}&=&
{\displaystyle \frac {2357947691}{754833945600}} \,\sqrt{572033}\,i\,\left(v^{9} - {\displaystyle \frac {218}{15}} \,v^{11}\right), \\
{\hat{H}_{11, \,10}}&=&
{\displaystyle \frac {1953125}{973028133}} \,\sqrt{104006}\,v^{10},\\ 
{\hat{H}_{11, \,9}}&=&
-{\displaystyle \frac {129140163}{102508313600}} \,\sqrt{22287}\,i\,\left(v^{9} - {\displaystyle \frac {66}{5}} \,v^{11}\right), \\
{\hat{H}_{11, \,8}}&=&
 - {\displaystyle \frac {8388608}{24325703325}} \,\sqrt{37145}\,v^{10}, \\
{\hat{H}_{11, \,7}}&=&
{\displaystyle \frac {40353607}{169452518400}} \,\sqrt{1955}\,i\,\left(v^{9} - {\displaystyle \frac {182}{15}} \,v^{11}\right), \\
{\hat{H}_{11, \,6}}&=&
{\displaystyle \frac {2187}{20021155}} \,\sqrt{782}\,v^{10}, \\
{\hat{H}_{11, \,5}}&=&
-{\displaystyle \frac {9765625}{332126936064}} \,\sqrt{69}\,i\,\left(v^{9} - {\displaystyle \frac {34}{3}} \,v^{11}\right),\\
{\hat{H}_{11, \,4}}&=&
 - {\displaystyle \frac {8192}{4865140665}} \,\sqrt{483}\,v^{10}, \\
{\hat{H}_{11, \,3}}&=&
{\displaystyle \frac {2187}{102508313600}} \,\sqrt{1610}\,i\,\left(v^{9} - {\displaystyle \frac {54}{5}} \,v^{11}\right), \\
{\hat{H}_{11, \,2}}&=&
{\displaystyle \frac {2}{1158366825}} \,\sqrt{115}\,v^{10}, \\
{\hat{H}_{11, \,1}}&=&
-{\displaystyle \frac {\sqrt{598}}{5140059724800}} \,i\,\left(v^{9} - {\displaystyle \frac {158}{15}} \,v^{11}\right), 
\end{eqnarray}
\end{subequations}
\begin{subequations}
\begin{eqnarray}
{\hat{H}_{12, \,12}}&=&
 - {\displaystyle \frac {1492992}{786545375}} \,\sqrt{2451570}\,v^{10}, \\
{\hat{H}_{12, \,11}}&=&
{\displaystyle \frac {25937424601}{21027517056000}} \,\sqrt{408595}\,i\,v^{11}, \\
{\hat{H}_{12, \,10}}&=&
{\displaystyle \frac {1953125}{1529044209}} \,\sqrt{35530}\,v^{10}, \\
{\hat{H}_{12, \,9}}&=&
-{\displaystyle \frac {1162261467}{951862912000}} \,\sqrt{4845}\,i\,v^{11}, \\
{\hat{H}_{12, \,8}}&=&
 - {\displaystyle \frac {16777216}{121628516625}} \,\sqrt{11305}\,v^{10}, \\
{\hat{H}_{12, \,7}}&=&
{\displaystyle \frac {282475249}{6608648217600}} \,\sqrt{11305}\,i\,v^{11}, \\
{\hat{H}_{12, \,6}}&=&
{\displaystyle \frac {729}{100105775}} \,\sqrt{3570}\,v^{10}, \\
{\hat{H}_{12, \,5}}&=&
-{\displaystyle \frac {9765625}{1850421500928}} \,\sqrt{255}\,i\,v^{11}, \\
{\hat{H}_{12, \,4}}&=&
 - {\displaystyle \frac {2048}{3475100475}} \,\sqrt{30}\,v^{10}, \\
{\hat{H}_{12, \,3}}&=&
{\displaystyle \frac {6561}{190372582400}} \,\sqrt{30}\,i\,v^{11}, \\
{\hat{H}_{12, \,2}}&=&
{\displaystyle \frac {2}{5791834125}} \,\sqrt{5}\,v^{10}, \\
{\hat{H}_{12, \,1}}&=&
-{\displaystyle \frac {\sqrt{770}}{77100895872000}} \,i\,v^{11}, 
\end{eqnarray}
\end{subequations}
\begin{subequations}
\begin{eqnarray}
{\hat{H}_{13, \,13}}&=&
-{\displaystyle \frac {1792160394037}{256212207820800}} \,\sqrt{289731}\,i\,v^{11},\\
{\hat{H}_{13, \,11}}&=&
{\displaystyle \frac {3138428376721}{1513981228032000}} \,\sqrt{22287}\,i\,v^{11},\\
{\hat{H}_{13, \,9}}&=&
-{\displaystyle \frac {3486784401}{7614903296000}} \,\sqrt{1938}\,i\,v^{11},\\
{\hat{H}_{13, \,7}}&=&
{\displaystyle \frac {1977326743}{2379113358336000}} \,\sqrt{746130}\,i\,v^{11},\\
{\hat{H}_{13, \,5}}&=&
-{\displaystyle \frac {48828125}{133230348066816}} \,\sqrt{561}\,i\,v^{11},\\
{\hat{H}_{13, \,3}}&=&
{\displaystyle \frac {6561}{7614903296000}} \,\sqrt{165}\,i\,v^{11},\\
{\hat{H}_{13, \,1}}&=&
-{\displaystyle \frac {\sqrt{3}}{252330204672000}} \,i\,v^{11}.
\end{eqnarray}
\end{subequations}
\end{widetext}
\section{5PN formulas for $\rho_{\ell m}$ for $5\le\ell\le 7$}
\label{sec:rho5PN}
In this appendix, we give the 5PN resummed gravitational waveforms 
for $5\le\ell\le 7$, which are not shown completely in Sec.~\ref{sec:rholm}. 
We do not give the phase factor 
$\delta_{\ell m}$ Eq.~(\ref{eq:psi2delta_45pn}) since one can derive it from 
the general 3PN formula  Eq.~(\ref{eq:psi2delta}). 
For $\ell\ge 8$, we do not show $\rho_{\ell m}$ since 
they are explicitly given by the general 2PN formula 
in Sec.~\ref{sec:rho2PN}, Eqs.~(\ref{eq:rho2PN}) to (\ref{eq:rho2PNeven}).
\begin{widetext}
\begin{subequations}
\begin{eqnarray}
{\rho _{5, \,5}}&=&
1 - {\displaystyle \frac {487}{390}} \,v^{2} - {\displaystyle \frac {3353747}{2129400}} \,v^{4}
 + \left( - {\displaystyle \frac {1546}{429}} \,{\rm eulerlog}\left(5, \,v\right) + {\displaystyle \frac {190606537999247}{11957879934000}} \right)\,v^{6} \cr
&&
 + \left( - {\displaystyle \frac {1213641959949291437}{118143853747920000}}  + {\displaystyle \frac {376451}{83655}} \,{\rm eulerlog}\left(5, \,v\right)\right)\,v^{8},\\
{\rho _{5, \,4}}&=&
1 - {\displaystyle \frac {2908}{2275}} \,v^{2} - {\displaystyle \frac {16213384}{15526875}} \,v^{4} 
 + \left( - {\displaystyle \frac {24736}{10725}} \,{\rm eulerlog}\left(4, \,v\right) + {\displaystyle \frac {6704294638171892}{653946558890625}} \right)\,v^{6},\\
{\rho _{5, \,3}}&=&
1 - {\displaystyle \frac {25}{26}} \,v^{2} - {\displaystyle \frac {410833}{709800}} \,v^{4} + \left( - {\displaystyle \frac {4638}{3575}} \,{\rm eulerlog}\left(3, \,v\right) + {\displaystyle \frac {7618462680967}{1328653326000}} \right)\,v^{6} \cr
&&
 + \left( - {\displaystyle \frac {77082121019870543}{39381284582640000}}  + {\displaystyle \frac {2319}{1859}} \,{\rm eulerlog}\left(3, \,v\right)\right)\,v^{8},\\
{\rho _{5, \,2}}&=&
1 - {\displaystyle \frac {2638}{2275}} \,v^{2} - {\displaystyle \frac {7187914}{15526875}} \,v^{4} + \left({\displaystyle \frac {1539689950126502}{653946558890625}}  - {\displaystyle \frac {6184}{10725}} \,{\rm eulerlog}\left(2, \,v\right)\right)\,v^{6},\\
{\rho _{5, \,1}}&=&
1 - {\displaystyle \frac {319}{390}} \,v^{2} - {\displaystyle \frac {31877}{304200}} \,v^{4} + \left( - {\displaystyle \frac {1546}{10725}} \,{\rm eulerlog}\left(1, \,v\right) + {\displaystyle \frac {7685351978519}{11957879934000}} \right)\,v^{6} \cr
&&
 + \left( - {\displaystyle \frac {821807362819271}{10740350340720000}}  + {\displaystyle \frac {22417}{190125}} \,{\rm eulerlog}\left(1, \,v\right)\right)\,v^{8},
\end{eqnarray}
\end{subequations}
\begin{subequations}
\begin{eqnarray}
{\rho _{6, \,6}}&=&
1 - {\displaystyle \frac {53}{42}} \,v^{2} - {\displaystyle \frac {1025435}{659736}} \,v^{4} + \left( - {\displaystyle \frac {3604}{1001}} \,{\rm eulerlog}\left(6, \,v\right) + {\displaystyle \frac {610931247213169}{36701493028200}} \right)\,v^{6},\\
{\rho _{6, \,5}}&=&
1 - {\displaystyle \frac {185}{144}} \,v^{2} - {\displaystyle \frac {59574065}{54286848}} \,v^{4} + \left( - {\displaystyle \frac {22525}{9009}} \,{\rm eulerlog}\left(5, \,v\right) + {\displaystyle \frac {67397117912549267}{5798416452820992}} \right)\,v^{6},\\
{\rho _{6, \,4}}&=&
1 - {\displaystyle \frac {43}{42}} \,v^{2} - {\displaystyle \frac {476887}{659736}} \,v^{4} + \left( - {\displaystyle \frac {14416}{9009}} \,{\rm eulerlog}\left(4, \,v\right) + {\displaystyle \frac {180067034480351}{24467662018800}} \right)\,v^{6},\\
{\rho _{6, \,3}}&=&
1 - {\displaystyle \frac {169}{144}} \,v^{2} - {\displaystyle \frac {152153941}{271434240}} \,v^{4} + \left( - {\displaystyle \frac {901}{1001}} \,{\rm eulerlog}\left(3, \,v\right) + {\displaystyle \frac {116042497264681103}{28992082264104960}} \right)\,v^{6},\\
{\rho _{6, \,2}}&=&
1 - {\displaystyle \frac {37}{42}} \,v^{2} - {\displaystyle \frac {817991}{3298680}} \,v^{4} + \left( - {\displaystyle \frac {3604}{9009}} \,{\rm eulerlog}\left(2, \,v\right) + {\displaystyle \frac {812992177581}{453104852200}} \right)\,v^{6},\\
{\rho _{6, \,1}}&=&
1 - {\displaystyle \frac {161}{144}} \,v^{2} - {\displaystyle \frac {79192261}{271434240}} \,v^{4} + \left( - {\displaystyle \frac {901}{9009}} \,{\rm eulerlog}\left(1, \,v\right) + {\displaystyle \frac {6277796663889319}{28992082264104960}} \right)\,v^{6},
\end{eqnarray}
\end{subequations}
\begin{subequations}
\begin{eqnarray}
{\rho _{7, \,7}}&=&
1 - {\displaystyle \frac {151}{119}} \,v^{2} - {\displaystyle \frac {32358125}{20986602}} \,v^{4} 
 + \left( - {\displaystyle \frac {11948}{3315}} \,{\rm eulerlog}\left(7, \,v\right) + {\displaystyle \frac {66555794049401803}{3856993267327200}} \right)\,v^{6},\\
{\rho _{7, \,6}}&=&
1 - {\displaystyle \frac {1072}{833}} \,v^{2} - {\displaystyle \frac {195441224}{171390583}} \,v^{4},\\
{\rho _{7, \,5}}&=&
1 - {\displaystyle \frac {127}{119}} \,v^{2} - {\displaystyle \frac {17354227}{20986602}} \,v^{4} + \left( - {\displaystyle \frac {59740}{32487}} \,{\rm eulerlog}\left(5, \,v\right) + {\displaystyle \frac {192862646381533}{22039961527584}} \right)\,v^{6},\\
{\rho _{7, \,4}}&=&
1 - {\displaystyle \frac {8878}{7497}} \,v^{2} - {\displaystyle \frac {2995755988}{4627545741}} \,v^{4},\\
{\rho _{7, \,3}}&=&
1 - {\displaystyle \frac {111}{119}} \,v^{2} - {\displaystyle \frac {7804375}{20986602}} \,v^{4} + \left( - {\displaystyle \frac {35844}{54145}} \,{\rm eulerlog}\left(3, \,v\right) + {\displaystyle \frac {1321461327981547}{428554807480800}} \right)\,v^{6},\\
{\rho _{7, \,2}}&=&
1 - {\displaystyle \frac {8416}{7497}} \,v^{2} - {\displaystyle \frac {1625746984}{4627545741}} \,v^{4},\\
{\rho _{7, \,1}}&=&
1 - {\displaystyle \frac {103}{119}} \,v^{2} - {\displaystyle \frac {1055091}{6995534}} \,v^{4} + \left( - {\displaystyle \frac {11948}{162435}} \,{\rm eulerlog}\left(1, \,v\right) + {\displaystyle \frac {142228318411021}{550999038189600}} \right)\,v^{6}.
\end{eqnarray}
\end{subequations}
\end{widetext}
\section{5.5PN formulas for polarization modes}
\label{sec:55PNpol}
In this appendix, we list the complete ready-to-use
 5.5PN gravitational wave polarizations
in the test-particle limit obtained using 
Eq.~(\ref{eq:hp_hx}) in Sec.~\ref{sec:zetalm}, 
\begin{subequations}
\begin{eqnarray}
h_{+} &=& -\left({\mu\over r}\right)\left({M\over r_0}\right)\sum_{\ell=2}^{13}\sum_{m=1}^{\ell} \zeta_{\ell m}^{+},\\
h_{\times} &=& -\left({\mu\over r}\right)\left({M\over r_0}\right)\sum_{\ell=2}^{13}\sum_{m=1}^{\ell} \zeta_{\ell m}^{\times},
\end{eqnarray}
\end{subequations}
where $\zeta_{\ell m}^{+,\times}$, Eqs.~(\ref{eq:zeta1}) and (\ref{eq:zeta2}), 
are functions of both the phase $\psi_{\ell m}$, Eq.~(\ref{eq:psi_lm}), 
and the angles $(\theta,\varphi)$ 
defining the observer relative to the source. 

Using the same notation and the same normalization
 $h_{+,\times}=2\m\,x\,H_{+,\times}/r$ 
as in Ref.~\cite{BFIS08}, where $x=v^2$, 
we consider the post-Newtonian expansion of $H_{+,\times}$ defined as 
\begin{eqnarray}
H_{+,\times} &=& \sum_{n=0}^{11} x^{n/2}\,H_{+,\times}^{(n/2)}.
\end{eqnarray}
For the computation of the PN coefficients $H_{+,\times}^{(n/2)}$, 
we change the phase variable $\psi_{\ell m}$ in each mode to 
1.5PN accurate $\psi_{2,\,2}$ as in ~\cite{BFIS08}.
Using $\psi_{2,\,2}$ at 1.5PN, Eq.~(\ref{eq:psi_lm}), 
we define the phase $\psi$ as\footnote{ Contrary to simple expectations, it is worth noting that a more accurate 4.5PN $\psi_{2,\,2}$, 
Eq.~(\ref{eq:psi_lm}), does not further simplify the expressions of 
$H_{+,\times}^{(n)}$ since the 3PN and 4.5PN terms in $\psi_{\ell m}$ are 
functions of $\ell$ and $m$ unlike the 1.5PN $\ln v$ terms of $\psi_{\ell m}$ 
which are the same for all modes.
Consequently, though 1.5PN $\psi_{2,\,2}$ simplifies 
the expressions of $H_{+,\times}^{(n)}$, no such simplification is obtained at higher PN orders. }
\begin{eqnarray}
\psi = \Omega(t-r^*)-\varphi+\left( {\frac {17}{6}}-2\,\gamma-3\,\ln \left( 4\,x \right) \right) {x}^{3/2}.
\label{ptm}
\end{eqnarray}
The above expression for $\psi$ can be rewritten as
\begin{equation}
\psi=\phi-3 x^{3/2}\ln \left(\frac{x}{x_0}\right)-\varphi
\end{equation}
by recalling that in the test-particle case
(see Sec.~\ref{sec:hlm}) 
$\ln x_0 = 17/18-2\ln 2-2\gamma/3$.
 
We list the 5.5PN $H_{+}^{(n/2)}$ and $H_{\times}^{(n/2)}$ in 
Appendix subsections~\ref{sec:55PNpol_plus} and \ref{sec:55PNpol_cross} respectively. 
In the following subsections, we use shorthand notations  
$c_\theta=\cos\theta$ and $s_\theta=\sin\theta$. 
As mentioned earlier, to compare to standard PN expressions~\cite{BFIS08} 
in the test-particle limit, where $\Delta=-1$, 
we need to replace $(\theta, \varphi)$ by $(i,\pi/2)$.
With this replacement our phase variable $\psi$ Eq.~(\ref{ptm}) is related
to the $\psi_{\rm BFIS}$ used in~\cite{BFIS08} as $\psi=\psi_{\rm BFIS}-\pi/2$ and the
polarizations agree\footnote{Since we do not compute the $m=0$ modes in this
work, we will not recover the ``direct current'' terms in~\cite{BFIS08,Favata09}.}  modulo an   overall sign for reasons 
discussed in Sec.~\ref{sec:zetalm}. 
\subsection{Plus modes}
\label{sec:55PNpol_plus}
\begin{widetext}
\begin{subequations}
\begin{eqnarray}
{H_{+}^{(0)}}&=&
 - \cos\left(2\,\psi \right)\,\left(1 + {c_\theta}^{2}\right),
\end{eqnarray}
\begin{eqnarray}
{H_{+}^{(0.5)}}&=&
{s_\theta}\,\sin\left(\psi \right)\,\left({\displaystyle \frac {5}{8}}  + {\displaystyle \frac {1}{8}} \,{c_\theta}^{2}\right) + {\displaystyle \frac {9}{8}} \,{s_\theta}\,\left(1 + {c_\theta}^{2}\right)\,\sin\left(3\,\psi \right),
\end{eqnarray}
\begin{eqnarray}
{H_{+}^{(1)}}&=&
\cos\left(2\,\psi \right)\,\left({\displaystyle \frac {19}{6}}  + {\displaystyle \frac {3}{2}} \,{c_\theta}^{2} - {\displaystyle \frac {1}{3}} \,{c_\theta}^{4}\right) + {\displaystyle \frac {4}{3}} \,{s_\theta}^{2}\,\cos\left(4\,\psi \right)\,\left(1 + {c_\theta}^{2}\right),
\end{eqnarray}
\begin{eqnarray}
{H_{+}^{(1.5)}}&=&
 - 2\,\pi \,\cos\left(2\,\psi \right)\,\left(1 + {c_\theta}^{2}\right) + {s_\theta}\,\sin\left(\psi \right)\,\left( - {\displaystyle \frac {19}{64}}  - {\displaystyle \frac {5}{16}} \,{c_\theta}^{2} + {\displaystyle \frac {1}{192}} \,{c_\theta}^{4}\right) \cr
&&
 + {s_\theta}\,\sin\left(3\,\psi \right)\,\left( - {\displaystyle \frac {657}{128}}  - {\displaystyle \frac {45}{16}} \,{c_\theta}^{2} + {\displaystyle \frac {81}{128}} \,{c_\theta}^{4}\right) - {\displaystyle \frac {625}{384}} \,{s_\theta}^{3}\,\left(1 + {c_\theta}^{2}\right)\,\sin\left(5\,\psi \right) ,
\end{eqnarray}
\begin{eqnarray}
{H_{+}^{(2)}}&=&
{s_\theta}\,\cos\left(\psi \right)\,\left({\displaystyle \frac {5}{4}} \,\ln\left(2\right) + {\displaystyle \frac {11}{40}}  + \left({\displaystyle \frac {1}{4}} \,\ln\left(2\right) + {\displaystyle \frac {7}{40}} \right)\,{c_\theta}^{2}\right) \cr
&&
 + \cos\left(2\,\psi \right)\,\left({\displaystyle \frac {11}{60}}  + {\displaystyle \frac {33}{10}} \,{c_\theta}^{2} + {\displaystyle \frac {29}{24}} \,{c_\theta}^{4} - {\displaystyle \frac {1}{24}} \,{c_\theta}^{6}\right) \cr
&&
 + {s_\theta}\,\cos\left(3\,\psi \right)\,\left(1 + {c_\theta}^{2}\right)\,\left( - {\displaystyle \frac {27}{4}} \,\ln\left({\displaystyle \frac {3}{2}} \right) + {\displaystyle \frac {189}{40}} \right) \cr
&&
 + {s_\theta}^{2}\,\cos\left(4\,\psi \right)\,\left( - {\displaystyle \frac {118}{15}}  - {\displaystyle \frac {14}{3}} \,{c_\theta}^{2} + {\displaystyle \frac {16}{15}} \,{c_\theta}^{4}\right) - {\displaystyle \frac {81}{40}} \,{s_\theta}^{4}\,\cos\left(6\,\psi \right)\,\left(1 + {c_\theta}^{2}\right) \cr
&&
 + \pi \,{s_\theta}\,\sin\left(\psi \right)\,\left({\displaystyle \frac {5}{8}}  + {\displaystyle \frac {1}{8}} \,{c_\theta}^{2}\right) + {\displaystyle \frac {27}{8}} \,\pi \,{s_\theta}\,\left(1 + {c_\theta}^{2}\right)\,\sin\left(3\,\psi \right) ,
\end{eqnarray}
\begin{eqnarray}
{H_{+}^{(2.5)}}&=&
\pi \,\cos\left(2\,\psi \right)\,\left({\displaystyle \frac {19}{3}}  + 3\,{c_\theta}^{2} - {\displaystyle \frac {2}{3}} \,{c_\theta}^{4}\right) + {\displaystyle \frac {16}{3}} \,\pi \,{s_\theta}^{2}\,\cos\left(4\,\psi \right)\,\left(1 + {c_\theta}^{2}\right) \cr
&&
 + {s_\theta}\,\sin\left(\psi \right)\,\left( - {\displaystyle \frac {1771}{5120}}  + {\displaystyle \frac {1667}{5120}} \,{c_\theta}^{2} - {\displaystyle \frac {217}{9216}} \,{c_\theta}^{4} + {\displaystyle \frac {1}{9216}} \,{c_\theta}^{6}\right) \cr
&&
 + \sin\left(2\,\psi \right)\,\left( - {\displaystyle \frac {9}{5}}  + {\displaystyle \frac {14}{5}} \,{c_\theta}^{2} + {\displaystyle \frac {7}{5}} \,{c_\theta}^{4}\right) \cr
&&
 + {s_\theta}\,\sin\left(3\,\psi \right)\,\left({\displaystyle \frac {3537}{1024}}  - {\displaystyle \frac {22977}{5120}} \,{c_\theta}^{2} - {\displaystyle \frac {15309}{5120}} \,{c_\theta}^{4} + {\displaystyle \frac {729}{5120}} \,{c_\theta}^{6}\right) \cr
&&
 + {s_\theta}^{2}\,\left(1 + {c_\theta}^{2}\right)\,\sin\left(4\,\psi \right)\,\left( - {\displaystyle \frac {56}{5}}  + {\displaystyle \frac {32}{3}} \,\ln\left(2\right)\right) \cr
&&
 + {s_\theta}^{3}\,\sin\left(5\,\psi \right)\,\left({\displaystyle \frac {108125}{9216}}  + {\displaystyle \frac {1875}{256}} \,{c_\theta}^{2} - {\displaystyle \frac {15625}{9216}} \,{c_\theta}^{4}\right) \cr
&&
 + {\displaystyle \frac {117649}{46080}} \,{s_\theta}^{5}\,\left(1 + {c_\theta}^{2}\right)\,\sin\left(7\,\psi \right) ,
\end{eqnarray}
\begin{eqnarray}
{H_{+}^{(3)}}&=&
{s_\theta}\,\cos\left(\psi \right) 
\left[ - {\displaystyle \frac {2159}{40320}}  - {\displaystyle \frac {19}{32}} \,\ln\left(2\right) + \left( - {\displaystyle \frac {5}{8}} \,\ln\left(2\right) - {\displaystyle \frac {95}{224}} \right)\,{c_\theta}^{2} + \left({\displaystyle \frac {1}{96}} \,\ln\left(2\right) + {\displaystyle \frac {181}{13440}} \right)\,{c_\theta}^{4}\right] \cr
&&
+ \cos\left(2\,\psi \right)\left[ - {\displaystyle \frac {465497}{11025}}  - {\displaystyle \frac {2}{3}} \,\pi ^{2} + {\displaystyle \frac {856}{105}} \,{\rm eulerlog}\left(2, \,v\right) \right.\cr
&&
\left. + \left( - {\displaystyle \frac {2}{3}} \,\pi ^{2} + {\displaystyle \frac {856}{105}} \,{\rm eulerlog}\left(2, \,v\right) - {\displaystyle \frac {3561541}{88200}} \right)\,{c_\theta}^{2} - {\displaystyle \frac {943}{720}} \,{c_\theta}^{4} + {\displaystyle \frac {169}{720}} \,{c_\theta}^{6}  - {\displaystyle \frac {1}{360}} \,{c_\theta}^{8}\right]\cr
&&
 + {s_\theta}\,\cos\left(3\,\psi \right)\left[{\displaystyle \frac {1971}{64}} \,\ln\left({\displaystyle \frac {3}{2}} \right) - {\displaystyle \frac {205119}{8960}} + \left( - {\displaystyle \frac {1917}{224}}  + {\displaystyle \frac {135}{8}} \,\ln\left({\displaystyle \frac {3}{2}} \right)\right)\,{c_\theta}^{2}\right.  
\cr
&&
\left. + \left( - {\displaystyle \frac {243}{64}} \,\ln\left({\displaystyle \frac {3}{2}} \right) + {\displaystyle \frac {43983}{8960}} \right)\,{c_\theta}^{4}\right] \cr
&&
 + {s_\theta}^{2}\,\cos\left(4\,\psi \right)\,\left[{\displaystyle \frac {2189}{210}}  - {\displaystyle \frac {1123}{210}} \,{c_\theta}^{2} - {\displaystyle \frac {56}{9}} \,{c_\theta}^{4} + {\displaystyle \frac {16}{45}} \,{c_\theta}^{6}\right] \cr
&&
 + {s_\theta}^{3}\,\cos\left(5\,\psi \right)\,\left(1 + {c_\theta}^{2}\right)\,\left[ - {\displaystyle \frac {113125}{5376}}  + {\displaystyle \frac {3125}{192}} \,\ln\left({\displaystyle \frac {5}{2}} \right)\right] \cr
&&
 + {s_\theta}^{4}\,\cos\left(6\,\psi \right)\,\left[{\displaystyle \frac {1377}{80}}  + {\displaystyle \frac {891}{80}} \,{c_\theta}^{2} - {\displaystyle \frac {729}{280}} \,{c_\theta}^{4}\right] \cr
&&
 + {\displaystyle \frac {1024}{315}} \,{s_\theta}^{6}\,\cos\left(8\,\psi \right)\,\left[1 + {c_\theta}^{2}\right] + \pi \,{s_\theta}\,\sin\left(\psi \right)\,\left[ - {\displaystyle \frac {19}{64}}  - {\displaystyle \frac {5}{16}} \,{c_\theta}^{2} + {\displaystyle \frac {1}{192}} \,{c_\theta}^{4}\right] \cr
&&
 - {\displaystyle \frac {428}{105}} \,\pi \,\left[1 + {c_\theta}^{2}\right]\,\sin\left(2\,\psi \right) + \pi \,{s_\theta}\,{\rm sin}\left(3\,\psi \right)\,\left[ - {\displaystyle \frac {1971}{128}}  - {\displaystyle \frac {135}{16}} \,{c_\theta}^{2} + {\displaystyle \frac {243}{128}} \,{c_\theta}^{4}\right] \cr
&&
 - {\displaystyle \frac {3125}{384}} \,\pi \,{s_\theta}^{3}\,\left[1 + {c_\theta}^{2}\right]\,\sin\left(5\,\psi \right) ,
\end{eqnarray}
\begin{eqnarray}
{H_{+}^{(3.5)}}&=&
\pi \,{s_\theta}\,\cos\left(\psi \right)\,\left[{\displaystyle \frac {5}{4}} \,\ln\left(2\right) - {\displaystyle \frac {53}{140}}  + \left({\displaystyle \frac {1}{4}} \,\ln\left(2\right) + {\displaystyle \frac {41}{420}} \right)\,{c_\theta}^{2}\right] \cr
&&
 + \pi \,\cos\left(2\,\psi \right)\,\left[{\displaystyle \frac {11}{30}}  + {\displaystyle \frac {33}{5}} \,{c_\theta}^{2} + {\displaystyle \frac {29}{12}} \,{c_\theta}^{4} - {\displaystyle \frac {1}{12}} \,{c_\theta}^{6}\right] 
 + \pi \,{s_\theta}\,\cos\left(3\,\psi \right)\,\left(1 + {c_\theta}^{2}\right)\,\left[ - {\displaystyle \frac {81}{4}} \,{\rm ln}\left({\displaystyle \frac {3}{2}} \right) + {\displaystyle \frac {1107}{140}} \right] \cr
&&
 + \pi \,{s_\theta}^{2}\,\cos\left(4\,\psi \right)\,\left[ - {\displaystyle \frac {472}{15}}  - {\displaystyle \frac {56}{3}} \,{c_\theta}^{2} + {\displaystyle \frac {64}{15}} \,{c_\theta}^{4}\right] 
 - {\displaystyle \frac {243}{20}} \,\pi \,{s_\theta}^{4}\,\cos\left(6\,\psi \right)\,\left[1 + {c_\theta}^{2}\right] \cr
&&
+ {s_\theta}\,\sin\left(\psi \right)\left[ - {\displaystyle \frac {11}{20}} \,\ln\left(2\right) + {\displaystyle \frac {5}{48}} \,\pi ^{2} - {\displaystyle \frac {5}{4}} \,\ln\left(2\right)^{2} \right.
 - {\displaystyle \frac {183}{140}} \,{\rm eulerlog}\left(1, \,v\right) + {\displaystyle \frac {3614800949}{541900800}}  \cr
&&
 + \left( - {\displaystyle \frac {13}{84}} \,{\rm eulerlog}\left(1, \,v\right) - {\displaystyle \frac {7}{20}} \,\ln\left(2\right) - {\displaystyle \frac {1}{4}} \,\ln\left(2\right)^{2} + {\displaystyle \frac {16709489}{18063360}}  + {\displaystyle \frac {1}{48}} \,\pi ^{2}\right)\,{c_\theta}^{2} \cr
&&
\left. + {\displaystyle \frac {6169}{143360}} \,{c_\theta}^{4} - {\displaystyle \frac {29}{40960}} \,{c_\theta}^{6} + {\displaystyle \frac {1}{737280}} \,{c_\theta}^{8}\right] \cr
&&
 + \sin\left(2\,\psi \right)\,\left[{\displaystyle \frac {17957}{5040}}  - {\displaystyle \frac {7583}{1680}} \,{c_\theta}^{2} - {\displaystyle \frac {1201}{240}} \,{c_\theta}^{4} + {\displaystyle \frac {83}{336}} \,{c_\theta}^{6}\right]\cr
&&
 + {s_\theta}\,\sin\left(3\,\psi \right)\left[{\displaystyle \frac {27}{16}} \,\pi ^{2} - {\displaystyle \frac {81}{4}} \,\ln\left({\displaystyle \frac {3}{2}} \right)^{2} - {\displaystyle \frac {351}{28}} \,{\rm eulerlog}\left(3, \,v\right) + {\displaystyle \frac {154720053}{2007040}}  + {\displaystyle \frac {567}{20}} \,\ln\left({\displaystyle \frac {3}{2}} \right) \right.\cr
&&
 + \left({\displaystyle \frac {27}{16}} \,\pi ^{2} - {\displaystyle \frac {351}{28}} \,{\rm eulerlog}\left(3, \,v\right) - {\displaystyle \frac {81}{4}} \,\ln\left({\displaystyle \frac {3}{2}} \right)^{2} + {\displaystyle \frac {567}{20}} \,{\rm ln}\left({\displaystyle \frac {3}{2}} \right) + {\displaystyle \frac {79851141}{1003520}} \right)\,{c_\theta}^{2} \cr
&&
\left. + {\displaystyle \frac {562761}{143360}} \,{c_\theta}^{4} - {\displaystyle \frac {19683}{20480}} \,{c_\theta}^{6} + {\displaystyle \frac {729}{40960}} \,{c_\theta}^{8}\right] \cr
&&
 + {s_\theta}^{2}\,\sin\left(4\,\psi \right)\left[ - {\displaystyle \frac {944}{15}} \,\ln\left(2\right) + {\displaystyle \frac {7076}{105}}  + \left( - {\displaystyle \frac {112}{3}} \,\ln\left(2\right) + {\displaystyle \frac {1212}{35}} \right)\,{c_\theta}^{2} + \left({\displaystyle \frac {128}{15}} \,\ln\left(2\right) - {\displaystyle \frac {1328}{105}} \right)\,{c_\theta}^{4}\right] \cr
&&
 + {s_\theta}^{3}\,\sin\left(5\,\psi \right)\,\left[ - {\displaystyle \frac {3953125}{172032}}  + {\displaystyle \frac {2696875}{516096}} \,{c_\theta}^{2} + {\displaystyle \frac {6078125}{516096}} \,{c_\theta}^{4} - {\displaystyle \frac {390625}{516096}} \,{c_\theta}^{6}\right] \cr
&&
 + {s_\theta}^{4}\,\left(1 + {c_\theta}^{2}\right)\,\sin\left(6\,\psi \right)\,\left[{\displaystyle \frac {20169}{560}}  - {\displaystyle \frac {243}{10}} \,\ln\left(3\right)\right] \cr
&&
 + {s_\theta}^{5}\,\sin\left(7\,\psi \right)\,\left[ - {\displaystyle \frac {36824137}{1474560}}  - {\displaystyle \frac {1529437}{92160}} \,{c_\theta}^{2} + {\displaystyle \frac {5764801}{1474560}} \,{c_\theta}^{4}\right] 
 - {\displaystyle \frac {4782969}{1146880}} \,{s_\theta}^{7}\,\left[1 + {c_\theta}^{2}\right]\,\sin\left(9\,\psi \right) ,
\end{eqnarray}
\begin{eqnarray}
{H_{+}^{(4)}}&=&
{s_\theta}\,\cos\left(\psi \right)\left[ - {\displaystyle \frac {1771}{2560}} \,\ln\left(2\right) - {\displaystyle \frac {5290289}{19353600}}  + \left({\displaystyle \frac {8167913}{19353600}}  + {\displaystyle \frac {1667}{2560}} \,\ln\left(2\right)\right)\,{c_\theta}^{2} \right.\cr
&&
\left. + \left( - {\displaystyle \frac {217}{4608}} \,\ln\left(2\right) - {\displaystyle \frac {141173}{2322432}} \right)\,{c_\theta}^{4} + \left({\displaystyle \frac {1}{4608}} \,\ln\left(2\right) + {\displaystyle \frac {4129}{11612160}} \right)\,{c_\theta}^{6}\right] \cr
&&
 + \cos\left(2\,\psi \right)\left[{\displaystyle \frac {19}{9}} \,\pi ^{2} - {\displaystyle \frac {49684}{2079}} \,{\rm eulerlog}\left(2, \,v\right) + {\displaystyle \frac {390473238599}{4610390400}}  \right.\cr
&&
 + \left({\displaystyle \frac {52743420953}{1536796800}}  + \pi ^{2} - {\displaystyle \frac {52676}{3465}} \,{\rm eulerlog}\left(2, \,v\right)\right)\,{c_\theta}^{2} \cr
&&
 + \left({\displaystyle \frac {12568}{10395}} \,{\rm eulerlog}\left(2, \,v\right) - {\displaystyle \frac {2903305411}{461039040}}  - {\displaystyle \frac {2}{9}} \,\pi ^{2}\right)\,{c_\theta}^{4}  \cr
&&
\left. - {\displaystyle \frac {1028339}{1995840}} \,{c_\theta}^{6} + {\displaystyle \frac {61}{2880}} \,{c_\theta}^{8} - {\displaystyle \frac {1}{8640}} \,{c_\theta}^{10}\right] \cr
&&
+ {s_\theta}\,\cos\left(3\,\psi \right)\left[ - {\displaystyle \frac {10611}{512}} \,\ln\left({\displaystyle \frac {3}{2}} \right) + {\displaystyle \frac {13690863}{716800}}  + \left({\displaystyle \frac {68931}{2560}} \,\ln\left({\displaystyle \frac {3}{2}} \right) - {\displaystyle \frac {612441}{20480}} \right)\,{c_\theta}^{2}\right.\cr
&&
\left.
 + \left( - {\displaystyle \frac {16433847}{716800}}  + {\displaystyle \frac {45927}{2560}} \,\ln\left({\displaystyle \frac {3}{2}} \right)\right)\,{c_\theta}^{4} + \left({\displaystyle \frac {1003347}{716800}}  - {\displaystyle \frac {2187}{2560}} \,\ln\left({\displaystyle \frac {3}{2}} \right)\right)\,{c_\theta}^{6}\right]\cr
&&
 + {s_\theta}^{2}\,\cos\left(4\,\psi \right)\left[{\displaystyle \frac {3196066957}{28814940}} \right. 
 - {\displaystyle \frac {201088}{10395}} \,{\rm eulerlog}\left(4, \,v\right) + {\displaystyle \frac {32}{9}} \,\pi ^{2} + {\displaystyle \frac {448}{5}} \,\ln\left(2\right) - {\displaystyle \frac {128}{3}} \,\ln\left(2\right)^{2}  \cr
&&
 +\left({\displaystyle \frac {448}{5}} \,\ln\left(2\right) + {\displaystyle \frac {3689222389}{28814940}}  - {\displaystyle \frac {201088}{10395}} \,{\rm eulerlog}\left(4, \,v\right) + {\displaystyle \frac {32}{9}} \,\pi ^{2} - {\displaystyle \frac {128}{3}} \,\ln\left(2\right)^{2}\right){c_\theta}^{2} \cr
&&
\left. + {\displaystyle \frac {320458}{31185}} \,{c_\theta}^{4} - {\displaystyle \frac {2648}{945}} \,{c_\theta}^{6} + {\displaystyle \frac {64}{945}} \,{c_\theta}^{8}\right] \cr
&&
+ {s_\theta}^{3}\,\cos\left(5\,\psi \right)\left[ - {\displaystyle \frac {540625}{4608}} \,\ln\left({\displaystyle \frac {5}{2}} \right)\right.
 + {\displaystyle \frac {355593125}{2322432}}  + \left({\displaystyle \frac {17029375}{193536}}  - {\displaystyle \frac {9375}{128}} \,\ln\left({\displaystyle \frac {5}{2}} \right)\right)\,{c_\theta}^{2} \cr
&&
\left. + \left( - {\displaystyle \frac {64515625}{2322432}}  + {\displaystyle \frac {78125}{4608}} \,\ln\left({\displaystyle \frac {5}{2}} \right)\right)\,{c_\theta}^{4}\right] \cr
&&
 + {s_\theta}^{4}\,\cos\left(6\,\psi \right)\,\left[ - {\displaystyle \frac {2193561}{49280}}  + {\displaystyle \frac {153333}{49280}} \,{c_\theta}^{2} + {\displaystyle \frac {94041}{4480}} \,{c_\theta}^{4} - {\displaystyle \frac {6561}{4480}} \,{c_\theta}^{6}\right] \cr
&&
 + {s_\theta}^{5}\,\cos\left(7\,\psi \right)\,\left(1 + {c_\theta}^{2}\right)\,\left[ - {\displaystyle \frac {823543}{23040}} \,{\rm ln}\left({\displaystyle \frac {7}{2}} \right) + {\displaystyle \frac {485772721}{8294400}} \right] \cr
&&
 + {s_\theta}^{6}\,\cos\left(8\,\psi \right)\,\left[ - {\displaystyle \frac {101888}{2835}}  - {\displaystyle \frac {512}{21}} \,{c_\theta}^{2} + {\displaystyle \frac {16384}{2835}} \,{c_\theta}^{4}\right] 
 - {\displaystyle \frac {390625}{72576}} \,{s_\theta}^{8}\,\cos\left(10\,\psi \right)\,\left[1 + {c_\theta}^{2}\right] \cr
&&
 + \pi \,{s_\theta}\,\sin\left(\psi \right)\,\left[ - {\displaystyle \frac {1771}{5120}}  + {\displaystyle \frac {1667}{5120}} \,{c_\theta}^{2} - {\displaystyle \frac {217}{9216}} \,{c_\theta}^{4} + {\displaystyle \frac {1}{9216}} \,{c_\theta}^{6}\right] \cr
&&
 + \pi \,\sin\left(2\,\psi \right)\,\left[{\displaystyle \frac {86788}{10395}}  + {\displaystyle \frac {45742}{3465}} \,{c_\theta}^{2} + {\displaystyle \frac {22822}{10395}} \,{c_\theta}^{4}\right] \cr
&&
 + \pi \,{s_\theta}\,\sin\left(3\,\psi \right)\,\left[{\displaystyle \frac {10611}{1024}}  - {\displaystyle \frac {68931}{5120}} \,{c_\theta}^{2} - {\displaystyle \frac {45927}{5120}} \,{c_\theta}^{4} + {\displaystyle \frac {2187}{5120}} \,{c_\theta}^{6}\right] \cr
&&
 + \pi \,{s_\theta}^{2}\,\left(1 + {c_\theta}^{2}\right)\,\sin\left(4\,\psi \right)\,\left[{\displaystyle \frac {128}{3}} \,\ln\left(2\right) - {\displaystyle \frac {365152}{10395}} \right] \cr
&&
 + \pi \,{s_\theta}^{3}\,\sin\left(5\,\psi \right)\,\left[{\displaystyle \frac {540625}{9216}}  + {\displaystyle \frac {9375}{256}} \,{c_\theta}^{2} - {\displaystyle \frac {78125}{9216}} \,{c_\theta}^{4}\right] 
 + {\displaystyle \frac {823543}{46080}} \,\pi \,{s_\theta}^{5}\,\left[1 + {c_\theta}^{2}\right]\,\sin\left(7\,\psi \right) ,
\end{eqnarray}
\begin{eqnarray}
{H_{+}^{(4.5)}}&=&
\pi \,{s_\theta}\,\cos\left(\psi \right)\left[ - {\displaystyle \frac {19}{32}} \,\ln\left(2\right) + {\displaystyle \frac {5044511}{17297280}}  + \left( - {\displaystyle \frac {5}{8}} \,\ln\left(2\right) - {\displaystyle \frac {36451}{160160}} \right)\,{c_\theta}^{2} \right.\cr
&&
\left. + \left({\displaystyle \frac {1}{96}} \,\ln\left(2\right) + {\displaystyle \frac {66827}{5765760}} \right)\,{c_\theta}^{4}\right] \cr
&&
+ \pi \,\cos\left(2\,\psi \right)\left[ - {\displaystyle \frac {930994}{11025}}  \right.
 + {\displaystyle \frac {1712}{105}} \,{\rm eulerlog}\left(2, \,v\right) + {\displaystyle \frac {4}{3}} \,\pi ^{2} \cr
&&
 + \left({\displaystyle \frac {1712}{105}} \,{\rm eulerlog}\left(2, \,v\right) - {\displaystyle \frac {3561541}{44100}}  + {\displaystyle \frac {4}{3}} \,\pi ^{2}\right)\,{c_\theta}^{2} - {\displaystyle \frac {943}{360}} \,{c_\theta}^{4} + {\displaystyle \frac {169}{360}} \,{c_\theta}^{6} \left. - {\displaystyle \frac {1}{180}} \,{c_\theta}^{8}\right]\cr
&&
 + \pi \,{s_\theta}\,\cos\left(3\,\psi \right)\left[ - {\displaystyle \frac {7502679}{183040}}  + {\displaystyle \frac {5913}{64}} \,\ln\left({\displaystyle \frac {3}{2}} \right)  + \left({\displaystyle \frac {405}{8}} \,\ln\left({\displaystyle \frac {3}{2}} \right) - {\displaystyle \frac {1244619}{160160}} \right)\,{c_\theta}^{2}\right.\cr
&&
\left. + \left( - {\displaystyle \frac {729}{64}} \,\ln\left({\displaystyle \frac {3}{2}} \right) + {\displaystyle \frac {16238961}{1281280}} \right)\,{c_\theta}^{4}\right] \cr
&&
 + \pi \,{s_\theta}^{2}\,\cos\left(4\,\psi \right)\,\left[{\displaystyle \frac {4378}{105}}  - {\displaystyle \frac {2246}{105}} \,{c_\theta}^{2} - {\displaystyle \frac {224}{9}} \,{c_\theta}^{4} + {\displaystyle \frac {64}{45}} \,{c_\theta}^{6}\right] \cr
&&
 + \pi \,{s_\theta}^{3}\,\cos\left(5\,\psi \right)\,\left(1 + {c_\theta}^{2}\right)\,\left[ - {\displaystyle \frac {208834375}{2306304}}  + {\displaystyle \frac {15625}{192}} \,\ln\left({\displaystyle \frac {5}{2}} \right)\right] \cr
&&
 + \pi \,{s_\theta}^{4}\,\cos\left(6\,\psi \right)\,\left[{\displaystyle \frac {4131}{40}}  + {\displaystyle \frac {2673}{40}} \,{c_\theta}^{2} - {\displaystyle \frac {2187}{140}} \,{c_\theta}^{4}\right]  + {\displaystyle \frac {8192}{315}} \,\pi \,{s_\theta}^{6}\,\cos\left(8\,\psi \right)\,\left[1 + {c_\theta}^{2}\right]\cr
&&
 + {s_\theta}\,\sin\left(\psi \right)\left[{\displaystyle \frac {2985361}{4324320}} \,{\rm eulerlog}\left(1, \,v\right) \right.
 + {\displaystyle \frac {2725185940926109}{797855721062400}}  - {\displaystyle \frac {19}{384}} \,\pi ^{2} + {\displaystyle \frac {19}{32}} \,\ln\left(2\right)^{2} + {\displaystyle \frac {2159}{20160}} \,\ln\left(2\right) \cr
&&
 + \left({\displaystyle \frac {15737}{40040}} \,{\rm eulerlog}\left(1, \,v\right) - {\displaystyle \frac {117462416679523}{88650635673600}}  + {\displaystyle \frac {5}{8}} \,\ln\left(2\right)^{2} - {\displaystyle \frac {5}{96}} \,\pi ^{2} + {\displaystyle \frac {95}{112}} \,\ln\left(2\right)\right){c_\theta}^{2} \cr
&&
 + \left( - {\displaystyle \frac {181}{6720}} \,\ln\left(2\right) + {\displaystyle \frac {1}{1152}} \,\pi ^{2} - {\displaystyle \frac {773}{205920}} \,{\rm eulerlog}\left(1, \,v\right) - {\displaystyle \frac {1}{96}} \,\ln\left(2\right)^{2}  - {\displaystyle \frac {66829188883}{3799312957440}} \right){c_\theta}^{4}\cr
&&
\left. + {\displaystyle \frac {23406169}{12076646400}} \,{c_\theta}^{6} - {\displaystyle \frac {341}{29491200}} \,{c_\theta}^{8}  + {\displaystyle \frac {1}{88473600}} \,{c_\theta}^{10}\right]\cr
&&
 + \sin\left(2\,\psi \right)\left[{\displaystyle \frac {64}{3}} \,\zeta \left(3\right) + {\displaystyle \frac {236081}{50400}}  - {\displaystyle \frac {856}{63}} \,\pi ^{2} + \left( - {\displaystyle \frac {856}{63}} \,\pi ^{2} - {\displaystyle \frac {250211}{151200}}  + {\displaystyle \frac {64}{3}} \,\zeta \left(3\right)\right)\,{c_\theta}^{2}\right.\cr
&&
 + {\displaystyle \frac {2416699}{453600}} \,{c_\theta}^{4} - {\displaystyle \frac {630719}{453600}} \,{c_\theta}^{6} \left. + {\displaystyle \frac {319}{16200}} \,{c_\theta}^{8}\right]\cr
&&
 + {s_\theta}\,\sin\left(3\,\psi \right)\left[{\displaystyle \frac {5913}{64}} \,\ln\left({\displaystyle \frac {3}{2}} \right)^{2} - {\displaystyle \frac {615357}{4480}} \,\ln\left({\displaystyle \frac {3}{2}} \right) \right.\cr
&&
 + {\displaystyle \frac {17738649}{320320}} \,{\rm eulerlog}\left(3, \,v\right) - {\displaystyle \frac {1971}{256}} \,\pi ^{2} - {\displaystyle \frac {863088610282869}{3283356876800}} \cr
&&
 + \left( - {\displaystyle \frac {497738577409209}{3283356876800}}  + {\displaystyle \frac {1433673}{40040}} \,{\rm eulerlog}\left(3, \,v\right) - {\displaystyle \frac {5751}{112}} \,\ln\left({\displaystyle \frac {3}{2}} \right) - {\displaystyle \frac {135}{32}} \,\pi ^{2} + {\displaystyle \frac {405}{8}} \,\ln\left({\displaystyle \frac {3}{2}} \right)^{2}\right){c_\theta}^{2}\cr
&&
 + \left({\displaystyle \frac {131949}{4480}} \,\ln\left({\displaystyle \frac {3}{2}} \right) - {\displaystyle \frac {729}{64}} \,\ln\left({\displaystyle \frac {3}{2}} \right)^{2} + {\displaystyle \frac {831522561921}{46905098240}}  + {\displaystyle \frac {243}{256}} \,\pi ^{2} - {\displaystyle \frac {187839}{45760}} \,{\rm eulerlog}\left(3, \,v\right)\right){c_\theta}^{4}\cr
&&
\left. + {\displaystyle \frac {370480473}{149094400}} \,{c_\theta}^{6} - {\displaystyle \frac {713691}{4587520}} \,{c_\theta}^{8} + {\displaystyle \frac {6561}{4587520}} \,{c_\theta}^{10}\right]\cr
&&
 + {s_\theta}^{2}\,\sin\left(4\,\psi \right)\left[ - {\displaystyle \frac {1332067}{14175}}  + {\displaystyle \frac {8756}{105}} \,\ln\left(2\right) + \left( - {\displaystyle \frac {4492}{105}} \,\ln\left(2\right) + {\displaystyle \frac {971317}{14175}} \right)\,{c_\theta}^{2}\right.\cr
&&
 + \left( - {\displaystyle \frac {448}{9}} \,\ln\left(2\right) + {\displaystyle \frac {147944}{2025}} \right)\,{c_\theta}^{4}\left. + \left({\displaystyle \frac {128}{45}} \,\ln\left(2\right) - {\displaystyle \frac {10208}{2025}} \right)\,{c_\theta}^{6}\right] \cr
&&
 + {s_\theta}^{3}\,\sin\left(5\,\psi \right)\left[ - {\displaystyle \frac {131672845144375}{1013150121984}} \right. 
 - {\displaystyle \frac {565625}{2688}} \,\ln\left({\displaystyle \frac {5}{2}} \right) + {\displaystyle \frac {15625}{192}} \,\ln\left({\displaystyle \frac {5}{2}} \right)^{2} \cr
&&
+ {\displaystyle \frac {2415625}{82368}} \,{\rm eulerlog}\left(5, \,v\right) - {\displaystyle \frac {15625}{2304}} \,\pi ^{2} \cr
&&
 + \left({\displaystyle \frac {15625}{192}} \,\ln\left({\displaystyle \frac {5}{2}} \right)^{2} + {\displaystyle \frac {2415625}{82368}} \,{\rm eulerlog}\left(5, \,v\right) - {\displaystyle \frac {91729414920625}{506575060992}}  - {\displaystyle \frac {565625}{2688}} \,\ln\left({\displaystyle \frac {5}{2}} \right) \right.\cr
&&
\left.\left. - {\displaystyle \frac {15625}{2304}} \,\pi ^{2}\right){c_\theta}^{2} - {\displaystyle \frac {2916390625}{120766464}} \,{c_\theta}^{4} + {\displaystyle \frac {169140625}{24772608}} \,{c_\theta}^{6} - {\displaystyle \frac {9765625}{49545216}} \,{c_\theta}^{8}\right] \cr
&&
 + {s_\theta}^{4}\,\sin\left(6\,\psi \right)\left[ - {\displaystyle \frac {1723599}{5600}}  + {\displaystyle \frac {4131}{20}} \,\ln\left(3\right) + \left( - {\displaystyle \frac {1053891}{5600}}  + {\displaystyle \frac {2673}{20}} \,\ln\left(3\right)\right)\,{c_\theta}^{2} \right.\cr
&&
\left. + \left({\displaystyle \frac {77517}{1400}}  - {\displaystyle \frac {2187}{70}} \,\ln\left(3\right)\right)\,{c_\theta}^{4}\right]\cr
&&
 + {s_\theta}^{5}\,\sin\left(7\,\psi \right)\left[{\displaystyle \frac {184464102431}{2300313600}}  + {\displaystyle \frac {6079747373}{2300313600}} \,{c_\theta}^{2} - {\displaystyle \frac {3810533461}{106168320}} \,{c_\theta}^{4} + {\displaystyle \frac {282475249}{106168320}} \,{c_\theta}^{6}\right] \cr
&&
 + {s_\theta}^{6}\,\left(1 + {c_\theta}^{2}\right)\,\sin\left(8\,\psi \right)\,\left[ - {\displaystyle \frac {1306624}{14175}}  + {\displaystyle \frac {32768}{315}} \,\ln\left(2\right)\right] \cr
&&
 + {s_\theta}^{7}\,\sin\left(9\,\psi \right)\,\left[{\displaystyle \frac {2358003717}{45875200}}  + {\displaystyle \frac {81310473}{2293760}} \,{c_\theta}^{2} - {\displaystyle \frac {387420489}{45875200}} \,{c_\theta}^{4}\right] \cr
&&
 + {\displaystyle \frac {25937424601}{3715891200}} \,{s_\theta}^{9}\,\left[1 + {c_\theta}^{2}\right]\,\sin\left(11\,\psi \right) ,
\end{eqnarray}
\begin{eqnarray}
{H_{+}^{(5)}}&=&
{s_\theta}\,\cos\left(\psi \right)\left[ - {\displaystyle \frac {1753}{1680}} \,\pi ^{2} - {\displaystyle \frac {11}{20}} \,\ln\left(2\right)^{2} + {\displaystyle \frac {5}{24}} \,\ln\left(2\right)\,\pi ^{2} + {\displaystyle \frac {3614800949}{270950400}} \,\ln\left(2\right) + {\displaystyle \frac {5}{3}} \,\zeta \left(3\right)\right.\cr
&&
 - {\displaystyle \frac {17}{28}} \,{\rm eulerlog}\left(1, \,v\right) + {\displaystyle \frac {2808855689}{851558400}}  - {\displaystyle \frac {5}{6}} \,\ln\left(2\right)^{3} - {\displaystyle \frac {183}{70}} \,\ln\left(2\right)\,{\rm eulerlog}\left(1, \,v\right) \cr
&&
 + \left( - {\displaystyle \frac {13}{60}} \,{\rm eulerlog}\left(1, \,v\right) + {\displaystyle \frac {1376391589}{993484800}}  - {\displaystyle \frac {7}{20}} \,\ln\left(2\right)^{2} - {\displaystyle \frac {13}{42}} \,\ln\left(2\right)\,{\rm eulerlog}\left(1, \,v\right) + {\displaystyle \frac {1}{3}} \,\zeta \left(3\right) \right.\cr
&&
 - {\displaystyle \frac {1}{6}} \,\ln\left(2\right)^{3} - {\displaystyle \frac {503}{5040}} \,\pi ^{2} + {\displaystyle \frac {1}{24}} \,\ln\left(2\right)\,\pi ^{2} \left. + {\displaystyle \frac {16709489}{9031680}} \,\ln\left(2\right)\right){c_\theta}^{2}\cr
&&
 + \left({\displaystyle \frac {6169}{71680}} \,\ln\left(2\right) + {\displaystyle \frac {247872791}{2235340800}} \right)\,{c_\theta}^{4} + \left( - {\displaystyle \frac {423289}{182476800}}  - {\displaystyle \frac {29}{20480}} \,\ln\left(2\right)\right)\,{c_\theta}^{6} \cr
&&
\left. + \left({\displaystyle \frac {13093}{2554675200}}  + {\displaystyle \frac {1}{368640}} \,\ln\left(2\right)\right)\,{c_\theta}^{8}\right] \cr
&&
 + \cos\left(2\,\psi \right)\left[{\displaystyle \frac {15970130661967}{222615993600}}  + {\displaystyle \frac {11}{90}} \,\pi ^{2} - {\displaystyle \frac {713942}{135135}} \,{\rm eulerlog}\left(2, \,v\right) \right.\cr
&&
 + \left( - {\displaystyle \frac {332896}{15015}} \,{\rm eulerlog}\left(2, \,v\right) + {\displaystyle \frac {643219288717}{4947022080}}  + {\displaystyle \frac {11}{5}} \,\pi ^{2}\right)\,{c_\theta}^{2} \cr
&&
 + \left({\displaystyle \frac {1458499764979}{77915597760}}  + {\displaystyle \frac {29}{36}} \,\pi ^{2} - {\displaystyle \frac {599887}{135135}} \,{\rm eulerlog}\left(2, \,v\right)\right)\,{c_\theta}^{4} \cr
&&
 + \left({\displaystyle \frac {901}{9009}} \,{\rm eulerlog}\left(2, \,v\right) - {\displaystyle \frac {1}{36}} \,\pi ^{2} + {\displaystyle \frac {13211394139}{51943731840}} \right)\,{c_\theta}^{6} \cr
&&
\left. - {\displaystyle \frac {449833}{6652800}} \,{c_\theta}^{8} + {\displaystyle \frac {673}{604800}} \,{c_\theta}^{10} - {\displaystyle \frac {1}{302400}} \,{c_\theta}^{12}\right]\cr
&&
 + {s_\theta}\,\cos\left(3\,\psi \right)\left[{\displaystyle \frac {37689199239}{110387200}} \right. 
 - {\displaystyle \frac {1701}{20}} \,\ln\left({\displaystyle \frac {3}{2}} \right)^{2} - {\displaystyle \frac {1053}{20}} \,{\rm eulerlog}\left(3, \,v\right) + 81\,\zeta \left(3\right) + {\displaystyle \frac {81}{2}} \,\ln\left({\displaystyle \frac {3}{2}} \right)^{3} \cr
&&
 - {\displaystyle \frac {464160159}{1003520}} \,{\rm ln}\left({\displaystyle \frac {3}{2}} \right) - {\displaystyle \frac {13581}{560}} \,\pi ^{2} - {\displaystyle \frac {81}{8}} \,\pi ^{2}\,\ln\left({\displaystyle \frac {3}{2}} \right) + {\displaystyle \frac {1053}{14}} \,{\rm eulerlog}\left(3, \,v\right)\,\ln\left({\displaystyle \frac {3}{2}} \right) \cr
&&
 + \left( - {\displaystyle \frac {13581}{560}} \,\pi ^{2} - {\displaystyle \frac {239553423}{501760}} \,\ln\left({\displaystyle \frac {3}{2}} \right) + {\displaystyle \frac {1053}{14}} \,{\rm eulerlog}\left(3, \,v\right)\,\ln\left({\displaystyle \frac {3}{2}} \right) - {\displaystyle \frac {1053}{20}} \,{\rm eulerlog}\left(3, \,v\right) \right.\cr
&&
 - {\displaystyle \frac {81}{8}} \,\pi ^{2}\,\ln\left({\displaystyle \frac {3}{2}} \right) + {\displaystyle \frac {704592069}{1971200}}  - {\displaystyle \frac {1701}{20}} \,\ln\left({\displaystyle \frac {3}{2}} \right)^{2} + 81\,\zeta \left(3\right) \left. + {\displaystyle \frac {81}{2}} \,\ln\left({\displaystyle \frac {3}{2}} \right)^{3}\right){c_\theta}^{2}\cr
&&
 + \left( - {\displaystyle \frac {1688283}{71680}} \,\ln\left({\displaystyle \frac {3}{2}} \right) + {\displaystyle \frac {32368167}{1103872}} \right)\,{c_\theta}^{4} + \left( - {\displaystyle \frac {37208403}{3942400}}  + {\displaystyle \frac {59049}{10240}} \,\ln\left({\displaystyle \frac {3}{2}} \right)\right)\,{c_\theta}^{6} \cr
&&
\left. + \left( - {\displaystyle \frac {2187}{20480}} \,\ln\left({\displaystyle \frac {3}{2}} \right) + {\displaystyle \frac {3181599}{15769600}} \right)\,{c_\theta}^{8}\right]  \cr
&&
 + {s_\theta}^{2}\,\cos\left(4\,\psi \right)\left[{\displaystyle \frac {3776}{15}} \,\ln\left(2\right)^{2} - {\displaystyle \frac {944}{45}} \,\pi ^{2} - {\displaystyle \frac {101708443628797}{194788994400}}  - {\displaystyle \frac {56608}{105}} \,{\rm ln}\left(2\right) + {\displaystyle \frac {15158848}{135135}} \,{\rm eulerlog}\left(4, \,v\right) \right.\cr
&&
 + \left( - {\displaystyle \frac {388693669747}{1113079968}}  + {\displaystyle \frac {10079296}{135135}} \,{\rm eulerlog}\left(4, \,v\right)  - {\displaystyle \frac {9696}{35}} \,\ln\left(2\right) - {\displaystyle \frac {112}{9}} \,\pi ^{2} + {\displaystyle \frac {448}{3}} \,\ln\left(2\right)^{2}\right){c_\theta}^{2}\cr
&&
 + \left({\displaystyle \frac {128}{45}} \,\pi ^{2} + {\displaystyle \frac {58122585749}{2029052025}} 
 + {\displaystyle \frac {10624}{105}} \,\ln\left(2\right) - {\displaystyle \frac {461312}{45045}} \,{\rm eulerlog}\left(4, \,v\right) - {\displaystyle \frac {512}{15}} \,\ln\left(2\right)^{2}\right){c_\theta}^{4}  \cr
&&
\left. + {\displaystyle \frac {438602}{51975}} \,{c_\theta}^{6}
- {\displaystyle \frac {632}{945}} \,{c_\theta}^{8} + {\displaystyle \frac {8}{945}} \,{c_\theta}^{10}\right]\cr
&&
 + {s_\theta}^{3}\,\cos\left(5\,\psi \right)\left[ - {\displaystyle \frac {36498184375}{119218176}}  + {\displaystyle \frac {19765625}{86016}} \,\ln\left({\displaystyle \frac {5}{2}} \right) \right.
 + \left({\displaystyle \frac {4506809375}{39739392}}  - {\displaystyle \frac {13484375}{258048}} \,\ln\left({\displaystyle \frac {5}{2}} \right)\right)\,{c_\theta}^{2} \cr
&&
 + \left({\displaystyle \frac {68434046875}{357654528}}  - {\displaystyle \frac {30390625}{258048}} \,\ln\left({\displaystyle \frac {5}{2}} \right)\right)\,{c_\theta}^{4} 
\left. + \left( - {\displaystyle \frac {5114453125}{357654528}}  + {\displaystyle \frac {1953125}{258048}} \,\ln\left({\displaystyle \frac {5}{2}} \right)\right)\,{c_\theta}^{6}\right]\cr
&& 
+ {s_\theta}^{4}\,\cos\left(6\,\psi \right)\left[{\displaystyle \frac {729}{5}} \,\ln\left(3\right)^{2} \right.
 - {\displaystyle \frac {60507}{140}} \,\ln\left(3\right) - {\displaystyle \frac {243}{20}} \,\pi ^{2} - {\displaystyle \frac {672377033403}{6412806400}}  + {\displaystyle \frac {218943}{5005}} \,{\rm eulerlog}\left(6, \,v\right) \cr
&&
 + \left({\displaystyle \frac {729}{5}} \,\ln\left(3\right)^{2} + {\displaystyle \frac {218943}{5005}} \,{\rm eulerlog}\left(6, \,v\right) - {\displaystyle \frac {60507}{140}} \,\ln\left(3\right) - {\displaystyle \frac {1440739832607}{6412806400}}  - {\displaystyle \frac {243}{20}} \,\pi ^{2}\right){c_\theta}^{2} \cr
&&
\left. - {\displaystyle \frac {25725681}{492800}} \,{c_\theta}^{4} + {\displaystyle \frac {133407}{8960}} \,{c_\theta}^{6} - {\displaystyle \frac {2187}{4480}} \,{c_\theta}^{8}\right] \cr
&&
 + {s_\theta}^{5}\cos\left(7\,\psi \right)\left[{\displaystyle \frac {257768959}{737280}} \,\ln\left({\displaystyle \frac {7}{2}} \right) - {\displaystyle \frac {139854425207}{243302400}}  \right.
 + \left({\displaystyle \frac {10706059}{46080}} \,\ln\left({\displaystyle \frac {7}{2}} \right) - {\displaystyle \frac {11134654307}{30412800}} \right)\,{c_\theta}^{2} \cr
&&
\left. + \left({\displaystyle \frac {75478539493}{729907200}}  - {\displaystyle \frac {40353607}{737280}} \,\ln\left({\displaystyle \frac {7}{2}} \right)\right)\,{c_\theta}^{4}\right] \cr
&&
 + {s_\theta}^{6}\,\cos\left(8\,\psi \right)\,\left[{\displaystyle \frac {4298368}{31185}}  + {\displaystyle \frac {763264}{51975}} \,{c_\theta}^{2} - {\displaystyle \frac {843776}{14175}} \,{c_\theta}^{4} + {\displaystyle \frac {65536}{14175}} \,{c_\theta}^{6}\right] \cr
&&
 + {s_\theta}^{7}\,\cos\left(9\,\psi \right)\,\left(1 + {c_\theta}^{2}\right)\left[ - {\displaystyle \frac {62623413117}{441548800}}  + {\displaystyle \frac {43046721}{286720}} \,\ln\left(3\right) - {\displaystyle \frac {43046721}{573440}} \,\ln\left(2\right)\right] \cr
&&
 + {s_\theta}^{8}\,\cos\left(10\,\psi \right)\,\left[{\displaystyle \frac {116796875}{1596672}}  + {\displaystyle \frac {7421875}{145152}} \,{c_\theta}^{2} - {\displaystyle \frac {9765625}{798336}} \,{c_\theta}^{4}\right] \cr
&&
 + {\displaystyle \frac {17496}{1925}} \,{s_\theta}^{10}\,\cos\left(12\,\psi \right)\,\left[1 + {c_\theta}^{2}\right]\cr
&&
 + \pi \,{s_\theta}\,\sin\left(\psi \right)\left[{\displaystyle \frac {3779306549}{541900800}}  - {\displaystyle \frac {5}{48}} \,\pi ^{2} \right.
 - {\displaystyle \frac {5}{4}} \,\ln\left(2\right)^{2} - {\displaystyle \frac {183}{140}} \,{\rm eulerlog}\left(1, \,v\right) + {\displaystyle \frac {53}{70}} \,\ln\left(2\right) \cr
&&
 + \left( - {\displaystyle \frac {41}{210}} \,\ln\left(2\right) - {\displaystyle \frac {13}{84}} \,{\rm eulerlog}\left(1, \,v\right) - {\displaystyle \frac {1}{4}} \,\ln\left(2\right)^{2} + {\displaystyle \frac {18666353}{18063360}}  - {\displaystyle \frac {1}{48}} \,\pi ^{2}\right)\,{c_\theta}^{2} \cr
&&
\left. + {\displaystyle \frac {6169}{143360}} \,{c_\theta}^{4} - {\displaystyle \frac {29}{40960}} \,{c_\theta}^{6} + {\displaystyle \frac {1}{737280}} \,{c_\theta}^{8}\right] \cr
&&
 + \pi \,\sin\left(2\,\psi \right)\,\left[{\displaystyle \frac {10559321}{1081080}}  + {\displaystyle \frac {16481}{8008}} \,{c_\theta}^{2} - {\displaystyle \frac {8420261}{1081080}} \,{c_\theta}^{4} + {\displaystyle \frac {32003}{72072}} \,{c_\theta}^{6}\right]  \cr
&&
 + \pi {s_\theta}\,\sin\left(3\,\psi \right)\left[ - {\displaystyle \frac {1053}{28}} \,{\rm eulerlog}\left(3, \,v\right) - {\displaystyle \frac {243}{4}} \,\ln\left({\displaystyle \frac {3}{2}} \right)^{2} - {\displaystyle \frac {81}{16}} \,\pi ^{2} + {\displaystyle \frac {516995487}{2007040}} + {\displaystyle \frac {3321}{70}} \,\ln\left({\displaystyle \frac {3}{2}} \right)  \right.\cr
&&
 + \left({\displaystyle \frac {3321}{70}} \,\ln\left({\displaystyle \frac {3}{2}} \right) + {\displaystyle \frac {265971087}{1003520}}  - {\displaystyle \frac {81}{16}} \,\pi ^{2} - {\displaystyle \frac {1053}{28}} \,{\rm eulerlog}\left(3, \,v\right) - {\displaystyle \frac {243}{4}} \,\ln\left({\displaystyle \frac {3}{2}} \right)^{2}\right)\,{c_\theta}^{2} \cr
&&
\left. + {\displaystyle \frac {1688283}{143360}} \,{c_\theta}^{4} - {\displaystyle \frac {59049}{20480}} \,{c_\theta}^{6} + {\displaystyle \frac {2187}{40960}} \,{c_\theta}^{8}\right] \cr
&&
+ \pi \,{s_\theta}^{2}\,\sin\left(4\,\psi \right)\left[ - {\displaystyle \frac {3776}{15}} \,\ln\left(2\right)\right.  + {\displaystyle \frac {28847824}{135135}}  + \left( - {\displaystyle \frac {448}{3}} \,\ln\left(2\right) + {\displaystyle \frac {2735696}{27027}} \right)\,{c_\theta}^{2}\cr
&&
\left. + \left({\displaystyle \frac {512}{15}} \,\ln\left(2\right) - {\displaystyle \frac {2048192}{45045}} \right)\,{c_\theta}^{4}\right] \cr
&&
 + \pi \,{s_\theta}^{3}\,\sin\left(5\,\psi \right)\left[ - {\displaystyle \frac {19765625}{172032}}  + {\displaystyle \frac {13484375}{516096}} \,{c_\theta}^{2} + {\displaystyle \frac {30390625}{516096}} \,{c_\theta}^{4} - {\displaystyle \frac {1953125}{516096}} \,{c_\theta}^{6}\right] \cr
&&
 + \pi \,{s_\theta}^{4}\,\left(1 + {c_\theta}^{2}\right)\,\sin\left(6\,\psi \right)\,\left[ - {\displaystyle \frac {729}{5}} \,\ln\left(3\right) + {\displaystyle \frac {7776729}{40040}} \right] \cr
&&
 + \pi \,{s_\theta}^{5}\,\sin\left(7\,\psi \right)\,\left[ - {\displaystyle \frac {257768959}{1474560}}  - {\displaystyle \frac {10706059}{92160}} \,{c_\theta}^{2} + {\displaystyle \frac {40353607}{1474560}} \,{c_\theta}^{4}\right] \cr
&&
 - {\displaystyle \frac {43046721}{1146880}} \,\pi \,{s_\theta}^{7}\,\left[1 + {c_\theta}^{2}\right]\,\sin\left(9\,\psi \right) ,
\end{eqnarray}
\begin{eqnarray}
{H_{+}^{(5.5)}}&=&
\pi \,{s_\theta}\,\cos\left(\psi \right)\left[{\displaystyle \frac {1561379273}{47048601600}}  - {\displaystyle \frac {1771}{2560}} \,\ln\left(2\right) \right. + \left({\displaystyle \frac {370608237}{1742540800}}  + {\displaystyle \frac {1667}{2560}} \,\ln\left(2\right)\right)\,{c_\theta}^{2}\cr
&&
 + \left( - {\displaystyle \frac {1475977183}{28229160960}}  - {\displaystyle \frac {217}{4608}} \,\ln\left(2\right)\right)\,{c_\theta}^{4} \left. + \left({\displaystyle \frac {840821}{2566287360}}  + {\displaystyle \frac {1}{4608}} \,\ln\left(2\right)\right)\,{c_\theta}^{6}\right]\cr
&&
 + \pi \,\cos\left(2\,\psi \right)\left[ - {\displaystyle \frac {99368}{2079}} \,{\rm eulerlog}\left(2, \,v\right) \right.
 - {\displaystyle \frac {38}{9}} \,\pi ^{2} + {\displaystyle \frac {80210663899}{461039040}}  \cr
&&
 + \left( - {\displaystyle \frac {105352}{3465}} \,{\rm eulerlog}\left(2, \,v\right) + {\displaystyle \frac {46804865561}{768398400}}  - 2\,\pi ^{2}\right)\,{c_\theta}^{2} \cr
&&
 + \left( - {\displaystyle \frac {17442960719}{1152597600}}  + {\displaystyle \frac {25136}{10395}} \,{\rm eulerlog}\left(2, \,v\right) + {\displaystyle \frac {4}{9}} \,\pi ^{2}\right)\,{c_\theta}^{4} \cr
&&
\left. - {\displaystyle \frac {1028339}{997920}} \,{c_\theta}^{6} + {\displaystyle \frac {61}{1440}} \,{c_\theta}^{8} - {\displaystyle \frac {1}{4320}} \,{c_\theta}^{10}\right] \cr
&&
+ \pi \,{s_\theta}\,\cos\left(3\,\psi \right)\left[{\displaystyle \frac {71755246683}{1742540800}}  - {\displaystyle \frac {31833}{512}} \,\ln\left({\displaystyle \frac {3}{2}} \right) \right.
 + \left({\displaystyle \frac {206793}{2560}} \,\ln\left({\displaystyle \frac {3}{2}} \right) - {\displaystyle \frac {125895720231}{1742540800}} \right)\,{c_\theta}^{2} \cr
&&
\left. + \left({\displaystyle \frac {137781}{2560}} \,\ln\left({\displaystyle \frac {3}{2}} \right) - {\displaystyle \frac {102720365379}{1742540800}} \right)\,{c_\theta}^{4} + \left( - {\displaystyle \frac {6561}{2560}} \,\ln\left({\displaystyle \frac {3}{2}} \right) + {\displaystyle \frac {612958509}{158412800}} \right)\,{c_\theta}^{6} \right]\cr
&&
 + \pi \,{s_\theta}^{2}\,\cos\left(4\,\psi \right)\left[{\displaystyle \frac {18906768449}{36018675}}  + {\displaystyle \frac {2921216}{10395}} \,\ln\left(2\right) - {\displaystyle \frac {512}{3}} \,\ln\left(2\right)^{2} - {\displaystyle \frac {128}{9}} \,\pi ^{2} \right. - {\displaystyle \frac {804352}{10395}} \,{\rm eulerlog}\left(4, \,v\right)\cr
&&
 + \left( - {\displaystyle \frac {512}{3}} \,{\rm ln}\left(2\right)^{2} - {\displaystyle \frac {128}{9}} \,\pi ^{2} + {\displaystyle \frac {21372545609}{36018675}}  \right.
\left. + {\displaystyle \frac {2921216}{10395}} \,\ln\left(2\right) - {\displaystyle \frac {804352}{10395}} \,{\rm eulerlog}\left(4, \,v\right)\right){c_\theta}^{2} \cr
&&
\left. + {\displaystyle \frac {1281832}{31185}} \,{c_\theta}^{4} - {\displaystyle \frac {10592}{945}} \,{c_\theta}^{6} + {\displaystyle \frac {256}{945}} \,{c_\theta}^{8}\right]\cr
&&
 + \pi \,{s_\theta}^{3}\,\cos\left(5\,\psi \right)\left[{\displaystyle \frac {3731798959375}{5645832192}} \right. 
 - {\displaystyle \frac {2703125}{4608}} \,\ln\left({\displaystyle \frac {5}{2}} \right) + \left( - {\displaystyle \frac {46875}{128}} \,\ln\left({\displaystyle \frac {5}{2}} \right) + {\displaystyle \frac {173466753125}{470486016}} \right)\,{c_\theta}^{2} \cr
&&
\left. + \left( - {\displaystyle \frac {65689140625}{513257472}}  + {\displaystyle \frac {390625}{4608}} \,\ln\left({\displaystyle \frac {5}{2}} \right)\right)\,{c_\theta}^{4}\right] \cr
&&
 + \pi \,{s_\theta}^{4}\,\cos\left(6\,\psi \right)\,\left[ - {\displaystyle \frac {6580683}{24640}}  + {\displaystyle \frac {459999}{24640}} \,{c_\theta}^{2} + {\displaystyle \frac {282123}{2240}} \,{c_\theta}^{4} - {\displaystyle \frac {19683}{2240}} \,{c_\theta}^{6}\right] \cr
&&
 + \pi \,{s_\theta}^{5}\,\cos\left(7\,\psi \right)\,\left(1 + {c_\theta}^{2}\right)\,\left[{\displaystyle \frac {692452248803}{1833062400}}  - {\displaystyle \frac {5764801}{23040}} \,{\rm ln}\left({\displaystyle \frac {7}{2}} \right)\right] \cr
&&
 + \pi \,{s_\theta}^{6}\,\cos\left(8\,\psi \right)\,\left[ - {\displaystyle \frac {815104}{2835}}  - {\displaystyle \frac {4096}{21}} \,{c_\theta}^{2} + {\displaystyle \frac {131072}{2835}} \,{c_\theta}^{4}\right] \cr
&&
 - {\displaystyle \frac {1953125}{36288}} \,\pi \,{s_\theta}^{8}\,\cos\left(10\,\psi \right)\,\left[1 + {c_\theta}^{2}\right]\cr
&&
 + {s_\theta}\,\sin\left(\psi \right)\left[{\displaystyle \frac {5290289}{9676800}} \,\ln\left(2\right) \right.
 + {\displaystyle \frac {61540628400739667719}{9223212135481344000}}  + {\displaystyle \frac {1771}{2560}} \,\ln\left(2\right)^{2} \cr
&&
 + {\displaystyle \frac {257536997}{420076800}} \,{\rm eulerlog}\left(1, \,v\right) - {\displaystyle \frac {1771}{30720}} \,\pi ^{2}\cr
&&
 + \left({\displaystyle \frac {1667}{30720}} \,\pi ^{2} + {\displaystyle \frac {376667450178313183}{144112689616896000}}  \right.
\left. - {\displaystyle \frac {1231221763}{2940537600}} \,{\rm eulerlog}\left(1, \,v\right) - {\displaystyle \frac {8167913}{9676800}} \,\ln\left(2\right) - {\displaystyle \frac {1667}{2560}} \,\ln\left(2\right)^{2}\right){c_\theta}^{2}\cr
&&
 + \left({\displaystyle \frac {141173}{1161216}} \,\ln\left(2\right) + {\displaystyle \frac {9999193}{588107520}} \,{\rm eulerlog}\left(1, \,v\right) - {\displaystyle \frac {74897190722604377}{1844642427096268800}}  \right.
\left. - {\displaystyle \frac {217}{55296}} \,\pi ^{2} + {\displaystyle \frac {217}{4608}} \,\ln\left(2\right)^{2}\right){c_\theta}^{4}\cr
&&
 + \left( - {\displaystyle \frac {2987}{53464320}} \,{\rm eulerlog}\left(1, \,v\right)\right. 
\left. - {\displaystyle \frac {57193270974289}{20961845762457600}}  - {\displaystyle \frac {1}{4608}} \,\ln\left(2\right)^{2} - {\displaystyle \frac {4129}{5806080}} \,\ln\left(2\right) + {\displaystyle \frac {1}{55296}} \,\pi ^{2}\right){c_\theta}^{6} \cr
&&
\left. + {\displaystyle \frac {118483271}{2779486617600}} \,{c_\theta}^{8} - {\displaystyle \frac {221}{1857945600}} \,{c_\theta}^{10} + {\displaystyle \frac {1}{14863564800}} \,{c_\theta}^{12}\right] \cr
&&
 + \sin\left(2\,\psi \right)\left[ - {\displaystyle \frac {608}{9}} \,\zeta \left(3\right) + {\displaystyle \frac {159032}{17325}} \,{\rm eulerlog}\left(2, \,v\right) + {\displaystyle \frac {1204678}{31185}} \,\pi ^{2} \right.
 - {\displaystyle \frac {249409850407}{3841992000}}  \cr
&&
 + \left( - {\displaystyle \frac {12752}{825}} \,{\rm eulerlog}\left(2, \,v\right) + {\displaystyle \frac {54465777787}{640332000}}  - 32\,\zeta \left(3\right) + {\displaystyle \frac {282784}{10395}} \,\pi ^{2}\right)\,{c_\theta}^{2} \cr
&&
 + \left( - {\displaystyle \frac {12568}{2475}} \,{\rm eulerlog}\left(2, \,v\right) - {\displaystyle \frac {33734}{31185}} \,\pi ^{2} + {\displaystyle \frac {742766124271}{23051952000}}  + {\displaystyle \frac {64}{9}} \,\zeta \left(3\right)\right)\,{c_\theta}^{4} \cr
&&
\left. + {\displaystyle \frac {1280598643}{419126400}} \,{c_\theta}^{6} - {\displaystyle \frac {2997143}{19958400}} \,{c_\theta}^{8} + {\displaystyle \frac {18421}{19958400}} \,{c_\theta}^{10}\right] \cr
&&
+ {s_\theta}\,\sin\left(3\,\psi \right)\left[ 
 - {\displaystyle \frac {3511527147}{108908800}} \,{\rm eulerlog}\left(3, \,v\right) - {\displaystyle \frac {31833}{512}} \,{\rm ln}\left({\displaystyle \frac {3}{2}} \right)^{2} + {\displaystyle \frac {10611}{2048}} \,\pi ^{2} + {\displaystyle \frac {41072589}{358400}} \,\ln\left({\displaystyle \frac {3}{2}} \right)\right. \cr
&&
 + {\displaystyle \frac {3514307823600697989}{151822421983232000}}  + \left( - {\displaystyle \frac {68931}{10240}} \,\pi ^{2} - {\displaystyle \frac {9723416976987988977}{37955605495808000}}  \right.\cr
&&
\left. + {\displaystyle \frac {206793}{2560}} \,\ln\left({\displaystyle \frac {3}{2}} \right)^{2} - {\displaystyle \frac {1837323}{10240}} \,\ln\left({\displaystyle \frac {3}{2}} \right) + {\displaystyle \frac {3804113403}{108908800}} \,{\rm eulerlog}\left(3, \,v\right)\right){c_\theta}^{2}\cr
&&
 + \left( - {\displaystyle \frac {10835432671991490567}{151822421983232000}}  + {\displaystyle \frac {137781}{2560}} \,\ln\left({\displaystyle \frac {3}{2}} \right)^{2} + {\displaystyle \frac {2141460099}{108908800}} \,{\rm eulerlog}\left(3, \,v\right) \right.\cr
&&
\left. - {\displaystyle \frac {45927}{10240}} \,\pi ^{2} - {\displaystyle \frac {49301541}{358400}} \,\ln\left({\displaystyle \frac {3}{2}} \right)\right){c_\theta}^{4}\cr
&&
 + \left({\displaystyle \frac {3010041}{358400}} \,\ln\left({\displaystyle \frac {3}{2}} \right) + {\displaystyle \frac {2187}{10240}} \,\pi ^{2} \right.
\left. - {\displaystyle \frac {1088541343451439}{431313698816000}}  - {\displaystyle \frac {6532569}{9900800}} \,{\rm eulerlog}\left(3, \,v\right) - {\displaystyle \frac {6561}{2560}} \,\ln\left({\displaystyle \frac {3}{2}} \right)^{2}\right){c_\theta}^{6} \cr
&&
\left. + {\displaystyle \frac {77975093151}{137258598400}} \,{c_\theta}^{8} - {\displaystyle \frac {2814669}{183500800}} \,{c_\theta}^{10} + {\displaystyle \frac {59049}{734003200}} \,{c_\theta}^{12}\right]\cr
&&
 + {s_\theta}^{2} \sin\left(4\,\psi \right)\left[{\displaystyle \frac {256}{9}} \,{\rm ln}\left(2\right)\,\pi ^{2} - {\displaystyle \frac {1024}{9}} \,\ln\left(2\right)^{3} - {\displaystyle \frac {134057079091}{120062250}}  + {\displaystyle \frac {6392133914}{7203735}} \,\ln\left(2\right) \right.\cr
&&
 + {\displaystyle \frac {402176}{2475}} \,{\rm eulerlog}\left(4, \,v\right) - {\displaystyle \frac {2048}{9}} \,\zeta \left(3\right) + {\displaystyle \frac {1792}{5}} \,\ln\left(2\right)^{2} - {\displaystyle \frac {1608704}{10395}} \,\ln\left(2\right)\,{\rm eulerlog}\left(4, \,v\right) + {\displaystyle \frac {1079488}{31185}} \,\pi ^{2} \cr
&&
 + \left( - {\displaystyle \frac {17336400559}{13340250}} \right.
 + {\displaystyle \frac {1079488}{31185}} \,\pi ^{2} + {\displaystyle \frac {402176}{2475}} \,{\rm eulerlog}\left(4, \,v\right) - {\displaystyle \frac {2048}{9}} \,\zeta \left(3\right) - {\displaystyle \frac {1024}{9}} \,\ln\left(2\right)^{3} \cr
&&
 + {\displaystyle \frac {1792}{5}} \,\ln\left(2\right)^{2} + {\displaystyle \frac {256}{9}} \,\ln\left(2\right)\,\pi ^{2} + {\displaystyle \frac {7378444778}{7203735}} \,\ln\left(2\right) 
\left. - {\displaystyle \frac {1608704}{10395}} \,\ln\left(2\right)\,{\rm eulerlog}\left(4, \,v\right)\right){c_\theta}^{2}\cr
&&
 + \left( - {\displaystyle \frac {34988014}{297675}}  + {\displaystyle \frac {2563664}{31185}} \,\ln\left(2\right)\right)\,{c_\theta}^{4} + \left({\displaystyle \frac {14435216}{363825}}  - {\displaystyle \frac {21184}{945}} \,\ln\left(2\right)\right)\,{c_\theta}^{6}\cr
&&
\left.  + \left({\displaystyle \frac {512}{945}} \,\ln\left(2\right) - {\displaystyle \frac {1178944}{1091475}} \right)\,{c_\theta}^{8}\right] \cr
&&
 + {s_\theta}^{3} \sin\left(5\,\psi \right)\left[ - {\displaystyle \frac {3514496875}{16803072}} \,{\rm eulerlog}\left(5, \,v\right) + {\displaystyle \frac {1240914560526518125}{1561602054684672}}  \right.\cr
&&
 + {\displaystyle \frac {2703125}{55296}} \,\pi ^{2} + {\displaystyle \frac {1777965625}{1161216}} \,\ln\left({\displaystyle \frac {5}{2}} \right) - {\displaystyle \frac {2703125}{4608}} \,\ln\left({\displaystyle \frac {5}{2}} \right)^{2}\cr
&&
 + \left({\displaystyle \frac {85146875}{96768}} \,\ln\left({\displaystyle \frac {5}{2}} \right) \right.
 - {\displaystyle \frac {46875}{128}} \,\ln\left({\displaystyle \frac {5}{2}} \right)^{2} - {\displaystyle \frac {349221875}{2450448}} \,{\rm eulerlog}\left(5, \,v\right)  + {\displaystyle \frac {60761394072551429375}{98380929445134336}} \cr
&&
\left. 
+ {\displaystyle \frac {15625}{512}} \,\pi ^{2}\right){c_\theta}^{2}\cr
&&
 + \left( - {\displaystyle \frac {390625}{55296}} \,\pi ^{2} - {\displaystyle \frac {322578125}{1161216}} \,\ln\left({\displaystyle \frac {5}{2}} \right) \right.
 + {\displaystyle \frac {233359375}{10692864}} \,{\rm eulerlog}\left(5, \,v\right) - {\displaystyle \frac {202353378607890625}{13415581287972864}} \cr
&&
\left.+ {\displaystyle \frac {390625}{4608}} \,\ln\left({\displaystyle \frac {5}{2}} \right)^{2}\right) 
\left.{c_\theta}^{4} - {\displaystyle \frac {15781558984375}{667076788224}} \,{c_\theta}^{6} + {\displaystyle \frac {5166015625}{2378170368}} \,{c_\theta}^{8} - {\displaystyle \frac {244140625}{7134511104}} \,{c_\theta}^{10}\right] \cr
&&
 + {s_\theta}^{4}\,\sin\left(6\,\psi \right)\left[ - {\displaystyle \frac {6580683}{12320}} \,\ln\left(3\right) + {\displaystyle \frac {2775122451}{3449600}}  \right.
 + \left( - {\displaystyle \frac {68404257}{492800}}  + {\displaystyle \frac {459999}{12320}} \,\ln\left(3\right)\right)\,{c_\theta}^{2}\cr
&&
 + \left({\displaystyle \frac {282123}{1120}} \,\ln\left(3\right) - {\displaystyle \frac {1528530021}{3449600}} \right)\,{c_\theta}^{4} 
\left. + \left( - {\displaystyle \frac {19683}{1120}} \,\ln\left(3\right) + {\displaystyle \frac {120860181}{3449600}} \right)\,{c_\theta}^{6}\right]\cr
&&
 + {s_\theta}^{5}\,\sin\left(7\,\psi \right)\left[ - {\displaystyle \frac {210540970605284389}{11407807217664000}}  + {\displaystyle \frac {5764801}{276480}} \,\pi ^{2} - {\displaystyle \frac {2459922941}{38188800}} \,{\rm eulerlog}\left(7, \,v\right) \right.\cr
&&
 + {\displaystyle \frac {3400409047}{4147200}} \,{\rm ln}\left({\displaystyle \frac {7}{2}} \right) - {\displaystyle \frac {5764801}{23040}} \,\ln\left({\displaystyle \frac {7}{2}} \right)^{2} \cr
&&
+ \left( - {\displaystyle \frac {5764801}{23040}} \,\ln\left({\displaystyle \frac {7}{2}} \right)^{2} \right.
 + {\displaystyle \frac {3400409047}{4147200}} \,{\rm ln}\left({\displaystyle \frac {7}{2}} \right) - {\displaystyle \frac {2459922941}{38188800}} \,{\rm eulerlog}\left(7, \,v\right) \cr
&&
\left. + {\displaystyle \frac {1312380850473448903}{5703903608832000}}  + {\displaystyle \frac {5764801}{276480}} \,\pi ^{2}\right){c_\theta}^{2} \cr
&&
 + {\displaystyle \frac {31513175135281}{297802137600}} \,{c_\theta}^{4}
\left. - {\displaystyle \frac {191800694071}{6370099200}} \,{c_\theta}^{6} + {\displaystyle \frac {13841287201}{12740198400}} \,{c_\theta}^{8}\right]\cr
&&
 + {s_\theta}^{6}\,\sin\left(8\,\psi \right) \left[ - {\displaystyle \frac {3260416}{2835}} \,\ln\left(2\right) + {\displaystyle \frac {133789696}{130977}}  + \left( - {\displaystyle \frac {16384}{21}} \,\ln\left(2\right) + {\displaystyle \frac {146262016}{218295}} \right)\,{c_\theta}^{2} \right.\cr
&&
\left. + \left({\displaystyle \frac {524288}{2835}} \,\ln\left(2\right) - {\displaystyle \frac {603619328}{3274425}} \right)\,{c_\theta}^{4}\right]\cr
&&
 + {s_\theta}^{7}\,\sin\left(9\,\psi \right)\left[ - {\displaystyle \frac {15729085317609}{68629299200}} - {\displaystyle \frac {232036175097}{6239027200}} \,{c_\theta}^{2} + {\displaystyle \frac {77871518289}{807403520}} \,{c_\theta}^{4} - {\displaystyle \frac {31381059609}{4037017600}} \,{c_\theta}^{6}\right] \cr
&&
 + {s_\theta}^{8}\,\left(1 + {c_\theta}^{2}\right)\,\sin\left(10\,\psi \right)\,\left[{\displaystyle \frac {7195703125}{33530112}}  - {\displaystyle \frac {1953125}{18144}} \,\ln\left(5\right)\right]\cr
&&
 + {s_\theta}^{9} \sin\left(11\,\psi \right)\left[ - {\displaystyle \frac {18493383740513}{178362777600}}  - {\displaystyle \frac {25937424601}{353894400}} \,{c_\theta}^{2} + {\displaystyle \frac {3138428376721}{178362777600}} \,{c_\theta}^{4}\right] \cr
&&
 - {\displaystyle \frac {23298085122481}{1961990553600}} \,{s_\theta}^{11}\,\left[1 + {c_\theta}^{2}\right]\,\sin\left(13\,\psi \right) .
\end{eqnarray}
\end{subequations}
\end{widetext}
\subsection{Cross modes}
\label{sec:55PNpol_cross}
\begin{widetext}
\begin{subequations}
\begin{eqnarray}
{H_{\times}^{(0)}}&=&
 - 2\,{c_\theta}\,\sin\left(2\,\psi \right), 
\end{eqnarray}
\begin{eqnarray}
{H_{\times}^{(0.5)}}&=&
 - {\displaystyle \frac {3}{4}} \,\cos\left(\psi \right)\,{s_\theta}\,{c_\theta} - {\displaystyle \frac {9}{4}} \,{s_\theta}\,{c_\theta}\,\cos\left(3\,\psi \right),
\end{eqnarray}
\begin{eqnarray}
{H_{\times}^{(1)}}&=&
{c_\theta}\,\sin\left(2\,\psi \right)\,\left({\displaystyle \frac {17}{3}}  - {\displaystyle \frac {4}{3}} \,{c_\theta}^{2}\right) + {\displaystyle \frac {8}{3}} \,{s_\theta}^{2}\,{c_\theta}\,\sin\left(4\,\psi \right),
\end{eqnarray}
\begin{eqnarray}
{H_{\times}^{(1.5)}}&=&
{s_\theta}\,{c_\theta}\,\cos\left(\psi \right)\,\left({\displaystyle \frac {21}{32}}  - {\displaystyle \frac {5}{96}} \,{c_\theta}^{2}\right) + {s_\theta}\,{c_\theta}\,\cos\left(3\,\psi \right)\,\left({\displaystyle \frac {603}{64}}  - {\displaystyle \frac {135}{64}} \,{c_\theta}^{2}\right) \cr
&&
 + {\displaystyle \frac {625}{192}} \,{s_\theta}^{3}\,{c_\theta}\,\cos\left(5\,\psi \right) - 4\,\pi \,{c_\theta}\,\sin\left(2\,\psi \right) ,
\end{eqnarray}
\begin{eqnarray}
{H_{\times}^{(2)}}&=&
 - {\displaystyle \frac {3}{4}}\,\pi \,{c_\theta}\,{s_\theta}\,\cos\left(\psi \right)  - {\displaystyle \frac {27}{4}} \,\pi \,{s_\theta}\,{c_\theta}\,\cos\left(3\,\psi \right) \cr
&&
 + {s_\theta}\,{c_\theta}\,\sin\left(\psi \right)\,\left({\displaystyle \frac {3}{2}} \,\ln\left(2\right) + {\displaystyle \frac {9}{20}} \right) + {c_\theta}\,\sin\left(2\,\psi \right)\,\left({\displaystyle \frac {17}{15}}  + {\displaystyle \frac {113}{30}} \,{c_\theta}^{2} - {\displaystyle \frac {1}{4}} \,{c_\theta}^{4}\right) \cr
&&
 + {s_\theta}\,{c_\theta}\,\sin\left(3\,\psi \right)\,\left( - {\displaystyle \frac {27}{2}} \,\ln\left({\displaystyle \frac {3}{2}} \right) + {\displaystyle \frac {189}{20}} \right) + {s_\theta}^{2}\,{c_\theta}\,\sin\left(4\,\psi \right)\,\left( - {\displaystyle \frac {44}{3}}  + {\displaystyle \frac {16}{5}} \,{c_\theta}^{2}\right) \cr
&&
 - {\displaystyle \frac {81}{20}} \,{s_\theta}^{4}\,{c_\theta}\,\sin\left(6\,\psi \right) ,
\end{eqnarray}
\begin{eqnarray}
{H_{\times}^{(2.5)}}&=&
{s_\theta}\,{c_\theta}\,\cos\left(\psi \right)\,\left( - {\displaystyle \frac {913}{7680}}  + {\displaystyle \frac {1891}{11520}} \,{c_\theta}^{2} - {\displaystyle \frac {7}{4608}} \,{c_\theta}^{4}\right) \cr
&&
 + {c_\theta}\,\cos\left(2\,\psi \right)\,\left(2 - {\displaystyle \frac {22}{5}} \,{c_\theta}^{2}\right) \cr
&&
 + {s_\theta}\,{c_\theta}\,\cos\left(3\,\psi \right)\,\left( - {\displaystyle \frac {12501}{2560}}  + {\displaystyle \frac {12069}{1280}} \,{c_\theta}^{2} - {\displaystyle \frac {1701}{2560}} \,{c_\theta}^{4}\right) \cr
&&
 + {s_\theta}^{2}\,{c_\theta}\,\cos\left(4\,\psi \right)\,\left({\displaystyle \frac {112}{5}}  - {\displaystyle \frac {64}{3}} \,\ln\left(2\right)\right) \cr
&&
 + {s_\theta}^{3}\,{c_\theta}\,\cos\left(5\,\psi \right)\,\left( - {\displaystyle \frac {101875}{4608}}  + {\displaystyle \frac {21875}{4608}} \,{c_\theta}^{2}\right) - {\displaystyle \frac {117649}{23040}} \,{s_\theta}^{5}\,{c_\theta}\,\cos\left(7\,\psi \right) \cr
&&
 + \pi \,{c_\theta}\,\sin\left(2\,\psi \right)\,\left({\displaystyle \frac {34}{3}}  - {\displaystyle \frac {8}{3}} \,{c_\theta}^{2}\right) + {\displaystyle \frac {32}{3}} \,\pi \,{s_\theta}^{2}\,{c_\theta}\,\sin\left(4\,\psi \right) ,
\end{eqnarray}
\begin{eqnarray}
{H_{\times}^{(3)}}&=&
\pi \,{s_\theta}\,{c_\theta}\,\cos\left(\psi \right)\,\left({\displaystyle \frac {21}{32}}  - {\displaystyle \frac {5}{96}} \,{c_\theta}^{2}\right) + {\displaystyle \frac {856}{105}} \,\pi \,{c_\theta}\,\cos\left(2\,\psi \right) \cr
&&
 + \pi \,{s_\theta}\,{c_\theta}\,\cos\left(3\,\psi \right)\,\left({\displaystyle \frac {1809}{64}}  - {\displaystyle \frac {405}{64}} \,{c_\theta}^{2}\right) + {\displaystyle \frac {3125}{192}} \,\pi \,{s_\theta}^{3}\,{c_\theta}\,\cos\left(5\,\psi \right) \cr
&&
 + {s_\theta}\,{c_\theta}\,\sin\left(\psi \right)\,\left( - {\displaystyle \frac {21}{16}} \,\ln\left(2\right) - {\displaystyle \frac {11617}{20160}}  + \left({\displaystyle \frac {5}{48}} \,\ln\left(2\right) + {\displaystyle \frac {251}{2240}} \right)\,{c_\theta}^{2}\right)\cr
&&
 + {c_\theta} \sin\left(2\,\psi \right) 
\left( - {\displaystyle \frac {3620761}{44100}}  + {\displaystyle \frac {1712}{105}} \,{\rm eulerlog}\left(2, \,v\right) - {\displaystyle \frac {4}{3}} \,\pi ^{2} - {\displaystyle \frac {3413}{1260}} \,{c_\theta}^{2} + {\displaystyle \frac {2909}{2520}} \,{c_\theta}^{4} - {\displaystyle \frac {1}{45}} \,{c_\theta}^{6}\right) \cr
&&
 + {s_\theta}\,{c_\theta}\,\sin\left(3\,\psi \right)\,\left( - {\displaystyle \frac {36801}{896}}  + {\displaystyle \frac {1809}{32}} \,\ln\left({\displaystyle \frac {3}{2}} \right) + \left( - {\displaystyle \frac {405}{32}} \,\ln\left({\displaystyle \frac {3}{2}} \right) + {\displaystyle \frac {65097}{4480}} \right)\,{c_\theta}^{2}\right) \cr
&&
 + {s_\theta}^{2}\,{c_\theta}\,\sin\left(4\,\psi \right)\,\left({\displaystyle \frac {1781}{105}}  - {\displaystyle \frac {1208}{63}} \,{c_\theta}^{2} + {\displaystyle \frac {64}{45}} \,{c_\theta}^{4}\right) \cr
&&
 + {s_\theta}^{3}\,{c_\theta}\,\sin\left(5\,\psi \right)\,\left({\displaystyle \frac {3125}{96}} \,\ln\left({\displaystyle \frac {5}{2}} \right) - {\displaystyle \frac {113125}{2688}} \right) \cr
&&
 + {s_\theta}^{4}\,{c_\theta}\,\sin\left(6\,\psi \right)\,\left({\displaystyle \frac {9153}{280}}  - {\displaystyle \frac {243}{35}} \,{c_\theta}^{2}\right) + {\displaystyle \frac {2048}{315}} \,{s_\theta}^{6}\,{c_\theta}\,\sin\left(8\,\psi \right) ,
\end{eqnarray}
\begin{eqnarray}
{H_{\times}^{(3.5)}}&=&
{s_\theta}\,{c_\theta}\,\cos\left(\psi \right)\left[{\displaystyle \frac {307}{210}} \,{\rm eulerlog}\left(1, \,v\right) + {\displaystyle \frac {9}{10}} \,\ln\left(2\right) - {\displaystyle \frac {1}{8}} \,\pi ^{2} + {\displaystyle \frac {3}{2}} \,\ln\left(2\right)^{2} \right.\cr
&&
\left. - {\displaystyle \frac {2025831067}{270950400}}  - {\displaystyle \frac {436553}{2580480}} \,{c_\theta}^{2} + {\displaystyle \frac {2291}{286720}} \,{c_\theta}^{4} - {\displaystyle \frac {1}{40960}} \,{c_\theta}^{6}\right] \cr
&&
 + {c_\theta}\,\cos\left(2\,\psi \right)\,\left[ - {\displaystyle \frac {1831}{280}}  + {\displaystyle \frac {3419}{252}} \,{c_\theta}^{2} - {\displaystyle \frac {1109}{840}} \,{c_\theta}^{4}\right]\cr
&&
 + {s_\theta}\,{c_\theta}\,\cos\left(3\,\psi \right)\left[ - {\displaystyle \frac {75787641}{501760}}  + {\displaystyle \frac {351}{14}} \,{\rm eulerlog}\left(3, \,v\right) - {\displaystyle \frac {567}{10}} \,\ln\left({\displaystyle \frac {3}{2}} \right) + {\displaystyle \frac {81}{2}} \,\ln\left({\displaystyle \frac {3}{2}} \right)^{2} - {\displaystyle \frac {27}{8}} \,\pi ^{2} \right.\cr
&&
\left. - {\displaystyle \frac {1794069}{143360}} \,{c_\theta}^{2} + {\displaystyle \frac {7209}{1792}} \,{c_\theta}^{4} - {\displaystyle \frac {2187}{20480}} \,{c_\theta}^{6}\right] \cr
&&
 + {s_\theta}^{2}\,{c_\theta}\,\cos\left(4\,\psi \right)\,\left[{\displaystyle \frac {352}{3}} \,\ln\left(2\right) - {\displaystyle \frac {13096}{105}}  + \left( - {\displaystyle \frac {128}{5}} \,\ln\left(2\right) + {\displaystyle \frac {3712}{105}} \right)\,{c_\theta}^{2}\right] \cr
&&
 + {s_\theta}^{3}\,{c_\theta}\,\cos\left(5\,\psi \right)\,\left[{\displaystyle \frac {90625}{2304}}  - {\displaystyle \frac {9115625}{258048}} \,{c_\theta}^{2} + {\displaystyle \frac {78125}{28672}} \,{c_\theta}^{4}\right] \cr
&&
 + {s_\theta}^{4}\,{c_\theta}\,\cos\left(6\,\psi \right)\,\left[ - {\displaystyle \frac {20169}{280}}  + {\displaystyle \frac {243}{5}} \,\ln\left(3\right)\right] 
 + {s_\theta}^{5}\,{c_\theta}\,\cos\left(7\,\psi \right)\,\left[{\displaystyle \frac {35177051}{737280}}  - {\displaystyle \frac {823543}{81920}} \,{c_\theta}^{2}\right]\cr
&&
 + {\displaystyle \frac {4782969}{573440}} \,{s_\theta}^{7}\,{c_\theta}\,\cos\left(9\,\psi \right) \cr
&&
 + \pi \,{s_\theta}\,{c_\theta}\,\sin\left(\psi \right)\,\left[ - {\displaystyle \frac {59}{210}}  + {\displaystyle \frac {3}{2}} \,\ln\left(2\right)\right]
 + \pi \,{c_\theta}\,\sin\left(2\,\psi \right)\,\left[{\displaystyle \frac {34}{15}}  + {\displaystyle \frac {113}{15}} \,{c_\theta}^{2} - {\displaystyle \frac {1}{2}} \,{c_\theta}^{4}\right] \cr
&&
 + \pi \,{s_\theta}\,{c_\theta}\,\sin\left(3\,\psi \right)\,\left[ - {\displaystyle \frac {81}{2}} \,\ln\left({\displaystyle \frac {3}{2}} \right) + {\displaystyle \frac {1107}{70}} \right] 
 + \pi \,{s_\theta}^{2}\,{c_\theta}\,\sin\left(4\,\psi \right)\,\left[ - {\displaystyle \frac {176}{3}}  + {\displaystyle \frac {64}{5}} \,{c_\theta}^{2}\right]\cr
&&
 - {\displaystyle \frac {243}{10}} \,\pi \,{s_\theta}^{4}\,{c_\theta}\,\sin\left(6\,\psi \right) ,
\end{eqnarray}
\begin{eqnarray}
{H_{\times}^{(4)}}&=&
\pi \,{s_\theta}\,{c_\theta}\,\cos\left(\psi \right)\,\left[ - {\displaystyle \frac {913}{7680}}  + {\displaystyle \frac {1891}{11520}} \,{c_\theta}^{2} - {\displaystyle \frac {7}{4608}} \,{c_\theta}^{4}\right] 
 + \pi \,{c_\theta}\,\cos\left(2\,\psi \right)\,\left[ - {\displaystyle \frac {187384}{10395}}  - {\displaystyle \frac {59452}{10395}} \,{c_\theta}^{2}\right] \cr
&&
 + \pi \,{s_\theta}\,{c_\theta}\,\cos\left(3\,\psi \right)\,\left[ - {\displaystyle \frac {37503}{2560}}  + {\displaystyle \frac {36207}{1280}} \,{c_\theta}^{2} - {\displaystyle \frac {5103}{2560}} \,{c_\theta}^{4}\right] \cr
&&
 + \pi \,{s_\theta}^{2}\,{c_\theta}\,\cos\left(4\,\psi \right)\,\left[ - {\displaystyle \frac {256}{3}} \,\ln\left(2\right) + {\displaystyle \frac {730304}{10395}} \right]
 + \pi \,{s_\theta}^{3}\,{c_\theta}\,\cos\left(5\,\psi \right)\,\left[ - {\displaystyle \frac {509375}{4608}}  + {\displaystyle \frac {109375}{4608}} \,{c_\theta}^{2}\right] \cr
&&
 - {\displaystyle \frac {823543}{23040}} \,\pi \,{s_\theta}^{5}\,{c_\theta}\,\cos\left(7\,\psi \right)\cr
&&
 + {s_\theta}\,{c_\theta}\,\sin\left(\psi \right)\left[{\displaystyle \frac {913}{3840}} \,\ln\left(2\right) + {\displaystyle \frac {4253489}{9676800}} + \left( - {\displaystyle \frac {5164843}{14515200}}  - {\displaystyle \frac {1891}{5760}} \,\ln\left(2\right)\right)\,{c_\theta}^{2}  \right.\cr
&&
\left. + \left({\displaystyle \frac {7}{2304}} \,\ln\left(2\right) + {\displaystyle \frac {26263}{5806080}} \right)\,{c_\theta}^{4}\right]\cr
&&
 + {c_\theta} \sin\left(2\,\psi \right)\left[{\displaystyle \frac {337599069919}{2305195200}}  - {\displaystyle \frac {457928}{10395}} \,{\rm eulerlog}\left(2, \,v\right) + {\displaystyle \frac {34}{9}} \,\pi ^{2} \right.\cr
&&
 + \left({\displaystyle \frac {64048}{10395}} \,{\rm eulerlog}\left(2, \,v\right) - {\displaystyle \frac {8}{9}} \,\pi ^{2} - {\displaystyle \frac {935502853}{28814940}} \right)\,{c_\theta}^{2} - {\displaystyle \frac {634357}{332640}} \,{c_\theta}^{4} \left. + {\displaystyle \frac {6611}{45360}} \,{c_\theta}^{6} - {\displaystyle \frac {1}{864}} \,{c_\theta}^{8}\right] \cr
&&
+ {s_\theta}\,{c_\theta}\,\sin\left(3\,\psi \right)\left[{\displaystyle \frac {10428183}{358400}}  - {\displaystyle \frac {37503}{1280}} \,\ln\left({\displaystyle \frac {3}{2}} \right) + \left({\displaystyle \frac {36207}{640}} \,\ln\left({\displaystyle \frac {3}{2}} \right) - {\displaystyle \frac {12089331}{179200}} \right)\,{c_\theta}^{2} \right.\cr
&&
\left. 
+ \left({\displaystyle \frac {2162943}{358400}}  - {\displaystyle \frac {5103}{1280}} \,\ln\left({\displaystyle \frac {3}{2}} \right)\right)\,{c_\theta}^{4}\right]\cr
&&
 + {s_\theta}^{2} {c_\theta}\,\sin\left(4\,\psi \right)\left[{\displaystyle \frac {64}{9}} \,\pi ^{2} - {\displaystyle \frac {402176}{10395}} \,{\rm eulerlog}\left(4, \,v\right) + {\displaystyle \frac {896}{5}} \,\ln\left(2\right) - {\displaystyle \frac {256}{3}} \,\ln\left(2\right)^{2} + {\displaystyle \frac {3160094713}{14407470}} \right.\cr
&&
\left. 
+ {\displaystyle \frac {166802}{4455}} \,{c_\theta}^{2} - {\displaystyle \frac {4304}{405}} \,{c_\theta}^{4} + {\displaystyle \frac {64}{189}} \,{c_\theta}^{6}\right]\cr
&&
 + {s_\theta}^{3}\,{c_\theta}\,\sin\left(5\,\psi \right) \left[{\displaystyle \frac {333911875}{1161216}}  - {\displaystyle \frac {509375}{2304}} \,\ln\left({\displaystyle \frac {5}{2}} \right) + \left( - {\displaystyle \frac {86196875}{1161216}}  + {\displaystyle \frac {109375}{2304}} \,\ln\left({\displaystyle \frac {5}{2}} \right)\right)\,{c_\theta}^{2}\right] \cr
&&
 + {s_\theta}^{4}\,{c_\theta}\,\sin\left(6\,\psi \right)\,\left[ - {\displaystyle \frac {1931607}{24640}}  + {\displaystyle \frac {68769}{1120}} \,{c_\theta}^{2} - {\displaystyle \frac {2187}{448}} \,{c_\theta}^{4}\right] \cr
&&
 + {s_\theta}^{5}\,{c_\theta}\,\sin\left(7\,\psi \right)\,\left[{\displaystyle \frac {485772721}{4147200}}  - {\displaystyle \frac {823543}{11520}} \,\ln\left({\displaystyle \frac {7}{2}} \right)\right] \cr
&&
 + {s_\theta}^{6}\,{c_\theta}\,\sin\left(8\,\psi \right)\,\left[ - {\displaystyle \frac {195584}{2835}}  + {\displaystyle \frac {8192}{567}} \,{c_\theta}^{2}\right] - {\displaystyle \frac {390625}{36288}} \,{s_\theta}^{8}\,{c_\theta}\,\sin\left(10\,\psi \right) ,
\end{eqnarray}
\begin{eqnarray}
{H_{\times}^{(4.5)}}&=&
{s_\theta}\,{c_\theta}\,\cos\left(\psi \right)\left[ - {\displaystyle \frac {11617}{10080}} \,\ln\left(2\right) - {\displaystyle \frac {2435639}{2162160}} \,{\rm eulerlog}\left(1, \,v\right) + {\displaystyle \frac {7}{64}} \,\pi ^{2} \right.
 - {\displaystyle \frac {21}{16}} \,\ln\left(2\right)^{2} - {\displaystyle \frac {731286776414651}{398927860531200}} \cr
&&
  + \left({\displaystyle \frac {5}{48}} \,\ln\left(2\right)^{2} - {\displaystyle \frac {5}{576}} \,\pi ^{2} + {\displaystyle \frac {341}{7280}} \,{\rm eulerlog}\left(1, \,v\right) - {\displaystyle \frac {136508869499}{604436152320}}  + {\displaystyle \frac {251}{1120}} \,\ln\left(2\right)\right){c_\theta}^{2} \cr
&&
\left. - {\displaystyle \frac {358093861}{22140518400}} \,{c_\theta}^{4} + {\displaystyle \frac {5927}{33177600}} \,{c_\theta}^{6} - {\displaystyle \frac {11}{44236800}} \,{c_\theta}^{8}\right]\cr
&&
 + {c_\theta} \cos\left(2\,\psi \right) 
\left[{\displaystyle \frac {1712}{63}} \,\pi ^{2} - {\displaystyle \frac {128}{3}} \,\zeta \left(3\right) + {\displaystyle \frac {7787}{25200}}  - {\displaystyle \frac {377687}{28350}} \,{c_\theta}^{2} + {\displaystyle \frac {200021}{32400}} \,{c_\theta}^{4} - {\displaystyle \frac {16619}{113400}} \,{c_\theta}^{6}\right] \cr
&&
 + {s_\theta}\,{c_\theta}\,\cos\left(3\,\psi \right)\left[{\displaystyle \frac {779604098799429}{1641678438400}}  + {\displaystyle \frac {1809}{128}} \,\pi ^{2} \right.
 - {\displaystyle \frac {16500051}{160160}} \,{\rm eulerlog}\left(3, \,v\right) + {\displaystyle \frac {110403}{448}} \,\ln\left({\displaystyle \frac {3}{2}} \right) - {\displaystyle \frac {5427}{32}} \,\ln\left({\displaystyle \frac {3}{2}} \right)^{2}\cr
&&
 + \left( - {\displaystyle \frac {5929269115101}{82083921920}}  - {\displaystyle \frac {195291}{2240}} \,\ln\left({\displaystyle \frac {3}{2}} \right) + {\displaystyle \frac {2553471}{160160}} \,{\rm eulerlog}\left(3, \,v\right) - {\displaystyle \frac {405}{128}} \,\pi ^{2} \right.\left. + {\displaystyle \frac {1215}{32}} \,\ln\left({\displaystyle \frac {3}{2}} \right)^{2}\right){c_\theta}^{2} \cr
&&
\left.
- {\displaystyle \frac {7442184177}{820019200}} \,{c_\theta}^{4} + {\displaystyle \frac {2413233}{2867200}} \,{c_\theta}^{6} - {\displaystyle \frac {24057}{2293760}} \,{c_\theta}^{8}\right] \cr
&&
 + {s_\theta}^{2}\,{c_\theta}\,\cos\left(4\,\psi \right)\left[ - {\displaystyle \frac {14248}{105}} \,\ln\left(2\right) + {\displaystyle \frac {87718}{567}}  + \left({\displaystyle \frac {9664}{63}} \,\ln\left(2\right) - {\displaystyle \frac {204416}{945}} \right)\,{c_\theta}^{2} \right.\left. + \left({\displaystyle \frac {269888}{14175}}  - {\displaystyle \frac {512}{45}} \,\ln\left(2\right)\right)\,{c_\theta}^{4}\right] \cr
&&
+ {s_\theta}^{3}\,{c_\theta}\,\cos\left(5\,\psi \right)\left[ - {\displaystyle \frac {2415625}{41184}} \,{\rm eulerlog}\left(5, \,v\right) \right.
 + {\displaystyle \frac {565625}{1344}} \,\ln\left({\displaystyle \frac {5}{2}} \right) + {\displaystyle \frac {15625}{1152}} \,\pi ^{2} - {\displaystyle \frac {15625}{96}} \,{\rm ln}\left({\displaystyle \frac {5}{2}} \right)^{2} \cr
&&
\left. + {\displaystyle \frac {132078703275625}{506575060992}} + {\displaystyle \frac {968461990625}{10627448832}} \,{c_\theta}^{2} - {\displaystyle \frac {1795859375}{74317824}} \,{c_\theta}^{4} + {\displaystyle \frac {21484375}{24772608}} \,{c_\theta}^{6}\right] \cr
&&
 + {s_\theta}^{4}\,{c_\theta}\,\cos\left(6\,\psi \right)\,\left[{\displaystyle \frac {1633689}{2800}}  - {\displaystyle \frac {27459}{70}} \,\ln\left(3\right) + \left({\displaystyle \frac {2916}{35}} \,\ln\left(3\right) - {\displaystyle \frac {199989}{1400}} \right)\,{c_\theta}^{2}\right] \cr
&&
 + {s_\theta}^{5}\,{c_\theta}\,\cos\left(7\,\psi \right) 
\left[ - {\displaystyle \frac {165535790119}{1150156800}}  + {\displaystyle \frac {13627166021}{132710400}} \,{c_\theta}^{2} - {\displaystyle \frac {443889677}{53084160}} \,{c_\theta}^{4}\right] \cr
&&
 + {s_\theta}^{6}\,{c_\theta}\,\cos\left(8\,\psi \right)\,\left[{\displaystyle \frac {2613248}{14175}}  - {\displaystyle \frac {65536}{315}} \,\ln\left(2\right)\right] 
 + {s_\theta}^{7}\,{c_\theta}\,\cos\left(9\,\psi \right)\,\left[ - {\displaystyle \frac {90876411}{917504}}  + {\displaystyle \frac {473513931}{22937600}} \,{c_\theta}^{2}\right] \cr
&&
 - {\displaystyle \frac {25937424601}{1857945600}} \,{s_\theta}^{9}\,{c_\theta}\,\cos\left(11\,\psi \right) \cr
&&
 + \pi \,{s_\theta}\,{c_\theta}\,\sin\left(\psi \right)\,\left[ - {\displaystyle \frac {22483}{1729728}}  - {\displaystyle \frac {21}{16}} \,\ln\left(2\right) + \left({\displaystyle \frac {2581}{29120}}  + {\displaystyle \frac {5}{48}} \,\ln\left(2\right)\right)\,{c_\theta}^{2}\right] \cr
&&
 + \pi {c_\theta}\,\sin\left(2\,\psi \right) 
\left[{\displaystyle \frac {8}{3}} \,\pi ^{2} + {\displaystyle \frac {3424}{105}} \,{\rm eulerlog}\left(2, \,v\right) - {\displaystyle \frac {3620761}{22050}}  - {\displaystyle \frac {3413}{630}} \,{c_\theta}^{2} + {\displaystyle \frac {2909}{1260}} \,{c_\theta}^{4} - {\displaystyle \frac {2}{45}} \,{c_\theta}^{6}\right] \cr
&&
 + \pi \,{s_\theta}\,{c_\theta}\,\sin\left(3\,\psi \right) 
\left[ - {\displaystyle \frac {45938043}{640640}}  + {\displaystyle \frac {5427}{32}} \,\ln\left({\displaystyle \frac {3}{2}} \right) + \left( - {\displaystyle \frac {1215}{32}} \,\ln\left({\displaystyle \frac {3}{2}} \right) + {\displaystyle \frac {3259953}{91520}} \right)\,{c_\theta}^{2}\right] \cr
&&
 + \pi \,{s_\theta}^{2}\,{c_\theta}\,\sin\left(4\,\psi \right)\,\left[{\displaystyle \frac {7124}{105}}  - {\displaystyle \frac {4832}{63}} \,{c_\theta}^{2} + {\displaystyle \frac {256}{45}} \,{c_\theta}^{4}\right] \cr
&&
 + \pi \,{s_\theta}^{3}\,{c_\theta}\,\sin\left(5\,\psi \right)\,\left[ - {\displaystyle \frac {208834375}{1153152}}  + {\displaystyle \frac {15625}{96}} \,\ln\left({\displaystyle \frac {5}{2}} \right)\right] \cr
&&
 + \pi \,{s_\theta}^{4}\,{c_\theta}\,\sin\left(6\,\psi \right)\,\left[{\displaystyle \frac {27459}{140}}  - {\displaystyle \frac {1458}{35}} \,{c_\theta}^{2}\right] + {\displaystyle \frac {16384}{315}} \,\pi \,{s_\theta}^{6}\,{c_\theta}\,\sin\left(8\,\psi \right) ,
\end{eqnarray}
\begin{eqnarray}
{H_{\times}^{(5)}}&=&
\pi \,{s_\theta}\,{c_\theta}\,\cos\left(\psi \right)\left[{\displaystyle \frac {3}{2}} \,\ln\left(2\right)^{2} - {\displaystyle \frac {2137436827}{270950400}}  - {\displaystyle \frac {59}{105}} \,\ln\left(2\right) + {\displaystyle \frac {1}{8}} \,\pi ^{2} \right.+ {\displaystyle \frac {307}{210}} \,{\rm eulerlog}\left(1, \,v\right) \cr
&&
\left. 
- {\displaystyle \frac {436553}{2580480}} \,{c_\theta}^{2} + {\displaystyle \frac {2291}{286720}} \,{c_\theta}^{4} - {\displaystyle \frac {1}{40960}} \,{c_\theta}^{6}\right] \cr
&&
 + \pi \,{c_\theta}\,\cos\left(2\,\psi \right)\,\left[ - {\displaystyle \frac {11456867}{540540}}  + {\displaystyle \frac {5135479}{270270}} \,{c_\theta}^{2} - {\displaystyle \frac {412037}{180180}} \,{c_\theta}^{4}\right] \cr
&&
 + \pi \,{s_\theta}\,{c_\theta}\cos\left(3\,\psi \right)\left[{\displaystyle \frac {243}{2}} \,{\rm ln}\left({\displaystyle \frac {3}{2}} \right)^{2} + {\displaystyle \frac {81}{8}} \,\pi ^{2} - {\displaystyle \frac {3321}{35}} \,\ln\left({\displaystyle \frac {3}{2}} \right) + {\displaystyle \frac {1053}{14}} \,{\rm eulerlog}\left(3, \,v\right) \right.- {\displaystyle \frac {253780587}{501760}} \cr
&&
\left. 
- {\displaystyle \frac {5382207}{143360}} \,{c_\theta}^{2} + {\displaystyle \frac {21627}{1792}} \,{c_\theta}^{4} - {\displaystyle \frac {6561}{20480}} \,{c_\theta}^{6}\right]\cr
&&
 + \pi \,{s_\theta}^{2}\,{c_\theta} \cos\left(4\,\psi \right)\,\left[{\displaystyle \frac {1408}{3}} \,\ln\left(2\right) - {\displaystyle \frac {53183264}{135135}}  + \left( - {\displaystyle \frac {512}{5}} \,\ln\left(2\right) + {\displaystyle \frac {5600512}{45045}} \right)\,{c_\theta}^{2}\right] \cr
&&
 + \pi \,{s_\theta}^{3}\,{c_\theta}\,\cos\left(5\,\psi \right)\,\left[{\displaystyle \frac {453125}{2304}}  - {\displaystyle \frac {45578125}{258048}} \,{c_\theta}^{2} + {\displaystyle \frac {390625}{28672}} \,{c_\theta}^{4}\right] \cr
&&
 + \pi \,{s_\theta}^{4}\,{c_\theta}\,\cos\left(6\,\psi \right)\,\left[ - {\displaystyle \frac {7776729}{20020}}  + {\displaystyle \frac {1458}{5}} \,\ln\left(3\right)\right] 
 + \pi \,{s_\theta}^{5}\,{c_\theta}\,\cos\left(7\,\psi \right)\,\left[{\displaystyle \frac {246239357}{737280}}  - {\displaystyle \frac {5764801}{81920}} \,{c_\theta}^{2}\right] \cr
&&
 + {\displaystyle \frac {43046721}{573440}} \,\pi \,{s_\theta}^{7}\,{c_\theta}\,\cos\left(9\,\psi \right) \cr
&&
+ {s_\theta}\,{c_\theta}\,\sin\left(\psi \right)\left[2\,\zeta \left(3\right) - {\rm ln}\left(2\right)^{3} \right.
 - {\displaystyle \frac {2881}{2520}} \,\pi ^{2} - {\displaystyle \frac {307}{105}} \,\ln\left(2\right)\,{\rm eulerlog}\left(1, \,v\right) + {\displaystyle \frac {1470466037}{331161600}}  \cr
&&
 - {\displaystyle \frac {9}{10}} \,\ln\left(2\right)^{2}
 + {\displaystyle \frac {2025831067}{135475200}} \,\ln\left(2\right) - {\displaystyle \frac {173}{210}} \,{\rm eulerlog}\left(1, \,v\right) + {\displaystyle \frac {1}{4}} \,\ln\left(2\right)\,\pi ^{2} \cr
&&
 + \left({\displaystyle \frac {436553}{1290240}} \,\ln\left(2\right) + {\displaystyle \frac {1120328171}{2980454400}} \right)\,{c_\theta}^{2} 
 + \left( - {\displaystyle \frac {71012509}{2980454400}}  - {\displaystyle \frac {2291}{143360}} \,\ln\left(2\right)\right)\,{c_\theta}^{4} \cr
&&
\left.+ \left({\displaystyle \frac {36899}{425779200}}  + {\displaystyle \frac {1}{20480}} \,\ln\left(2\right)\right)\,{c_\theta}^{6}\right] \cr
&&
 + {c_\theta}\,\sin\left(2\,\psi \right)\left[{\displaystyle \frac {125273231828323}{779155977600}}  + {\displaystyle \frac {34}{45}} \,\pi ^{2} - {\displaystyle \frac {313384}{19305}} \,{\rm eulerlog}\left(2, \,v\right) \right.\cr
&&
 + \left({\displaystyle \frac {12035301795191}{194788994400}}  + {\displaystyle \frac {113}{45}} \,\pi ^{2} - {\displaystyle \frac {2198276}{135135}} \,{\rm eulerlog}\left(2, \,v\right)\right)\,{c_\theta}^{2} \cr
&&
 + \left( - {\displaystyle \frac {1}{6}} \,\pi ^{2} + {\displaystyle \frac {31862}{45045}} \,{\rm eulerlog}\left(2, \,v\right) - {\displaystyle \frac {198525061333}{129859329600}} \right)\,{c_\theta}^{4}\cr
&&
\left. - {\displaystyle \frac {24734341}{64864800}} \,{c_\theta}^{6} + {\displaystyle \frac {98573}{9979200}} \,{c_\theta}^{8} - {\displaystyle \frac {1}{25200}} \,{c_\theta}^{10}\right] \cr
&&
+ {s_\theta}\,{c_\theta}\,\sin\left(3\,\psi \right)\left[{\displaystyle \frac {18081675159}{27596800}} \right. 
 - {\displaystyle \frac {1701}{10}} \,\ln\left({\displaystyle \frac {3}{2}} \right)^{2} + {\displaystyle \frac {1053}{7}} \,{\rm eulerlog}\left(3, \,v\right)\,\ln\left({\displaystyle \frac {3}{2}} \right) + 162\,\zeta \left(3\right)\cr
&&
 - {\displaystyle \frac {13581}{280}} \,\pi ^{2} 
 - {\displaystyle \frac {1053}{10}} \,{\rm eulerlog}\left(3, \,v\right) - {\displaystyle \frac {227362923}{250880}} \,\ln\left({\displaystyle \frac {3}{2}} \right) + 81\,\ln\left({\displaystyle \frac {3}{2}} \right)^{3} - {\displaystyle \frac {81}{4}} \,\pi ^{2}\,\ln\left({\displaystyle \frac {3}{2}} \right) \cr
&&
 + \left({\displaystyle \frac {5493794949}{55193600}}  - {\displaystyle \frac {5382207}{71680}} \,\ln\left({\displaystyle \frac {3}{2}} \right)\right)\,{c_\theta}^{2} + \left({\displaystyle \frac {21627}{896}} \,\ln\left({\displaystyle \frac {3}{2}} \right) - {\displaystyle \frac {254813931}{6899200}} \right)\,{c_\theta}^{4} \cr
&&
\left. + \left({\displaystyle \frac {9024291}{7884800}}  - {\displaystyle \frac {6561}{10240}} \,\ln\left({\displaystyle \frac {3}{2}} \right)\right)\,{c_\theta}^{6}\right]\cr
&&
 + {s_\theta}^{2}\,{c_\theta}\,\sin\left(4\,\psi \right)\left[ - {\displaystyle \frac {352}{9}} \,\pi ^{2} \right. 
 + {\displaystyle \frac {28469888}{135135}} \,{\rm eulerlog}\left(4, \,v\right) + {\displaystyle \frac {1408}{3}} \,\ln\left(2\right)^{2} - {\displaystyle \frac {2667962875727}{2782699920}} 
 - {\displaystyle \frac {104768}{105}} \,\ln\left(2\right)  \cr
&&
+ \left( - {\displaystyle \frac {307712}{9009}} \,{\rm eulerlog}\left(4, \,v\right) - {\displaystyle \frac {512}{5}} \,\ln\left(2\right)^{2} + {\displaystyle \frac {29696}{105}} \,\ln\left(2\right) \right.
\left. + {\displaystyle \frac {128}{15}} \,\pi ^{2} + {\displaystyle \frac {195652128227}{2029052025}} \right){c_\theta}^{2} \cr
&&
\left.+ {\displaystyle \frac {20597648}{675675}} \,{c_\theta}^{4} - {\displaystyle \frac {69536}{22275}} \,{c_\theta}^{6} + {\displaystyle \frac {16}{315}} \,{c_\theta}^{8} \right]\cr
&&
 + {s_\theta}^{3}\,{c_\theta}\,\sin\left(5\,\psi \right)\left[{\displaystyle \frac {453125}{1152}} \,\ln\left({\displaystyle \frac {5}{2}} \right) - {\displaystyle \frac {23517203125}{44706816}}  \right.
 + \left( - {\displaystyle \frac {45578125}{129024}} \,{\rm ln}\left({\displaystyle \frac {5}{2}} \right) + {\displaystyle \frac {100077521875}{178827264}} \right)\,{c_\theta}^{2} \cr
&&
\left. + \left({\displaystyle \frac {390625}{14336}} \,\ln\left({\displaystyle \frac {5}{2}} \right) - {\displaystyle \frac {2938515625}{59609088}} \right)\,{c_\theta}^{4}\right]\cr
&&
 + {s_\theta}^{4}\,{c_\theta}\,\sin\left(6\,\psi \right)\left[ - {\displaystyle \frac {243}{10}} \,\pi ^{2} \right. 
 + {\displaystyle \frac {1458}{5}} \,\ln\left(3\right)^{2} - {\displaystyle \frac {140126622651}{641280640}}  + {\displaystyle \frac {437886}{5005}} \,{\rm eulerlog}\left(6, \,v\right) - {\displaystyle \frac {60507}{70}} \,\ln\left(3\right) \cr
&&
\left. - {\displaystyle \frac {45043209}{228800}} \,{c_\theta}^{2} + {\displaystyle \frac {12323259}{246400}} \,{c_\theta}^{4} - {\displaystyle \frac {2187}{1120}} \,{c_\theta}^{6}\right] \cr
&&
+ {s_\theta}^{5}\,{c_\theta}\,\sin\left(7\,\psi \right)
\left[{\displaystyle \frac {246239357}{368640}} \,\ln\left({\displaystyle \frac {7}{2}} \right) - {\displaystyle \frac {400350017431}{364953600}}  + \left({\displaystyle \frac {31563932561}{121651200}}  - {\displaystyle \frac {5764801}{40960}} \,{\rm ln}\left({\displaystyle \frac {7}{2}} \right)\right)\,{c_\theta}^{2} \right]\cr
&&
 + {s_\theta}^{6}\,{c_\theta}\,\sin\left(8\,\psi \right)\,\left[{\displaystyle \frac {39104768}{155925}}  - {\displaystyle \frac {26046464}{155925}} \,{c_\theta}^{2} + {\displaystyle \frac {65536}{4725}} \,{c_\theta}^{4}\right] \cr
&&
 + {s_\theta}^{7}\,{c_\theta}\,\sin\left(9\,\psi \right)\,\left[ - {\displaystyle \frac {62623413117}{220774400}}  + {\displaystyle \frac {43046721}{143360}} \,\ln\left(3\right) - {\displaystyle \frac {43046721}{286720}} \,\ln\left(2\right)\right] \cr
&&
 + {s_\theta}^{8}\,{c_\theta}\,\sin\left(10\,\psi \right)\,\left[{\displaystyle \frac {112890625}{798336}}  - {\displaystyle \frac {1953125}{66528}} \,{c_\theta}^{2}\right] \cr
&&
 + {\displaystyle \frac {34992}{1925}} \,{s_\theta}^{10}\,{c_\theta}\,\sin\left(12\,\psi \right) ,
\end{eqnarray}
\begin{eqnarray}
{H_{\times}^{(5.5)}}&=&
{s_\theta}\,{c_\theta}\,\cos\left(\psi \right)\left[ - {\displaystyle \frac {44966609055228224561}{4611606067740672000}}  - {\displaystyle \frac {913}{46080}} \,\pi ^{2} + {\displaystyle \frac {913}{3840}} \,\ln\left(2\right)^{2} \right.\cr&&
 - {\displaystyle \frac {13785043}{210038400}} \,{\rm eulerlog}\left(1, \,v\right) + {\displaystyle \frac {4253489}{4838400}} \,\ln\left(2\right) \cr
&&
+ \left( - {\displaystyle \frac {5164843}{7257600}} \,\ln\left(2\right) \right.
 - {\displaystyle \frac {1891}{5760}} \,\ln\left(2\right)^{2} + {\displaystyle \frac {2289618368599426463}{4611606067740672000}}  + {\displaystyle \frac {1891}{69120}} \,\pi ^{2} 
\left. - {\displaystyle \frac {3991213}{27227200}} \,{\rm eulerlog}\left(1, \,v\right)\right){c_\theta}^{2} \cr
&&
+ \left({\displaystyle \frac {267271}{294053760}} \,{\rm eulerlog}\left(1, \,v\right) - {\displaystyle \frac {7}{27648}} \,\pi ^{2}\right. 
\left. + {\displaystyle \frac {7}{2304}} \,\ln\left(2\right)^{2} + {\displaystyle \frac {5542529007807251}{461160606774067200}}  + {\displaystyle \frac {26263}{2903040}} \,\ln\left(2\right)\right){c_\theta}^{4} \cr
&&
\left. - {\displaystyle \frac {374959007}{694871654400}} \,{c_\theta}^{6} + {\displaystyle \frac {190121}{81749606400}} \,{c_\theta}^{8} - {\displaystyle \frac {13}{7431782400}} \,{c_\theta}^{10}\right]\cr
&&
 + {c_\theta} \cos\left(2\,\psi \right)\left[{\displaystyle \frac {3195371713}{48024900}}  + {\displaystyle \frac {1088}{9}} \,\zeta \left(3\right) - {\displaystyle \frac {449612}{6237}} \,\pi ^{2} - {\displaystyle \frac {29728}{3465}} \,{\rm eulerlog}\left(2, \,v\right) \right.\cr
&&
 + \left( - {\displaystyle \frac {646758699649}{5762988000}}  + {\displaystyle \frac {345376}{17325}} \,{\rm eulerlog}\left(2, \,v\right) - {\displaystyle \frac {256}{9}} \,\zeta \left(3\right) + {\displaystyle \frac {228764}{31185}} \,\pi ^{2}\right)\,{c_\theta}^{2} \cr
&&
\left. - {\displaystyle \frac {825921}{78400}} \,{c_\theta}^{4} + {\displaystyle \frac {10107527}{10478160}} \,{c_\theta}^{6} - {\displaystyle \frac {29171}{3326400}} \,{c_\theta}^{8}\right]\cr
&&
 + {s_\theta}\,{c_\theta}\,\cos\left(3\,\psi \right)
\left[ - {\displaystyle \frac {37503}{5120}} \,\pi ^{2} + {\displaystyle \frac {3926864534929543197}{75911210991616000}}  - {\displaystyle \frac {31284549}{179200}} \,\ln\left({\displaystyle \frac {3}{2}} \right) + {\displaystyle \frac {112509}{1280}} \,\ln\left({\displaystyle \frac {3}{2}} \right)^{2} \right.\cr
&&
 + {\displaystyle \frac {2354400351}{54454400}} \,{\rm eulerlog}\left(3, \,v\right) + \left({\displaystyle \frac {36207}{2560}} \,\pi ^{2} + {\displaystyle \frac {19408843768367516823}{75911210991616000}}  \right.\cr
&&
\left. + {\displaystyle \frac {36267993}{89600}} \,\ln\left({\displaystyle \frac {3}{2}} \right) - {\displaystyle \frac {372580641}{5445440}} \,{\rm eulerlog}\left(3, \,v\right) - {\displaystyle \frac {108621}{640}} \,\ln\left({\displaystyle \frac {3}{2}} \right)^{2}\right){c_\theta}^{2}\cr
&&
 + \left( - {\displaystyle \frac {6488829}{179200}} \,\ln\left({\displaystyle \frac {3}{2}} \right) - {\displaystyle \frac {5103}{5120}} \,\pi ^{2} + {\displaystyle \frac {190312011}{54454400}} \,{\rm eulerlog}\left(3, \,v\right) + {\displaystyle \frac {15309}{1280}} \,\ln\left({\displaystyle \frac {3}{2}} \right)^{2} \right.\cr
&&
\left. + {\displaystyle \frac {59325223140785601}{37955605495808000}} \right){c_\theta}^{4} - {\displaystyle \frac {18541319661}{6862929920}} \,{c_\theta}^{6} + {\displaystyle \frac {414504297}{4037017600}} \,{c_\theta}^{8} 
\left. - {\displaystyle \frac {255879}{367001600}} \,{c_\theta}^{10}\right]\cr
&& + {s_\theta}^{2}\,{c_\theta}\,\cos\left(4\,\psi \right)\left[{\displaystyle \frac {4096}{9}} \,\zeta \left(3\right) \right. 
 - {\displaystyle \frac {804352}{2475}} \,{\rm eulerlog}\left(4, \,v\right) - {\displaystyle \frac {3584}{5}} \,\ln\left(2\right)^{2} + {\displaystyle \frac {3217408}{10395}} \,\ln\left(2\right)\,{\rm eulerlog}\left(4, \,v\right) \cr
&&
 + {\displaystyle \frac {2048}{9}} \,\ln\left(2\right)^{3} - {\displaystyle \frac {2158976}{31185}} \,\pi ^{2} - {\displaystyle \frac {512}{9}} \,\ln\left(2\right)\,\pi ^{2} - {\displaystyle \frac {12640378852}{7203735}} \,\ln\left(2\right)  + {\displaystyle \frac {43812580277}{20010375}} \cr
&&
+ \left({\displaystyle \frac {290708048}{654885}}  - {\displaystyle \frac {1334416}{4455}} \,\ln\left(2\right)\right)\,{c_\theta}^{2} 
 + \left( - {\displaystyle \frac {469895744}{3274425}}  + {\displaystyle \frac {34432}{405}} \,\ln\left(2\right)\right)\,{c_\theta}^{4} \cr
&&
\left.+ \left( - {\displaystyle \frac {512}{189}} \,\ln\left(2\right) + {\displaystyle \frac {5637568}{1091475}} \right)\,{c_\theta}^{6}\right]\cr
&&
 + {s_\theta}^{3} {c_\theta}\,\cos\left(5\,\psi \right)\left[ - {\displaystyle \frac {2546875}{27648}} \,\pi ^{2} + {\displaystyle \frac {3326215625}{8401536}} \,{\rm eulerlog}\left(5, \,v\right) \right.\cr
&&
 - {\displaystyle \frac {10344364790975800625}{7027209246081024}}  - {\displaystyle \frac {1669559375}{580608}} \,\ln\left({\displaystyle \frac {5}{2}} \right) + {\displaystyle \frac {2546875}{2304}} \,\ln\left({\displaystyle \frac {5}{2}} \right)^{2} \cr
&&
+ \left( - {\displaystyle \frac {3884921875}{58810752}} \,{\rm eulerlog}\left(5, \,v\right) + {\displaystyle \frac {430984375}{580608}} \,\ln\left({\displaystyle \frac {5}{2}} \right) + {\displaystyle \frac {6095596325790625}{288225379233792}}  \right.\cr
&&
\left. + {\displaystyle \frac {546875}{27648}} \,\pi ^{2} - {\displaystyle \frac {546875}{2304}} \,\ln\left({\displaystyle \frac {5}{2}} \right)^{2}\right){c_\theta}^{2} + {\displaystyle \frac {445893828125}{5294260224}} \,{c_\theta}^{4} 
\left. - {\displaystyle \frac {44912890625}{4904976384}} \,{c_\theta}^{6} + {\displaystyle \frac {634765625}{3567255552}} \,{c_\theta}^{8}\right]
\cr
&& + {s_\theta}^{4}\,{c_\theta}\,\cos\left(6\,\psi \right)
\left[{\displaystyle \frac {5794821}{6160}} \,\ln\left(3\right) - {\displaystyle \frac {2447860257}{1724800}}  + \left( - {\displaystyle \frac {206307}{280}} \,\ln\left(3\right) + {\displaystyle \frac {6280821}{4928}} \right)\,{c_\theta}^{2} \right.\cr
&&
\left. + \left( - {\displaystyle \frac {194738499}{1724800}}  + {\displaystyle \frac {6561}{112}} \,\ln\left(3\right)\right)\,{c_\theta}^{4}\right]\cr
&&
 + {s_\theta}^{5}\,{c_\theta}\,\cos\left(7\,\psi \right)\left[ - {\displaystyle \frac {5764801}{138240}} \,\pi ^{2} \right. 
 - {\displaystyle \frac {3400409047}{2073600}} \,{\rm ln}\left({\displaystyle \frac {7}{2}} \right) + {\displaystyle \frac {2459922941}{19094400}} \,{\rm eulerlog}\left(7, \,v\right) \cr
&&
+ {\displaystyle \frac {68859193174529009}{5703903608832000}}  
 + {\displaystyle \frac {5764801}{11520}} \,\ln\left({\displaystyle \frac {7}{2}} \right)^{2} - {\displaystyle \frac {42612989808613}{108291686400}} \,{c_\theta}^{2} + {\displaystyle \frac {6799703840321}{70071091200}} \,{c_\theta}^{4} \cr
&&
\left. - {\displaystyle \frac {25705247659}{6370099200}} \,{c_\theta}^{6}\right]\cr
&&
 + {s_\theta}^{6}\,{c_\theta}\,\cos\left(8\,\psi \right)\left[ - {\displaystyle \frac {6415888384}{3274425}}  + {\displaystyle \frac {6258688}{2835}} \,\ln\left(2\right) + \left({\displaystyle \frac {1480835072}{3274425}}  - {\displaystyle \frac {262144}{567}} \,\ln\left(2\right)\right)\,{c_\theta}^{2}\right] \cr
&&
 + {s_\theta}^{7}\,{c_\theta}\,\cos\left(9\,\psi \right) \left[{\displaystyle \frac {14460154075971}{34314649600}}  - {\displaystyle \frac {1678822119}{6307840}} \,{c_\theta}^{2} + {\displaystyle \frac {45328197213}{2018508800}} \,{c_\theta}^{4}\right] \cr
&&
 + {s_\theta}^{8}\,{c_\theta}\,\cos\left(10\,\psi \right)\,\left[{\displaystyle \frac {1953125}{9072}} \,\ln\left(5\right) - {\displaystyle \frac {7195703125}{16765056}} \right] \cr
&&
 + {s_\theta}^{9}\,{c_\theta}\,\cos\left(11\,\psi \right)\,\left[{\displaystyle \frac {17922760399291}{89181388800}}  - {\displaystyle \frac {3709051717943}{89181388800}} \,{c_\theta}^{2}\right] \cr
&&
 + {\displaystyle \frac {23298085122481}{980995276800}} \,{s_\theta}^{11}\,{c_\theta}\,\cos\left(13\,\psi \right) \cr
&&
+ \pi \,{s_\theta}\,{c_\theta}\,\sin\left(\psi \right)\left[{\displaystyle \frac {11112194167}{23524300800}}  \right.
\left. + {\displaystyle \frac {913}{3840}} \,\ln\left(2\right) + \left( - {\displaystyle \frac {9969427309}{35286451200}}  - {\displaystyle \frac {1891}{5760}} \,\ln\left(2\right)\right)\,{c_\theta}^{2} \right.\cr
&&
\left. + \left({\displaystyle \frac {8204407}{2016368640}}  + {\displaystyle \frac {7}{2304}} \,\ln\left(2\right)\right)\,{c_\theta}^{4}\right]\cr
&&
 + \pi \,{c_\theta}\,\sin\left(2\,\psi \right)\left[{\displaystyle \frac {342543430879}{1152597600}} \right. 
 - {\displaystyle \frac {68}{9}} \,\pi ^{2} - {\displaystyle \frac {915856}{10395}} \,{\rm eulerlog}\left(2, \,v\right) \cr
&&
 + \left( - {\displaystyle \frac {5395550969}{72037350}}  + {\displaystyle \frac {16}{9}} \,\pi ^{2} + {\displaystyle \frac {128096}{10395}} \,{\rm eulerlog}\left(2, \,v\right)\right)\,{c_\theta}^{2} - {\displaystyle \frac {634357}{166320}} \,{c_\theta}^{4} 
\left. + {\displaystyle \frac {6611}{22680}} \,{c_\theta}^{6} - {\displaystyle \frac {1}{432}} \,{c_\theta}^{8}\right]\cr
&&
 + \pi \,{s_\theta}\,{c_\theta}\,\sin\left(3\,\psi \right)\left[{\displaystyle \frac {57217535811}{871270400}}\right.
 - {\displaystyle \frac {112509}{1280}} \,\ln\left({\displaystyle \frac {3}{2}} \right) + \left( - {\displaystyle \frac {73264265343}{435635200}}  + {\displaystyle \frac {108621}{640}} \,\ln\left({\displaystyle \frac {3}{2}} \right)\right)\,{c_\theta}^{2} \cr
&&
\left. + \left({\displaystyle \frac {2035978173}{124467200}}  - {\displaystyle \frac {15309}{1280}} \,\ln\left({\displaystyle \frac {3}{2}} \right)\right)\,{c_\theta}^{4}\right]\cr
&&
 + \pi \,{s_\theta}^{2}\,{c_\theta}\,\sin\left(4\,\psi \right)\left[ - {\displaystyle \frac {1024}{3}} \,\ln\left(2\right)^{2} \right.
 + {\displaystyle \frac {5842432}{10395}} \,\ln\left(2\right) - {\displaystyle \frac {256}{9}} \,\pi ^{2} - {\displaystyle \frac {1608704}{10395}} \,{\rm eulerlog}\left(4, \,v\right) \cr
&&
\left. + {\displaystyle \frac {37453814458}{36018675}} 
+ {\displaystyle \frac {667208}{4455}} \,{c_\theta}^{2} - {\displaystyle \frac {17216}{405}} \,{c_\theta}^{4} + {\displaystyle \frac {256}{189}} \,{c_\theta}^{6}\right]\cr
&&
 + \pi \,{s_\theta}^{3}\,{c_\theta}\,\sin\left(5\,\psi \right)
\left[ - {\displaystyle \frac {2546875}{2304}} \,\ln\left({\displaystyle \frac {5}{2}} \right) + {\displaystyle \frac {3499894615625}{2822916096}}  + \left({\displaystyle \frac {546875}{2304}} \,\ln\left({\displaystyle \frac {5}{2}} \right) - {\displaystyle \frac {136354984375}{403273728}} \right)\,{c_\theta}^{2} \right]\cr
&&
 + \pi \,{s_\theta}^{4}\,{c_\theta}\,\sin\left(6\,\psi \right)\,\left[ - {\displaystyle \frac {5794821}{12320}}  + {\displaystyle \frac {206307}{560}} \,{c_\theta}^{2} - {\displaystyle \frac {6561}{224}} \,{c_\theta}^{4}\right] \cr
&&
 + \pi \,{s_\theta}^{5}\,{c_\theta}\,\sin\left(7\,\psi \right)\,\left[ - {\displaystyle \frac {5764801}{11520}} \,\ln\left({\displaystyle \frac {7}{2}} \right) + {\displaystyle \frac {692452248803}{916531200}} \right] \cr
&&
 + \pi \,{s_\theta}^{6}\,{c_\theta}\,\sin\left(8\,\psi \right)\,\left[ - {\displaystyle \frac {1564672}{2835}}  + {\displaystyle \frac {65536}{567}} \,{c_\theta}^{2}\right] \cr
&&
 - {\displaystyle \frac {1953125}{18144}} \,\pi \,{s_\theta}^{8}\,{c_\theta}\,\sin\left(10\,\psi \right) .
\end{eqnarray}
\end{subequations}
\end{widetext}
Note that the $\cos 5\psi$ term in $H_{+}^{(2.5)}$ and 
the $\sin 5\psi$ term in $H_{\times}^{(2.5)}$ in \cite{BFIS08} have been 
further simplified in our presentation above and are equivalent. 

\end{document}